\documentclass[]{aa}  

\usepackage{graphicx}
\usepackage{booktabs}
\usepackage{multirow}
\usepackage[varg]{txfonts}
\usepackage{subfigure}

\usepackage{mathtools} 
\usepackage{url}

\usepackage[english]{babel}
\usepackage{subeqnarray}

\usepackage{tabularx} 
\usepackage{array} 
\usepackage{textcomp} 
\usepackage{threeparttable}  
\usepackage{enumitem} 

\graphicspath{{Pictures/}} 

\PassOptionsToPackage{hyphens}{url}\usepackage[colorlinks=true,citecolor=blue,linkcolor=blue,urlcolor=blue,breaklinks=True]{hyperref} 
\makeatletter
\renewcommand*\aa@pageof{, page \thepage{} of \pageref*{LastPage}}
\makeatother

\begin{document}

   \title{Full inversion of solar relativistic electron events\\ measured by the Helios spacecraft}

   \author{D. Pacheco\inst{1}
          \and
          N. Agueda\inst{1}
          \and
          A. Aran\inst{1}
          \and
          B. Heber\inst{2}
          \and
          D. Lario\inst{3},\inst{4}
          }

   \institute{Dep. F\'{i}sica Qu\`{a}ntica i Astrof\'{i}sica, Institut de Ci\`{e}ncies del Cosmos (ICCUB), Universitat de Barcelona, Barcelona, Spain.\\
              \email{\href{mailto:dpacheco@fqa.ub.edu}{dpacheco@fqa.ub.edu}}\
         \and
             Institut f\"{u}r Experimentelle und Angewandte Physik, University of Kiel, Germany.
         \and
             The Johns Hopkins University, Applied Physics Laboratory, Laurel, Maryland, USA.
             \and Now at NASA Goddard Space Flight Center, Heliophysics Science Division, Greenbelt, Maryland, USA\\
             }

   \date{Received 25 October 2018; Accepted 9 February 2019}
   \titlerunning{Inversion of Helios NR electron events}
 
  \abstract
   {The Parker Solar Probe and the incoming Solar Orbiter mission will provide measurements of solar energetic particle (SEP) events at close heliocentric distances from the Sun. Up to present, the largest data set of SEP events in the inner heliosphere are the observations by the two Helios spacecraft.}
   {We re-visit a sample of 15 solar relativistic electron events measured by the Helios mission with the goal of better characterising the injection histories of solar energetic particles and their interplanetary transport conditions at heliocentric distances <1~AU.} 
   {The measurements provided by the E6 instrument on board Helios provide us with the electron directional distributions in eight different sectors that we use to infer the detailed evolution of the electron pitch-angle distributions. The results of a Monte Carlo interplanetary transport model, combined with a full inversion procedure, were used to fit the observed directional intensities in the 300\,--\,800~keV nominal energy channel. Unlike previous studies, we have considered both the energy and angular responses of the detector. This method allowed us to infer the electron release time profile at the source and determine the electron interplanetary transport conditions.}
   {We discuss the duration of the release time profiles and the values of the radial mean free path, and compare them with the values reported previously in the literature using earlier approaches. Five of the events show short injection histories ($<$30~min) at the Sun and ten events show long-lasting ($>$30~min) injections. The values of mean free path range from 0.02~AU to 0.27~AU.}
   {The inferred injection histories match with the radio and soft x-ray emissions found in literature. We find no dependence of the radial mean free path on the radial distance. In addition, we find no apparent relation between the strength of interplanetary scattering and the size of the solar particle release.}

   \maketitle
%

\section{Introduction}

Solar energetic particle (SEP) events observed in the heliosphere show a wide variety of spatial and temporal extents, as well as a broad diversity of particle intensities and ion composition. The observed SEP intensities at a given point in the heliosphere are shaped by both the injection history of SEPs at their acceleration site, and the propagation processes undergone by the particles as they travel along the interplanetary magnetic field (IMF), from their source to the particle detectors on board spacecraft. Fortunately, both intensity-time profiles and particle angular distributions measured relative to the local direction of the magnetic field, that is pitch-angle distributions (PADs), provide us with the necessary imprint to unveil the contribution of the particle injection and transport processes in SEP events \citep[e.g.][and references therein]{Aran2018}. 

Particle experiments on board spacecraft use either multiple fields-of-view or the rotation of the spacecraft to measure PADs. With most experiments it is not possible to observe the whole range of directions and hence a portion of the PADs might be missed. To overcome this problem, a simultaneous analysis of the omni-directional particle intensities and the computed anisotropy time profiles was traditionally used in the analysis of SEP events \citep[e.g.][]{Droge93,Kallenrode92b,Kallenrode93b,Ruffolo98}. More recently \citep{Agueda08} the most direct form of data, that is the measured directional distributions, have been included in the analysis together with the angular response of the experiment.

Helios, launched in the 1970s, was the first solar mission to explore the inner heliosphere \citep{Porsche75} and it consisted of two spacecraft, Helios 1 and Helios 2. At present, it is still the mission that has provided the largest data set of SEP events measured between 0.29 AU and 0.98 AU \citep[e.g.][]{Heliosbook90,Heliosbook91,Re96,Lario06}. The data set is particularly interesting not only because it provides examples of SEP events observed close to the particle acceleration site (for which interplanetary transport could play a small role), but also because it is now possible to study relativistic electron events with detailed information about the instrumental response of the particle experiment, that is 1) the energy response of the detector, as the geometrical factor strongly depends on the particle energy \citep{Bialk91} and, 2) the angular response of a sector scanned by a spinning particle detector \citep{Agueda08}.

We have re-visited an extensive sample of relativistic electron events observed by Helios in order to get further insight on the solar release time histories of SEPs in the low corona and their interplanetary transport conditions. We used a Monte Carlo transport model \citep{Agueda08} that solves the focused transport of particles parallel to the IMF \citep{Roelof69} assuming a Parker spiral configuration. Since the two Helios spacecraft were in the inner heliosphere, where focusing effects dominate, and solar relativistic electron events evolve very fast, other processes such as perpendicular transport \citep[e.g.][]{Droge10,Droge16,StraussEtal17} are neglected. It is worth noting that the role of the different processes that may describe the particle transport across IMF lines during the evolution of SEP events remains unclear, as discussed by \cite{StraussEtal17}.

Previous works \citep{Kallenrode92b,Kallenrode93b,Agueda16} have made use of similar transport models to study electron events measured by the Helios spacecraft. However, in the time between these studies, the modelling techniques to analyse the data have evolved substantially. The first studies fitted omni-directional intensities and anisotropy profiles using a forward modelling approach, by assuming a prompt solar particle injection of a delta function form or a Reid-Axford profile \citep{Kallenrode92b,Kallenrode93b}. In contrast, \cite{Agueda16} fitted the observed PADs by using an inverse modelling approach, which implies that non a priori assumptions are made regarding the temporal profile of the solar particle release \citep{Agueda08}. This inversion model, highly versatile, has been also applied to study solar near-relativistic events observed by other spacecraft such as ACE, Wind, Ulysses, and STEREO \citep[e.g.][]{Agueda12,Agueda14,Pacheco17}. In this paper, for the first time, we consider both the angular and energy responses of the Cosmic Ray Experiment (E6) of Helios in order to model the SEP events. This update provides a complete full inversion methodology that can be further exploited for the interpretation of data from the Parker Solar Probe and Solar Orbiter.

In Sect. 2 we describe the Helios instrumentation used in this work. In Sect. 3 we present a systematic study of a sample of 15 solar relativistic electron events measured by the Helios spacecraft. The full inversion approach used to fit the observations is described in Sect. 4 and the results are presented and discussed in Sects. 5 and 6, respectively. Finally, Sect. 7 gives the conclusions of this work.

\section{Observations}
In this work we use electron measurements of the E03 channel of the Cosmic Ray Experiment (E6) on board Helios \citep{Kunow75, Kunow77}, whose nominal energy range is 0.3\,--\,0.8~MeV. So far, Helios measurements are the only source available of electron events detected at the inner heliosphere with sectored data allowing us to reconstruct the PADs.

E6 was designed at the University of Kiel to study solar energetic particles. The semiconductor detector configuration had a nominal geometrical factor of $\sim$0.48 cm$^{2}$~sr and a nominal full opening angle of $55^{\circ}$. \cite{Bialk91} presented a Monte Carlo simulation of the response of the E6 detectors to a wide energy-range of protons, $\gamma$-rays and electrons, taking into account the geometry, energy resolution and electronics of the detector. Their simulations show that the response function for protons was in good agreement with the nominal geometrical factors. On the other hand, the electron energy response function was much wider, considerably overlapping the different electron channels, whose nominal energies were 0.3\,--\,0.8~MeV (E03), 0.8\,--\,2.0~MeV (E08), 2.0\,--\,3.0~MeV (E2), 3.0\,--\,4.0~MeV (E3). \cite{Bialk91} showed that the energy response of the E03 channel extended up to $\sim$5~MeV. Furthermore, it was found that electron channels were sensitive to protons especially in the higher energy channels. The proton contamination in the E08 electron channel was found to be much more significant than for the E03, which can be easily avoided taking periods with low-proton intensities. For that reason, together with the fact that E08 was affected by more data gaps, we decided to use only E03 data in the present study.

The E6 experiment used the rotation of the spacecraft to measure the particle angular distributions relative to the local direction of the magnetic field (i.e. the PADs) of SEPs in interplanetary space. The particle pitch angle, $\alpha$, is defined as the angle between the magnetic field vector and the particle velocity. The cosine of $\alpha$, $\mu$, is then given by the scalar product of the unit vector along the magnetic field direction, $\hat{B}$, and the velocity unit vector, $\hat{v}$, of the particle, $\mu = \hat{B} \cdot \hat{v}$.

\begin{figure}[t]
\begin{centering}
\includegraphics[width=\columnwidth]{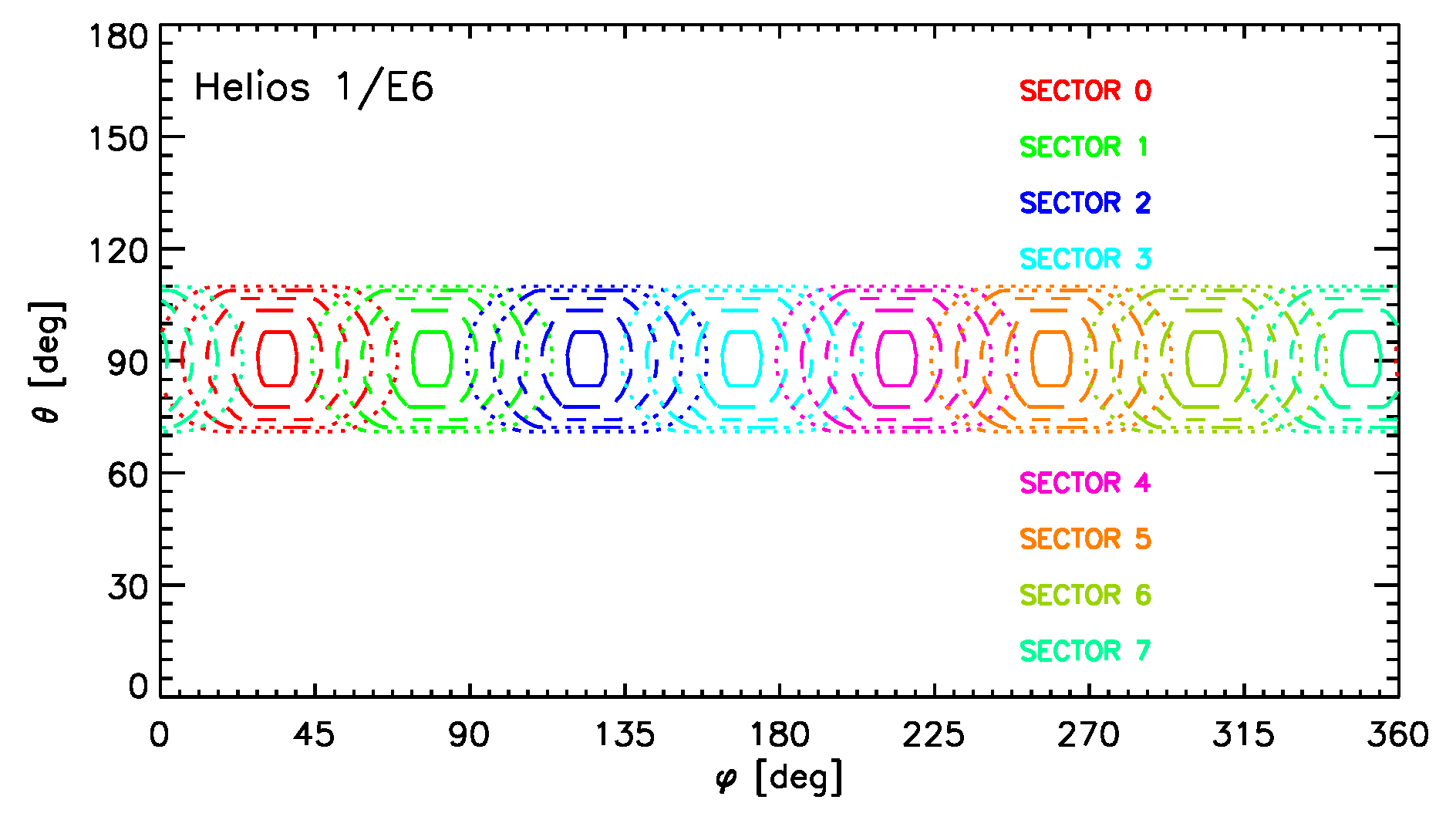}
\par\end{centering}
\caption{Angular response of the eight sectors of Helios 1/E6 defined in the SSE coordinate system. Each different contour level shows a decrease in 20\% of detection rate. For Helios 2 the configuration would be the same but shifted $23^{\circ}$.}
\label{5fig:sector_response}
\end{figure}

The spinning of the spacecraft allows a single detector to scan different directions of the sky. The region of space swept by E6 during a complete spin of the spacecraft was divided into eight sectors. We modelled the E6 particle detector on board Helios as a conical aperture of half-width $20^{\circ}$. The nominal aperture of the detector was $25^{\circ}$ but we assumed a smaller value because the detector response function decays linearly due to edge effects. We used the Monte Carlo technique to model a set of particle trajectories drawn from an isotropic particle distribution and recorded how they would be seen by a rotating detector sweeping a $45^{\circ}$-wide clock angle sector. We used the methodology presented by \citet{Agueda08Tesis} and \citet{Agueda08} to construct the angular response of the E6 sectors. Table~\ref{5tab:defSec} shows, for both Helios 1 and Helios 2, the coordinates of the sector centre unit vectors, $\hat{s}$, in the spacecraft solar ecliptic (SSE)\footnote{This coordinate system uses the Earth mean ecliptic plane as XY-plane, being the $X$ axis the projection over XY-plane of the spacecraft-Sun line and takes the $Z$ axis perpendicular to them. \cite{Franz02} mention that this coordinate system points towards the ecliptic south pole. Comparison with magnetic field data provided by the NASA's NSSDCA Archive (see below) in both the SSE and the radial tangential normal (RTN) coordinate system shows that $B_N$=$B_{Z_(\rm{SSE})}$, thus indicating that Z$_{\rm{SSE}}$ points towards the ecliptic north.} coordinate system \citep[e.g.][]{Franz02}, where $\theta$ is the colatitude and $\varphi$ is the azimuth. The $Z$ axis corresponds to $\theta = 0^{\circ}$ and is perpendicular to the ecliptic plane. The azimuth origin is the spacecraft-to-Sun line. Sectors are labelled from 0 to 7. Figure~\ref{5fig:sector_response} shows the angular response function of the eight sectors of E6 defined in the SSE coordinate system. Different style line for each sector shows a 20\% difference in detection rate. We note that the sector response is not a boxcar function, but it peaks at the zenith of the midpoint clock-angle of each sector. 

\begin{table}[t]
\centering
\caption{Pointing directions of the sectors of the E6 experiment on board the Helios probes.}
\vspace{1ex}
\label{5tab:defSec}
\begin{tabular}{cccccc}
\hline\hline
 & \multicolumn{2}{c}{\textbf{Helios 1}} & &\multicolumn{2}{c}{\textbf{Helios 2}}\\
\textbf{Sector} & \multicolumn{2}{c}{$\mathbf{\hat{s}}$} & &\multicolumn{2}{c}{$\mathbf{\hat{s}}$}\\
\cmidrule(lr){2-3} \cmidrule(l){5-6} 
\textbf{ID} & $\mathbf{\theta}$ & $\mathbf{\varphi}$ &  & $\mathbf{\theta}$ & $\mathbf{\varphi}$ \\ \hline 
0 & $90^{\circ}$ & $11^{\circ}$ &  & $90^{\circ}$ & $34^{\circ}$ \tabularnewline
1 & $90^{\circ}$ & $56^{\circ}$ &  & $90^{\circ}$ & $79^{\circ}$ \tabularnewline
2 & $90^{\circ}$ & $101^{\circ}$ &  & $90^{\circ}$ & $124^{\circ}$ \tabularnewline
3 & $90^{\circ}$ & $146^{\circ}$ &  & $90^{\circ}$ & $169^{\circ}$ \tabularnewline
4 & $90^{\circ}$ & $191^{\circ}$ &  & $90^{\circ}$ & $214^{\circ}$ \tabularnewline
5 & $90^{\circ}$ & $236^{\circ}$ &  & $90^{\circ}$ & $259^{\circ}$ \tabularnewline
6 & $90^{\circ}$ & $281^{\circ}$ &  & $90^{\circ}$ & $304^{\circ}$ \tabularnewline
7 & $90^{\circ}$ & $326^{\circ}$ &  & $90^{\circ}$ & $349^{\circ}$ \tabularnewline
\hline
\end{tabular}
\end{table}

The detector pitch-angle coverage depends on the direction of the local magnetic field. We can generally expect a good coverage in pitch angle when $ 60^{\circ} \lesssim \theta_{B} \lesssim 120^{\circ}$, where $\theta_{B}$ is the colatitude of the magnetic field in the SSE coordinate system. However, when the magnetic field vector is aligned with the spin axis of the spacecraft, all sectors scan the same pitch-angle range, being the observations worthless for the study of the particle PAD due to the restricted angular information available (just one point).

The E2 experiment \citep{Musmann75} was a three axes flux-gate magnetometer on board Helios. Its observations were combined with particle data from E6 by the team at the University of Kiel to create a single data set, with proton, alpha-particles and electron intensities and magnetic field vector data. E2 data originally presented a very good time resolution, of eight measurements per second at the highest resolution mode. On the other hand, E6 time resolution was around 40~s in spite of the fact that it presented strong irregularities in the time step. The resolution of the instruments used at each moment depended on the information transmission rate from the probe to the Earth ground-based stations, that unfortunately was often quite small. 

For the present study, data files of the whole mission were provided by the University of Kiel as daily files containing both particle sectored data and IMF strength and direction (in the SSE coordinate system) with the same time resolution. This time resolution varies from 1 min to 15 min although the precise time step is not totally uniform, for example the exact values can vary from 40~s to 90~s for the 1~min resolution files. We linearly-interpolated data of the E03 electron channel and of the IMF to obtain data in a regularly spaced time grid of 1~min (or occasionally 15~min, when 1~min data were unavailable). It is important to note that we preserved data gaps, which are quiet frequent and can extend for several days.

\begin{table*}[t]
\centering
\caption{Observational characteristics of the selected events.}
\label{5tab:obsevents}
\begin{tabular}{cccccccccc}
\hline\hline
\multirow{2}{*}{\textbf{Year}} & \multirow{2}{*}{\textbf{Date}} & \multirow{2}{*}{\textbf{DOY}} & \multirow{2}{*}{\textbf{S/C}} & \textbf{Onset} & \textbf{Rise} & \textbf{Reso.} & \textbf{Distance} & \textbf{$\mathbf{u}$} & \textbf{IMF} \\
 &  &  &  & \textbf{[UT]} & \textbf{[min]} & \textbf{[min]} & \textbf{[AU]} & \textbf{[km s$\mathbf{^{-1}}$]} & \textbf{Polarity} \\ 
\hline
1976 & Mar 21 & 81 & H1 & 12:52 & 18 & 1 & 0.36 & 450 & -1 \\
1978 & Jan 1 & 1 & H1 & 22:00 & 53 & 1 & 0.94 & 434 & -1 \\
1978 & Dec 11 & 345 & H2 & 20:00 & 74 & 1 & 0.73 & 317 & -1 \\
1980 & Apr 5 & 96 & H1 & 15:55 & 28 & 15 & 0.85 & 314 & +1 \\
1980 & Apr 26 & 117 & H1 & 13:40 & 19 & 1 & 0.66 & 389 & +1 \\
1980 & May 3 & 124 & H1 & 08:00 & 20 & 1 & 0.58 & 421 & -1 \\
1980 & May 12 & 133 & H1 & 02:51 & 30 & 1 & 0.46 & 328 & -1 \\
1980 & May 28 a & 149 & H1 & 15:44 & 5 & 1 & 0.31 & 278 & -1 \\
1980 & May 28 b & 149 & H1 & 17:04 & 10 & 1 & 0.31 & 278 & -1 \\
1980 & May 28 c & 149 & H1 & 19:38 & 7 & 1 & 0.31 & 278 & -1 \\
1980 & May 28 d & 149 & H1 & 23:34 & 7 & 1 & 0.31 & 278 & -1 \\
1981 & Jan 14 & 14 & H1 & 21:01 & 20 & 1 & 0.73 & 309 & +1 \\
1981 & May 8 & 128 & H1 & 22:50 & 86 & 15 & 0.69 & 369 & -1 \\
1981 & Jun 10 & 161 & H1 & 06:16 & 30 & 1 & 0.32 & 245 & -1 \\
1982 & Jun 2 & 153 & H1 & 15:44 & 59 & 1 & 0.59 & 467 & -1 \\
\hline
\end{tabular}%
\end{table*}

In order to analyse the SEP events, we also used the solar wind speed, density and temperature from the Plasma Detector Experiment, E1 \citep{Schwenn75}, available at NASA's National Space Science Data Center, Space Physics Data Facility.\footnote{\url{ftp://spdf.sci.gsfc.nasa.gov/pub/data/helios}}

\section{Event selection}\label{5sec:eventselection}

We selected relativistic electron events observed during the whole Helios mission in the E03 channel, with nominal energy range 0.3\,--\,0.8~MeV. We initially found more than 200 electron events detected by at least one of the two spacecraft. The criteria adopted to select the sample of best-observed events for modelling were: 
\begin{itemize}[noitemsep]
\item[i)] no data gaps during the rising phase of the electron event;
\item[ii)] electron event peak intensities at least one order of magnitude above the pre-event background;
\item[iii)] IMF as close as possible to an ideal Parker spiral;
\item[iv)] location of the parent solar activity associated with each event documented in the literature; 
\item[vi)] no significant cross-contamination by protons during the time interval selected for each event.
\end{itemize}

The first criterion was by far the most restrictive owing to the long periods when the coverage of one of the instruments was lost or presented numerous gaps. In particular, we found several cases showing data gaps during the rising phase or just at the peak of the event, either in the particle intensities or the magnetic field data. Moreover, many other events were discarded as being not intense enough or happening right after a larger event, in such a way that the pre-event intensity level masked the event onset. In addition, we required a smooth time profile of the solar wind plasma parameters and magnetic field strength in order to avoid events affected by the crossing of an interplanetary shock that would distort the IMF. Only few events were discarded due to this criterion (iii). We checked shock crossing at the spacecraft since a shock between the sun and the spacecraft or beyond the spacecraft could affect the transport conditions of electrons and/or increase the length of the field lines where electrons propagate. Furthermore, we required that the parent solar source of SEP events was reported in the literature, either in published articles, Solar-Geophysical Data (SGD) reports\footnote{\url{ftp://ftp.ngdc.noaa.gov/STP/SOLAR\_DATA/SGD\_PDFversion/}} or GOES soft x-ray (SXR) data available online\footnote{\url{ftp://ftp.ngdc.noaa.gov/STP/space-weather/solar-data/solar-features/solar-flares/x-rays/goes/xrs/}}. Finally, in the last selection step we evaluated the cross-contamination of protons in the electron channel E03. We compared the E03 electron intensities with the proton intensities for channels P4 and P13 and discarded events showing a combined intensity of both proton channels higher than the 10\% of the E03 electron intensity. 

The final sample of events is comprises 15 events. In Table~\ref{5tab:obsevents} we list the year, date and day of the year (DOY) when the onset of the event occurred, the spacecraft used in the analysis, the onset time and rise time of the event, that is the time from the onset to the E03 peak intensity, the time resolution of the data used in the study, the helioradius of the spacecraft, the average solar wind speed during the six hours prior to the event onset, and the modal polarity of the IMF. The polarity of the IMF up to 1 AU can be defined as
\begin{equation}
    \text{sign}(\vec{B}) = \text{sign}(B_{R}-B_{T}\,\tan{\Psi}),\label{5eq:polarity}
\end{equation}
\noindent
where $B_{R}$ and $B_{T}$ are the $R$ and $T$ components of the magnetic field vector in the radial tangential normal (RTN) coordinate system, respectively, and $\Psi$ is the angle between the nominal Parker spiral magnetic field vector and the radial vector from the Sun, that is $\tan{\Psi} = r\Omega/u$, being $r$ the radial distance to the Sun, $\Omega$ the solar rotation rate and $u$ the solar wind speed.

At first sight, the difference in the number of events observed by Helios 1 (14 events) and Helios 2 (only one event) is remarkable. The solar maximum period of Solar Cycle 21 started on the second half of 1977, 1980 being one of the years with larger number of SEP events. Unfortunately, Helios 2 stopped operating in early March of that year, explaining the larger number of selected events detected by Helios 1. Also, Helios 2 instruments underwent more problems reflected in numerous data gaps.

The sample of selected SEP events was observed during streams of solar wind with speeds between 245~km~s$^{-1}$ and 467~km~s$^{-1}$. The whole range of radial distances is well covered, ranging from 0.31~AU to 0.94~AU. We can divide the locations into three different groups: events observed at radial distances $<\,$0.40~AU (six events), events observed between 0.40\,--\,0.70~AU (five events) and events observed at radial distances $>\,$0.70~AU (four events). It is also important to notice that as a result of the different time-resolutions in the SEP data files, we studied 13 events with 1~min time resolution and two events (1980 April 5 and 1981 May 8) with 15~min time resolution as indicated in Table~\ref{5tab:obsevents}.

\begin{figure*}[t]
\begin{center}
\includegraphics [width=0.49\textwidth, trim={0 0 9.cm 0},clip]{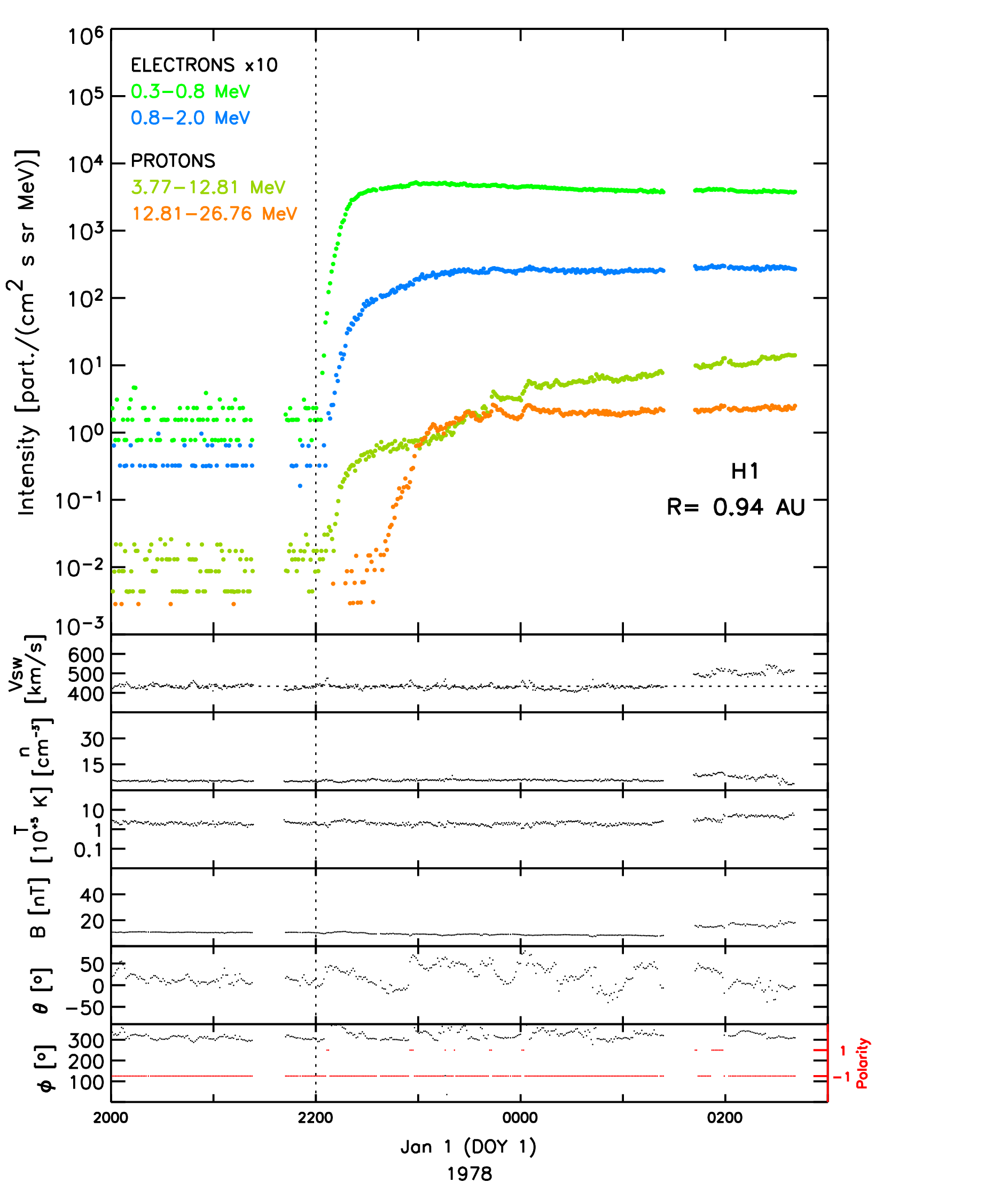}
\includegraphics [width=0.49\textwidth, trim={0 0 9.cm 0},clip]{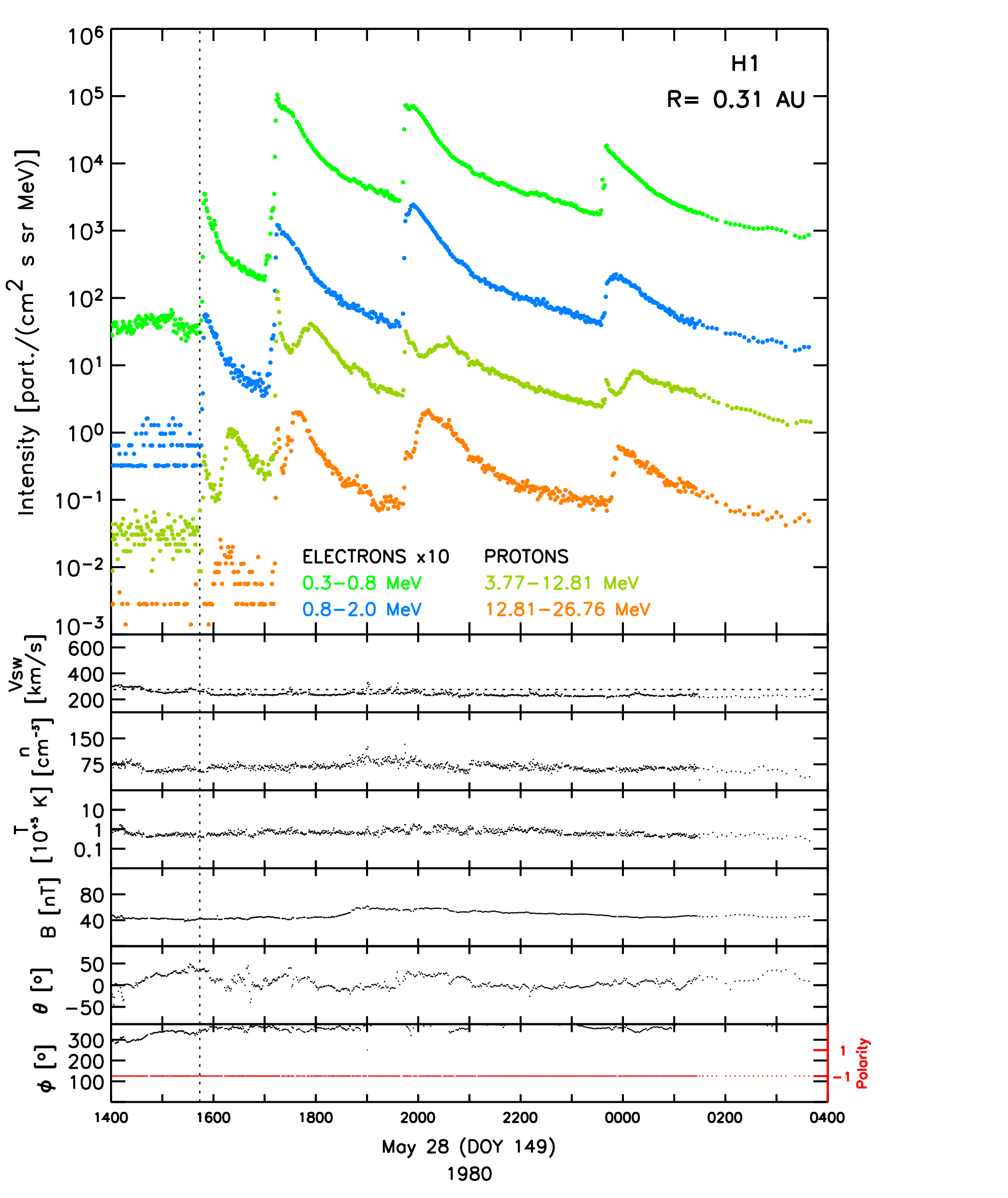}
\end{center}
\caption{In-situ measurements by Helios 1 on 1978 January 1 (left) and on 1980 May 29 (right). Top panel: particle intensities for electrons (dots, scaled by a factor of 10) for channels E03 (green) and E08 (blue), and protons (curves) from channels P4 (olive) and P13 (orange), measured by E6. Next panels from top to bottom: proton solar wind speed (horizontal dotted line shows the averaged pre-event speed), density and temperature; magnetic field strength and direction (RTN). Red dotted line in the last panel depict the IMF polarity. The vertical dotted line across all panels indicates the event onset.}
\label{5fig:eventimpulsivegradual}
\end{figure*}

The sample includes 11 events with rising times shorter than 35~min, and four events with longer rising times of up to $\sim$90~min. The different characteristics of the intensity-time profiles is also noticeable in the decaying phase: the first group of events at distances $<\,$0.4~AU shows a fast decay after the peak, while the second group of events (at distances 0.4\,--\,0.7~AU) shows a sustained intensity plateau extending for several hours. As an example, Figure~\ref{5fig:eventimpulsivegradual} shows a gradual event showing the aforementioned plateau (left) and a series of impulsive events with a short rise and a fast decay (right).

Table~\ref{5tab:EM} lists the location of the parent solar source associated with each event as observed from Earth in H$\alpha$ \citep[Solar-Geophysical Data reports]{Kallenrode92a,Kallenrode92b,Lario06,Agueda16,Gardini11}, the longitudinal distance between the H$\alpha$ source and the spacecraft and the connectivity of the source ($\Delta$), that is the longitudinal distance between the solar source and the footpoint of the Archimedean magnetic field line connecting the spacecraft to the Sun, calculated by using the solar wind speed measured in situ. The magnetic connectivity is positive when the footpoint is further West than the solar source, and it is negative if the footpoint is further East.

Most of the selected events are well connected to the solar source, with $\Delta \le 20^\circ$. Only three events have larger values of $\Delta$, which correspond to one western event near the limb (1981 January 14) and two events associated with central meridian sources located $1^\circ$ East (1981 June 10) and $5^\circ$ East (1978 December 11). We noted that these locations of the solar sources are given as seen by the Helios spacecraft. The fact that most of the events are well connected can be attributed to the event selection criteria requiring intensity peaks of at least one order of magnitude above the pre-event background. Hence, our sample is biased to western events, owing to the large-scale bending of the IMF that favours the magnetic connection of the spacecraft to western and central meridian solar locations, as seen from the spacecraft.

Table~\ref{5tab:EM} also lists the electromagnetic emissions (EMs) associated with each event. In particular, we list the start, peak and end of the SXR emission observed by GOES, the x-ray flare class, and the timing and frequency of the solar radio emission listed in the SGD reports. For each event, we identified the SXR emission and the radio emission occurring closest to the onset time of the event. We took into account the travel time of 950~keV electrons from the Sun to the spacecraft along the IMF and the time needed for the EMs to reach 1 AU, where GOES and radio stations are. We found no SXR emission matching precisely the timing of the event on 1980 April 26, so we listed the closest SXR emission found in GOES reports, corresponding to a 15~min long C3-class flare. For several events, various stations reported radio emission periods at different frequencies. In these cases we chose the lower radio frequency (always $>\,$29~MHz) and the reported radio emission lasting less than 30~min, when possible. When the emission shows several peaks at the same frequency observed by the same station, we understood that they belong to the same radio burst, and we listed them together as an extended emission with several peaks with no duration. The sample includes radio emissions at frequencies larger than 29~MHz and up to 9400~MHz.

\begin{sidewaystable*}
\centering
\caption{Solar electromagnetic emissions detected from Earth associated with the selected events.}
\label{5tab:EM}
\resizebox{\textwidth}{!}{%
\begin{tabular}{cccccccccccccccc}
\hline\hline
&  & \textbf{Onset} & \textbf{H$\mathbf{\alpha}$} & \textbf{H$\mathbf{\alpha}$} & \textbf{$\mathbf{\Delta}$} &  & \multicolumn{4}{c}{\textbf{Soft x-rays\tablefootmark{a}}} &  & \multicolumn{4}{c}{\textbf{Radio\tablefootmark{b}}} \\ \cmidrule(lr){8-11} \cmidrule(l){13-16} 
\multirow{-2}{*}{\textbf{Year}} & \multirow{-2}{*}{\textbf{Date}} & \textbf{[UT]} & \textbf{Earth} & \textbf{observer} & \textbf{[$\mathbf{\circ}$]} &  & \textbf{Start [UT]} & \textbf{Peak [UT]} & \textbf{End [UT]} & \textbf{Class} &  & \textbf{Start [UT]} & \textbf{Peak [UT]} & \textbf{Duration\tablefootmark{c} [min]} & \textbf{Freq [MHz]} \\ \hline
1976 & Mar 21 & 12:52 & N03W33 & W08 & -10 &  & 13:06 & 13:10 & 13:23 & M1 &  & 12:42 & 12:51 & 29.0 & 2695 \\
 &  &  &  &  &  &  &  &  &  &  &  & 12:42 & 13:00 & - & 2695 \\
 &  &  &  &  &  &  &  &  &  &  &  & 13:11 & 13:11 & 15.0 & 2695 \\
1978 & Jan 1 & 22:00 & S21E06 & W33 & -17 &  & 21:47 & 21:58 & 22:30 & M3 &  & 21:48 & 22:07 & 19.1 & 9400 \\
 &  &  &  &  &  &  &  &  &  &  &  & 22:07 & 22:20 & 12.9 & 9400 \\
1978 & Dec 11 & 20:00 & S15E14 & E05 & -57 &  & 18:33 & 19:42 & 20:18 & X1 &  & 19:40 & 19:49 & 11.0 & 1420 \\
 &  &  &  &  &  &  &  &  &  &  &  & 19:51 & 19:57 & - & 1420 \\
 &  &  &  &  &  &  &  &  &  &  &  & 19:51 & 20:32 & 49.0 & 1420 \\
1980 & Apr 5 & 15:55 & N08E22 & W64 & 2 &  & 15:32 & 16:52 & 16:58 & M5 &  & 15:46 & 15:53 & 21.0 & 600 \\
1980 & Apr 26 & 13:40 & S21E23 & W59 & 20 &  & 12:50 & 12:52 & 13:05 & C3 &  & 13:40 & 13:43 & 11.8 & 29 \\
1980 & May 3 & 08:00 & N25E35 & W41 & 9 &  & 07:50 & 08:06 & 08:13 & C9.4 &  & 08:00 & 08:01 & 4.5 & 808 \\
1980 & May 12 & 02:51 & S14E11 & W51 & 18 &  & 02:54 & 02:55 & 03:03 & C4.3 &  & 02:54 & 02:54 & 11.0 & 500 \\
1980 & May 28 a & 15:44 & S24W28 & W23 & -3 &  & 15:53 & 15:58 & 16:19 & C9.1 &  & 15:45 & 15:53 & 35.0 & 5000 \\
1980 & May 28 b & 17:04 & S18W35 & W30 & 4 &  & 17:05 & 17:18 & 17:53 & M3.6 &  & 17:02 & 17:17 & 30.0 & 930 \\
1980 & May 28 c & 19:38 & S18W33 & W28 & 2 &  & 19:24 & 19:51 & 20:53 & X1.1 &  & 19:47 & 19:48 & 21.0 & 100 \\
1980 & May 28 d & 23:34 & S17W39 & W34 & 8 &  & 23:32 & 23:44 & 24:45 & M6.9 &  & 23:38 & 23:42 & 18.0 & 500 \\
1981 & Jan 14 & 21:01 & N12E02 & W80 & 26 &  & 20:58 & 21:33 & 21:33 & C7.2 &  & 20:59 & 21:00 & 3.3 & 245 \\
1981 & May 8 & 22:50 & N09E37 & W61 & 18 &  & 22:50 & 22:51 & 23:20 & M1.7 &  & 23:10 & 23:20 & 38.6 & 245 \\
1981 & Jun 10 & 06:16 & N13E24 & E01 & -31 &  & 06:22 & 06:24 & 07:08 & M2.2 &  & 06:21 & 06:24 & 24.0 & 200 \\
 &  &  &  &  &  &  &  &  &  &  &  & 06:21 & 06:26 & - & 200 \\
 &  &  &  &  &  &  &  &  &  &  &  & 06:21 & 06:32 & - & 200 \\
1982 & Jun 2 & 15:44 & S08E81 & W25 & -4 &  & 15:08 & 15:48 & 16:20 & M9.9 &  & 15:44 & 15:46 & 11.5 & 2800 \\ \hline
\end{tabular}%
}
\tablefoot{
	\tablefoottext{a}{GOES catalogue: \url{ftp://ftp.ngdc.noaa.gov/STP/space-weather/solar-data/solar-features/solar-flares/x-rays/goes/xrs/}}\\
	\tablefoottext{b}{Solar-Geophysical Data reports: \url{ftp://ftp.ngdc.noaa.gov/STP/SOLAR_DATA/SGD_PDFversion/}}
	\tablefoottext{c}{Radio emissions with no duration value indicate different peaks inside a long lasting emission.}}
\end{sidewaystable*}

The events in our sample are associated mostly with M- and C- class flares. Eight events are associated with M-class flares, five with C-class flares and only 2 with X-class flares. The duration of the SXR emission is less than 60~min for most of the events, except for the largest five flares in the sample (1978 December 11, 1980 April 5, 1980 May 28 c and 1980 May 28 d, and 1982 June 2) that show SXR emission lasting up to $\sim$2~h. The rise time of the SXR emission is less than 40~min in most cases (13 events), and only two events were associated with flares with longer rise times. These latter are the events on 1978 December 11 and 1980 April 5. In general, we observe that flares with shorter rise times have shorter durations.

The start of the radio emission is observed within 10~min of the beginning of the SXR emission for most of the events (seven events). For five events, the radio emission starts between 10~min and 30~min before/after the beginning of the SXR emission. And only for three events (1978 December 11, 1980 April 26 and 1982 June 2) the beginning of the radio emission is delayed between 30~min and 70~min with respect to the beginning of the SXR emission. The peak in SXR emission appears within 10~min of the radio emission for ten events, while for three events the time of the peak emission differs from 10~min to $\sim$30~min. One event of the sample (1980 April 5) shows a long delay of an hour between the peak in radio emission and the peak in SXRs. 

For four cases, we found complex radio emission consisting on several bursts and showing various peaks for the same frequency. On 1978 January 1, for example, we identified a complex radio emission divided into 2 different bursts of 19-min and $\sim$13~min length at 9400~MHz, showing two different peaks 9~min and 22~min after the SXR peak. Moreover, for the event on 1978 December 11 a complex radio emission was detected at 1420~MHz, consisting of a 11-min precursor with a peak 9~min after the onset, and a great burst with two different peaks 17~min and 52~min after the initial radio onset. The first two peaks fall inside the rising phase of the event while the last one is located in the decaying phase.
\section{Modelling}

In the absence of large-scale disturbances, the IMF can be described as an average field given by an Archimedean spiral with a superposed turbulent component. The propagation of charged particles along the IMF has then two components, adiabatic motion along the smooth field and pitch-angle scattering by magnetic turbulence. The quantitative treatment of the evolution of the particles' phase space density, $f(t,z,\mu,v)$, is described by the focused transport equation \citep{Roelof69},
\begin{equation}
    \frac{\partial f}{\partial t}+ v \mu \frac{\partial f}{\partial z} + \frac{1-\mu^2}{2L}v\ \frac{\partial f}{\partial \mu} -\frac{\partial}{\partial \mu}\left(D_{\mu\mu} \frac{\partial f}{\partial \mu}\right)=q(z,\mu,t),\label{5eq:focutransp}
\end{equation}
\noindent
where $t$ is the time, $z$ is the distance along the magnetic field line, $\mu$ is the particle pitch-angle cosine, and $v$ is the particle speed. The focusing effect is characterised by the focusing length, $L(z) = -B(z)/(\partial B/\partial z)$, in the diverging magnetic field, $B$, while the pitch-angle diffusion coefficient, $D_{\mu\mu}$, describes stochastic processes undergone by the particles modifying their pitch-angles. The injection of particles close to the Sun is given by $q(z,\mu,t)$. 

Equation~\ref{5eq:focutransp} neglects convection and adiabatic deceleration and it is a useful approximation for near-relativistic particles. We note also that in this equation the particle speed, $v$, acts only as a parameter, and it can be removed using an appropriate scaling factor. If instead of $f$, we consider the differential intensity, $dI/dE = p^2 f$, and multiply Eq.~\ref{5eq:focutransp} by $p^2/v$, we get that the focused transport equation for the scaled quantity $j = I c/v$ is valid regardless of the speed we use to obtain the solution\footnote{\url{http://www.ieap.uni-kiel.de/et/people/heber/summerschoool/GF-scaling.pdf}} \citep[e.g.][]{Heber18}. Therefore Green's functions computed for hypothetical relativistic particles ($v = c$) can be used to obtain the Green's functions for other mono-energetic particles. The Green's function computed for $v = c$, $J(t,z,\mu,c)$ with the time variable and the intensity scaled with the quantity $v/c$ provides the Green's function for mono-energetic particles with speed $v$, that is

\begin{equation}\label{5eq:scaling}
J(t,z,\mu,v) = \frac{v}{c}\ J\left(\frac{v}{c}t,z,\mu,c\right).
\end{equation}

In this study Green's functions were obtained for different interplanetary transport scenarios given by (1) the IMF characterised by the Parker spiral topology using the solar wind speed values listed in Table~\ref{5tab:obsevents}, and (2) pitch-angle diffusion coefficient given by
\begin{equation}\label{5eq:difcoef}
D_{\mu \mu} =\frac{\nu_{0}}{2} \left( \frac{|\mu |}{1-|\mu |} + \epsilon \right) \left( 1- \mu^{2} \right) ,
\end{equation}
\noindent
where $\nu_{0}$ is the scattering frequency and $\epsilon$ is a parameter that allows us to model a range of scattering conditions \citep{AguedaEtV13}. We used $\epsilon = 0.01$ to obtain an anisotropic $D_{\mu \mu}$ with different values of the radial mean free path, $\lambda_{r}$, logarithmically spaced between 0.01~AU and ~0.5~AU. $\lambda_{r}$ is assumed to be constant throughout the event and along the IMF field line where electrons propagate.

\subsection{Energy response of the detector}\label{5sec:energyresponse}

The Green's functions of the interplanetary transport for the electrons of the E03 channel were computed with the transport model by \citet{Agueda08Tesis}. The model assumes that the solar source is static at two solar radii. To obtain the Green's function in the nominal 300\,--\,800 keV energy range, we considered 20 discrete electron energies with a constant logarithmic step within the range. We scaled the Green's function obtained for $v=c$ for each electron energy according to Eq.~\ref{5eq:scaling} and then interpolated the results in order to obtain the intensities in a 1~min time resolution grid. Then the intensities were scaled according to the normalised solar source energy spectrum. We assumed a power-law dependence $N(E) \propto E^{-\gamma}$, where $\gamma$ is the spectral index assumed for the modelled solar injection. 

The differential intensities for the nominal 300\,--\,800~keV energy range were obtained according to
\begin{equation}
J_{c} = \frac{\int_{E_1}^{E_2}J(E)\ \ dE}{\Delta E},
\end{equation}
\noindent
where $J(E)$ are the differential intensities of electrons with kinetic energy $E$, $E_1=0.3$~MeV, $E_2=0.8$~MeV, $\Delta E = E_2 - E_1$, and $J_c$ are the differential intensities of the nominal channel.

The procedure described so far to construct the Green's function of the nominal energy channel based on mono-energetic Green's functions assumes a flat energy response within the energy range under consideration. However, \citet{Bialk91} showed that the electron channel E03 may respond to electrons with energies higher than 800~keV and that the energy response is not flat, but similar to a Gaussian, peaking at 950~keV.

A more accurate estimation of the Green's function of the E03 channel can be obtained by taking into account that
\begin{equation}\label{5eq:countrates}
C_c = \int J(E)\ R(E)\ dE,
\end{equation}
\noindent
where $C_c$ is the count rate of the channel in [counts/s], $J(E)$ is the differential intensity of electrons with kinetic energy $E$, and $R(E)$ is the energy response function of the E03 channel. Here we assume that proton contamination in the electron channel is negligible, as we selected for the study only events where this effect was small (see Sect.~\ref{5sec:eventselection}).

\begin{figure}[t]
\begin{center}
\includegraphics [width=\columnwidth, angle=0]{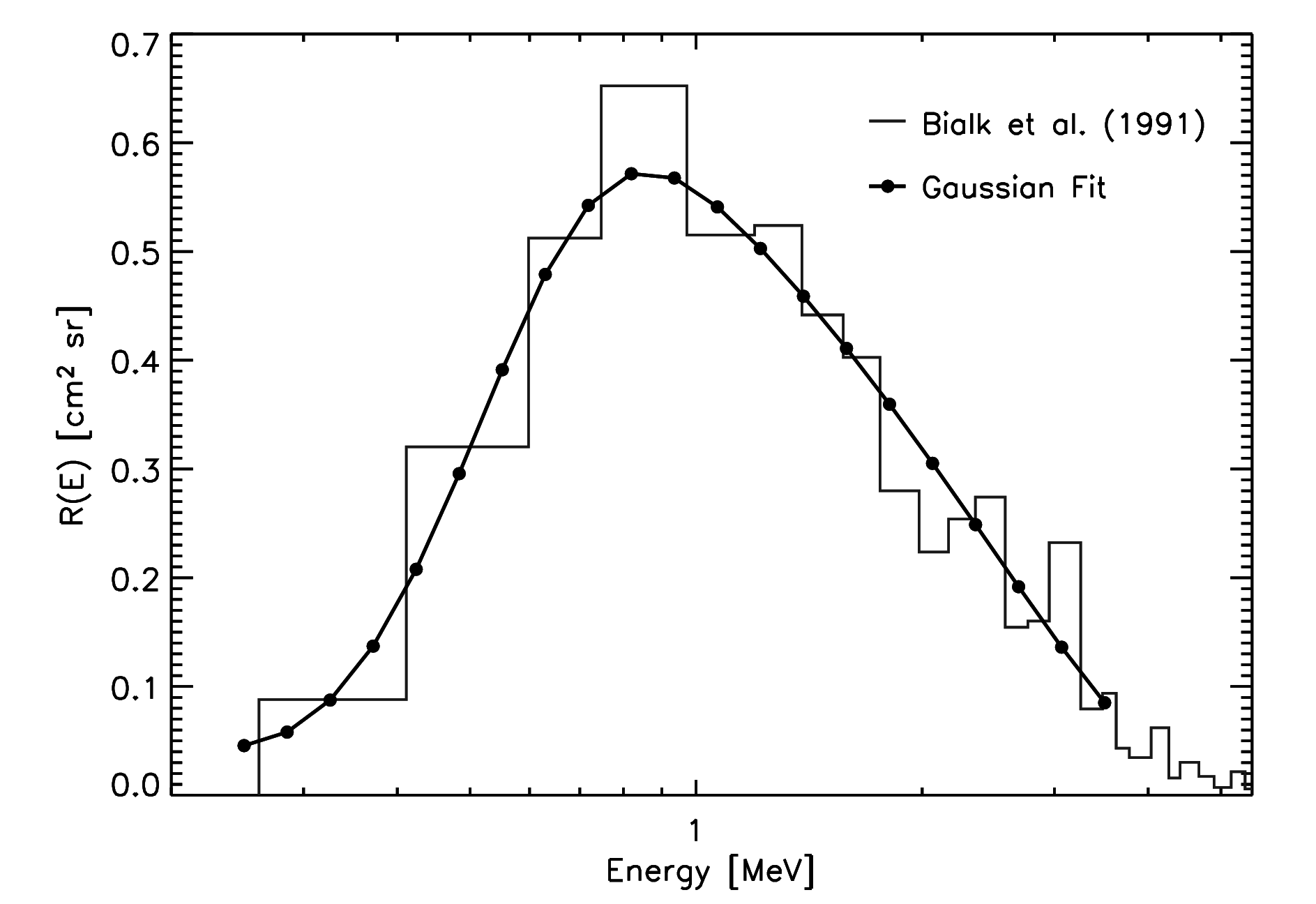}
\end{center}
\caption{Energy response for channel E03 of the E6 experiment on board Helios. The histogram (grey) shows the energy response computed by \citet{Bialk91}, the black solid curve shows a 6-parameter Gaussian fit, and dots display the response values for a grid of 20 logarithmically spaced energies.}\label{fig_gaussian}
\end{figure}

Figure~\ref{fig_gaussian} shows the E03 energy response computed by \citet{Bialk91} together with a 6-parameter Gaussian fit. It can be seen that the energy range from 0.25~MeV to 3.5~MeV covers the relevant part of the response. Neglecting energies above 3.5~MeV introduces a small error since the intensities decrease as a power-law (we noted that the values of $\gamma$ under consideration range from 2.4 to 4.6. See Table~\ref{5tab:results} in Sect.~\ref{Sec_results} for more details) and the values of the response function are very low ($<0.08$). A logarithmic grid of 20 energies between 0.25~MeV and 3.5~MeV (dots) is able to cast the main characteristics of the profile. 

\begin{figure}[th]
\begin{center}
\includegraphics [width=\columnwidth, angle=0]{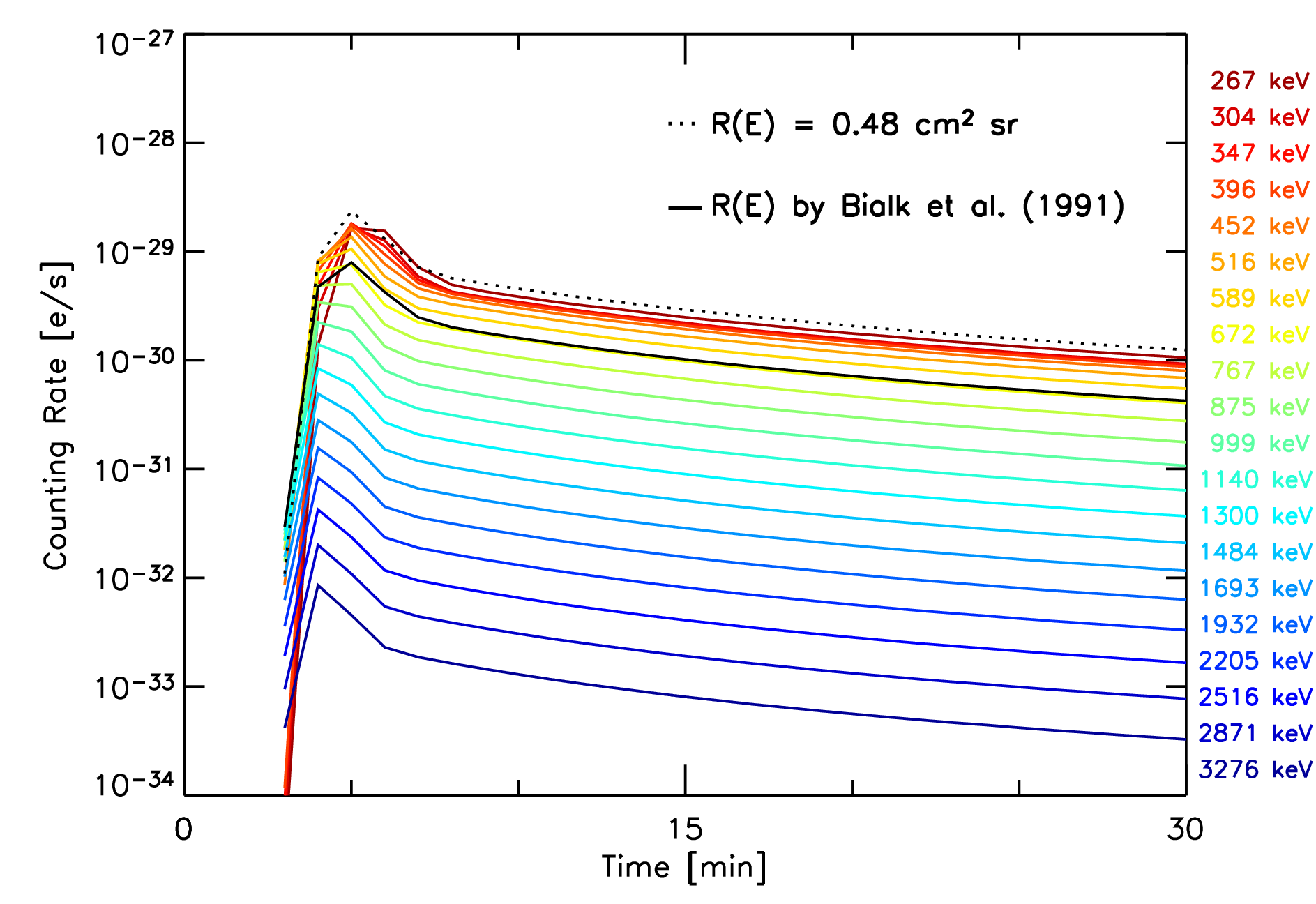}
\end{center}
\caption{Mono-energetic Green's functions from 0.25 MeV to 3.5 MeV (coloured curves) assuming an instantaneous release at the Sun at $t=0$ and assuming a spectral index of $\gamma=3.5$, a radial mean free path of $\lambda_r=0.14$~AU and taking into account the energy response of E03. The black curve shows the Green's function of the channel (in units of counts/s) assuming the energy response computed by \citet{Bialk91}. The dotted curve shows the Green's function of the nominal energy channel assuming a geometric factor of 0.48~cm$^2$~sr.}\label{fig_notnominal}
\end{figure}

As an example, Fig.~\ref{fig_notnominal} shows the Green's function (in units of counts/s) computed by taking into account the energy response of E03 and an energy spectra with $\gamma = 3.5$ (black solid curve). For comparison, Fig.~\ref{fig_notnominal} includes the Green's function for the nominal energy channel (300\,--\,800~keV) assuming a constant geometric factor of 0.48~cm~$^2$~sr (black dotted curve). It can be seen that the timing of the two Green's functions does not differ, meaning that the onset and the time of the peak of the count rate are the same within the 1-min time resolution. On the other hand, the peak is smaller by a factor $\sim$3 when the extended energy response is considered, since the instrument is mostly sensitive to particles with energies higher than the nominal energy range, for which there are lower intensities. The effect depends on the spectral index of the source. For simplicity, in this study we assumed that the spectral index of the electron source equals the observational value of the spectral index computed from in-situ data. As a first approximation, we computed the spectral index using the intensity measurements at the peak of the four electron channels of E6.

\subsection{Sectored Green's function and full inversion}\label{5sec:greenfunctions}

Simulated PADs were computed with a $9^{\circ}$ pitch-angle resolution. Once we know the angular response function of the sectors of the E6 telescope, it is possible to transform the simulated PADs into modelled sectored count rates which are directly comparable with observations. 

The simulated count rates observed in sector $s$ are given by
\begin{equation}\label{5eq:sectored}
G^{s}(t)=\sum_{jk}\ R_{jk}^{s}\ C_c(\mu_{jk}(t),t),\end{equation}
\noindent 
where $R_{jk}^{s}$ is the angular response of the E6 sector $s$, $C_c(\mu,t)$ are the PADs given by Eq.~\ref{5eq:countrates}, and $G^{s}(t)$ are the modelled count rates for each sector $s$ as a function of time. In Eq.~\ref{5eq:sectored} the matrix product is performed element by element and the sum extends over all $(\theta,\varphi)$ directions identified by the indices $j$,$k$, where $j$ is the azimuthal angle index from 0 to 360 and $k$ is the colatitude index, from 0 to 180. We note that observations of the local magnetic field vector show changes as a function of time (with the same time resolution as the particle data), varying the grid of pitch-angle cosines, $\mu_{jk}(t)$, at each time.

We used an inversion approach to fit the sectored observations, that is we computed the channel Green's function expected at the observer location for a set of multiple consecutive instantaneous injection episodes and then solve a least-square problem to find out what should be the relative weight of each injection episode to best fit the observed intensities \citep[see][for details]{Agueda08Tesis,Agueda08,Agueda14}. The best possible release time history was then obtained by evaluating, for each transport scenario, the goodness of the fit of the modelled data to the observations in each sector \citep[see details in][]{Agueda08}.

\section{Results}\label{Sec_results}

For each event, we selected fitting periods of 1 hour to 3 hours of duration approximately which include the onset, the rising phase and a sufficient part of the decaying phase of the electron event. Table~\ref{5tab:results} lists the fitting parameters and the results for each event in our sample. Columns 1 and 2 show the year and date of the event, column 3 shows the duration of the fitting period, column 4 lists the assumed source spectral index, column 5 and 6 list the inferred values of the maximum injection at the Sun and the type of injection: short ($<$30~min) or extended ($>$30~min). Columns 7 and 8 list the inferred value of the radial mean free path of the electrons and the ratio between the parallel mean free path and the focusing length, $L$, at the position of the spacecraft. The focusing length for the Archimedean spiral IMF is given by the expression $L= \frac{r(r^{2}+R^{2})^{3/2}}{R(r^{2}+2R^{2})}$, where $r$ is the radial distance from the Sun to the observer, and $R= v_{sw}/\Omega$, being $v_{sw}$ the solar wind speed and $\Omega$ the solar rotation rate. Following the work by \cite{Beeck86}, this ratio allows us to discriminate among transport regimes of the particles along the Archimedean spiral by giving a simple estimation of focused ($\lambda_{\parallel}/L \geq 1$), weak-focused ($1 > \lambda_{\parallel}/L > 0.1$) and diffusive ($\lambda_{\parallel}/L \leq 0.1$) propagation.

The values of the radial mean free path derived in this study range from $\lambda_{r} \sim 0.02$~AU to $\lambda_{r}\sim 0.27$~AU. These values are in general small compared to the distance between the Sun and the spacecraft, suggesting that the transport was not scatter-free for most of the events of the sample. For ten of the events, the electrons propagated in the weak-focused regime, for four of them in the focused and only for one of them the transport was clearly diffusive.

\begin{table*}[t]
\centering
\caption{Fitting parameters for the selected events.}
\label{5tab:results}

\begin{tabular}{cccccccc}
\hline\hline
\multirow{2}{*}{\textbf{Year}} & \multirow{2}{*}{\textbf{Date}} & \textbf{Fitting Period} & \textbf{Spectral} & \textbf{Max inj.} & \textbf{Injection} & \textbf{$\mathbf{\lambda_{r}}$} & \multicolumn{1}{c}{\multirow{2}{*}{\textbf{$\mathbf{\lambda_{\parallel}/L}$}}} \\
 &  & \textbf{[min]} & \textbf{Index} & \textbf{[e (s sr)$\mathbf{^{-1}}$]} & \textbf{Type} & \textbf{[AU]} & \multicolumn{1}{c}{} \\ \hline
1976 & Mar 21 & 55 & 3.4 & $1.7 \times 10^{28}$ & Short & 0.040 & 0.22 \\
1978 & Jan 1 & 100 & 2.4 & $2.2 \times 10^{30}$ & Extended & 0.106 & 0.24 \\
1978 & Dec 11 & 100 & 3.5 & $4.6 \times 10^{29}$ & Extended & 0.100 & 0.29 \\
1980 & Apr 5 & 130 & 4.6 & $1.9 \times 10^{29}$ & Short & 0.060 & 0.15 \\
1980 & Apr 26 & 60 & 4.0 & $6.0 \times 10^{29}$ & Short & 0.080 & 0.28 \\
1980 & May 3 & 150 & 3.7 & $7.3 \times 10^{29}$ & Extended & 0.080 & 0.28 \\
1980 & May 12 & 65 & 3.3 & $1.7 \times 10^{29}$ & Extended & 0.120 & 0.53 \\
1980 & May 28 a & 72 & 4.4 & $1.3 \times 10^{29}$ & Extended & 0.270 & 1.75 \\
1980 & May 28 b & 120 & 4.5 & $3.9 \times 10^{30}$ & Extended & 0.160 & 1.04 \\
1980 & May 28 c & 82 & 3.9 & $3.1 \times 10^{30}$ & Extended & 0.207 & 1.34 \\
1980 & May 28 d & 63 & 4.1 & $6.2 \times 10^{29}$ & Extended & 0.270 & 1.75 \\
1981 & Jan 14 & 60 & 2.9 & $3.0 \times 10^{28}$ & Short & 0.090 & 0.26 \\
1981 & May 8 & 195 & 3.1 & $3.0 \times 10^{30}$ & Extended & 0.070 & 0.21 \\
1981 & Jun 10 & 85 & 3.6 & $3.4 \times 10^{28}$ & Extended & 0.080 & 0.50 \\
1982 & Jun 2 & 162 & 2.9 & $7.7 \times 10^{28}$ & Short & 0.020 & 0.07 \\ \hline
\end{tabular}
\end{table*}

For most of the events we obtained a clear minimum of the goodness-of-fit estimator, which allowed us to identify the ranges of $\lambda_{r}$-values providing the best fit. However, there are five cases (1978 December 11, 1980 May 3, 1980 May 12, 1981 June 10, 1982 June 2) showing a plateau for low values of the radial mean free path (see Figure~\ref{5fig:GOF} in the Appendix). 

Noisy observational data may affect our inferred injection histories by adding either small precursors in the case of short injection profiles or intercalated gaps in the case of extended injection profiles \citep[see][for a discussion]{Agueda14}. In order to minimise these effects, we decided to select as best injection profile the simplest well-behaved profile.

\begin{figure}
\begin{centering}
\includegraphics[width=\columnwidth]{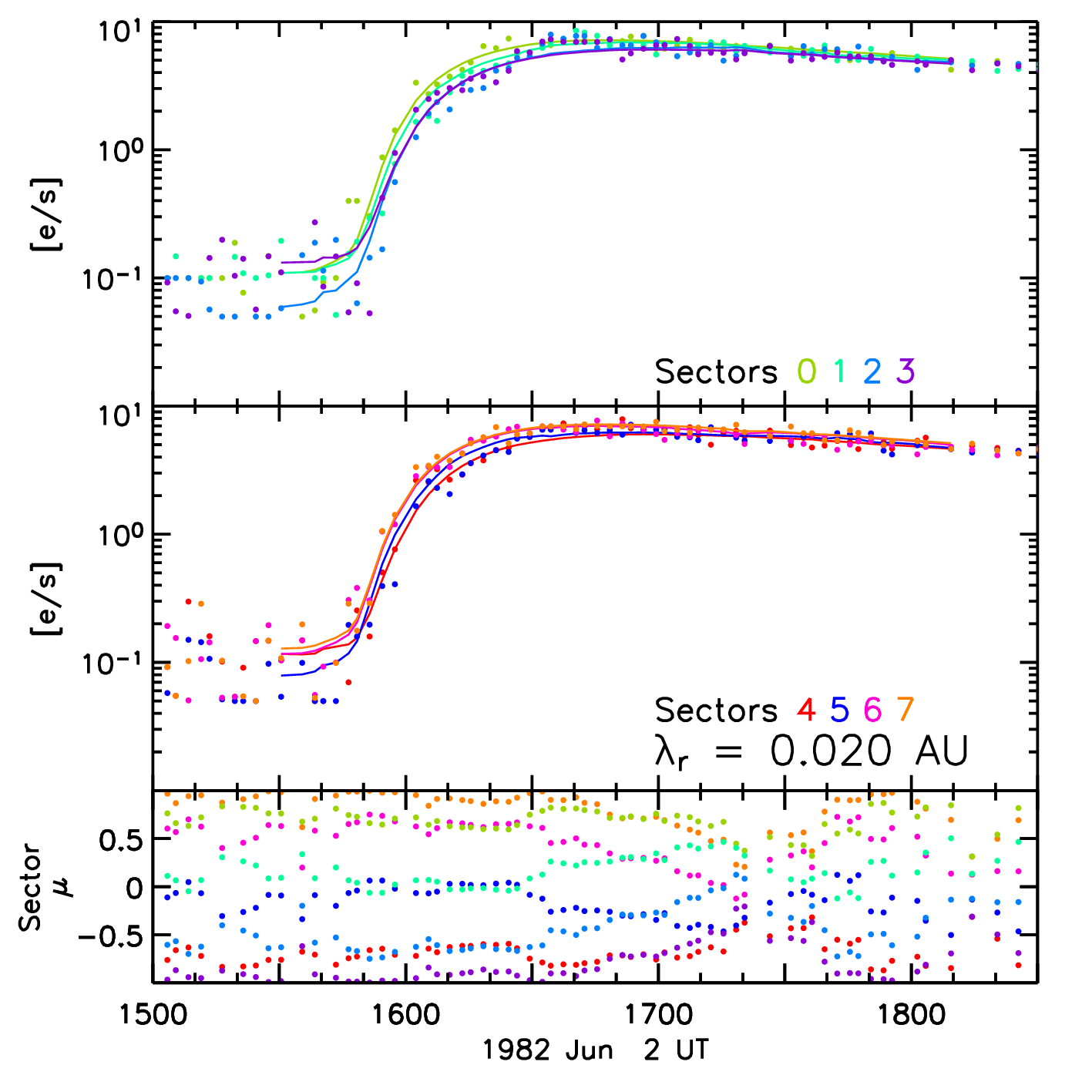}
\par\end{centering}
\caption{Top two panels: observational sectored data (dots) and model predictions (coloured curves) on 1982 June 2. Bottom panel: electron pitch-angle cosine observed by the midpoint clock-angle of each sector with the same colour code.}
\label{5fig:Results_1fit}
\end{figure}

Figure~\ref{5fig:Results_1fit} shows the results of the fit for the event on 1982 June 2. The first two panels show the data (dots) and the modelled (solid curve) counts per second in the sectors 0, 1, 2, 3 (top panel) and sectors 4, 5, 6, and 7 (middle panel). The third panel shows the pitch-angle cosine observed by the eight sectors throughout the event. In the Appendix, Figs.~\ref{5fig:Results_allfits1}, \ref{5fig:Results_allfits2} and \ref{5fig:Results_allfits3} show the fits for the other events in our sample. We obtained a good fit for most of the events in the list. However, for some of the events (e.g. 1978 January 1, 1980 April 26) some discrepancies are clear between the observations and the model for short periods of time. For the event on 1978 January 1 (Fig.~\ref{5fig:Results_allfits1}), the model is unable to reproduce the data hollow observed around $\sim$22:20 UT coinciding with a sudden local fluctuation of the magnetic field given by a rotation in latitude ${\theta}$ larger than $50^{\circ}$. For the event occurring on 1980 April 26 (see Fig.~\ref{5fig:Results_allfits1}), the model underestimates a double peak appearing in those sectors observing antisunward particles with $\mu \sim 1$. On the other hand, the model overestimates the observations for $\mu \sim 0.5$. For this event we find a sudden change in the latitude of the local magnetic field vector of $20^{\circ}$. For the case of the event on 1980 May 3 (Fig.~\ref{5fig:Results_allfits1}), we obtained a very good fit despite the data gap between 08:25\,--\,09:00~UT. The gap is not affecting the rising phase nor the peak of the event. Therefore it was possible to infer a reliable value of the mean free path and the injection time-profile. 

\begin{figure}
\begin{centering}
\includegraphics[width=\columnwidth]{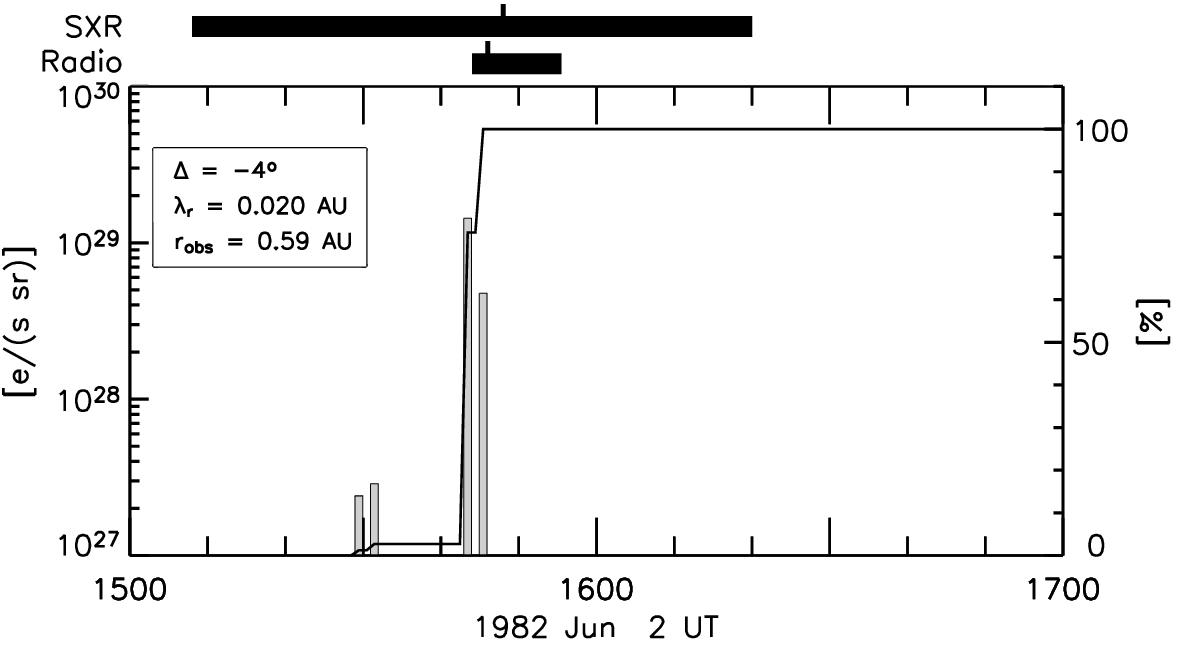}
\par\end{centering}
\caption{Release time profile inferred for the event on 1982 June 2. The histogram shows the inversion result with 1-min time resolution; the solid curve shows the accumulated percentage of injected electrons as a function of time. The profile has been shifted by +8~min to allow the comparison with EM emissions. Black thick horizontal bars on the top of the panel show the timing of the SXR and radio emissions. The time of the EM peaks are indicated with vertical lines. The legend shows the connectivity of the source ($\Delta$), the inferred radial mean free path and the radial distance of the spacecraft.}
\label{5fig:Results_1inj}
\end{figure}

For the event on 1982 June 2, the inferred injection profile at the Sun corresponding to the best fit mean free path is shown in Fig.~\ref{5fig:Results_1inj}. The bars indicate the rate of released particles per steradian for each 1-min time bin. The profile is shifted 8~min in order to directly compare the timing of the injection profile with the electromagnetic emissions detected at Earth, shown as horizontal thick bars on the top of the plot. The peak time of the EM emissions are indicated by small vertical lines in these bars. For this event, the injection profile is very short and the main release episode occurs between 15:40~UT and 15:45~UT in coincidence with the peaks in SXR and radio emission, which suggests that the release for this event was clearly associated with the well-connected M9.9 flare.

\begin{figure}
\begin{centering}
\includegraphics[width=\columnwidth]{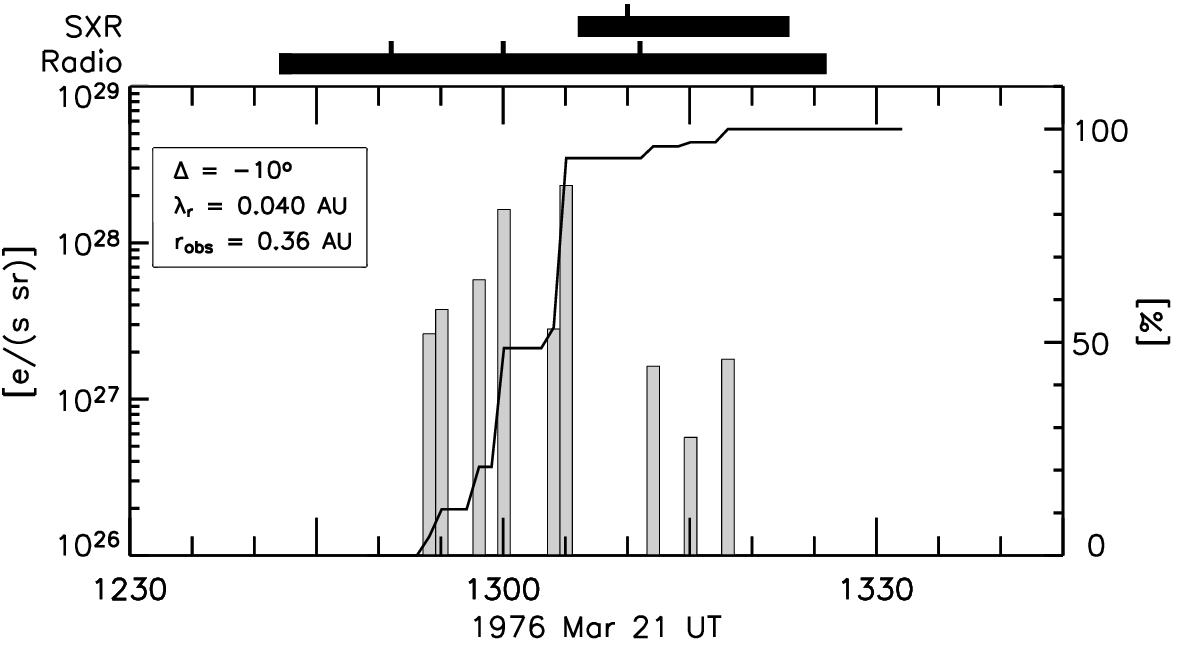}
\includegraphics[width=\columnwidth]{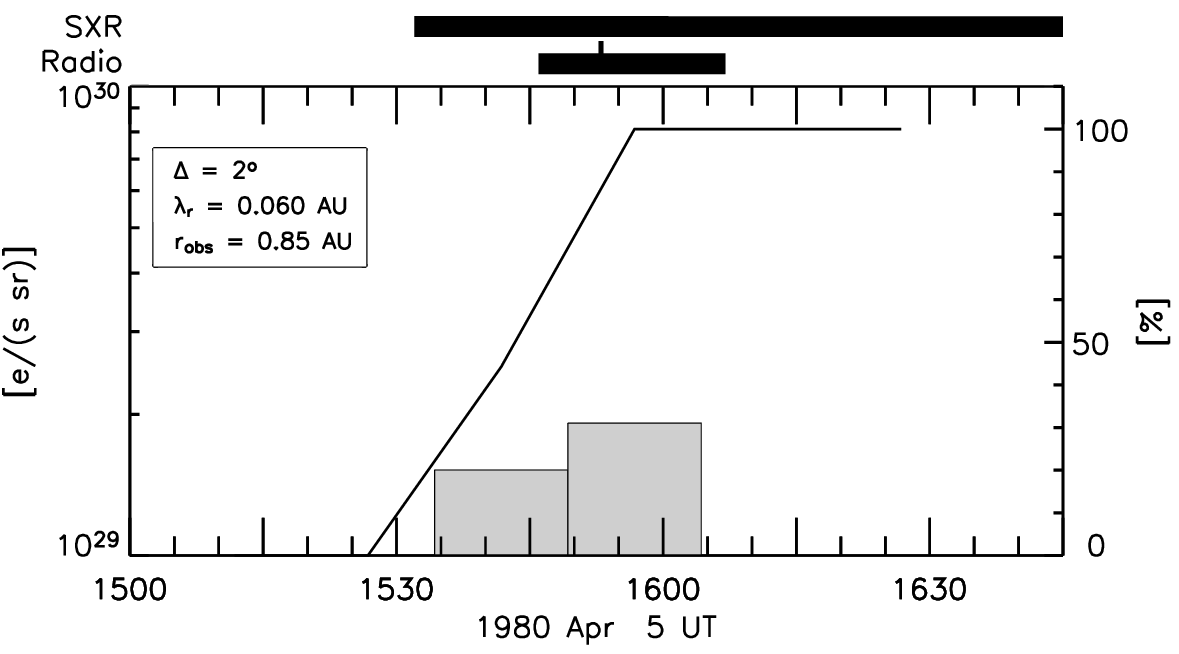}
\includegraphics[width=\columnwidth]{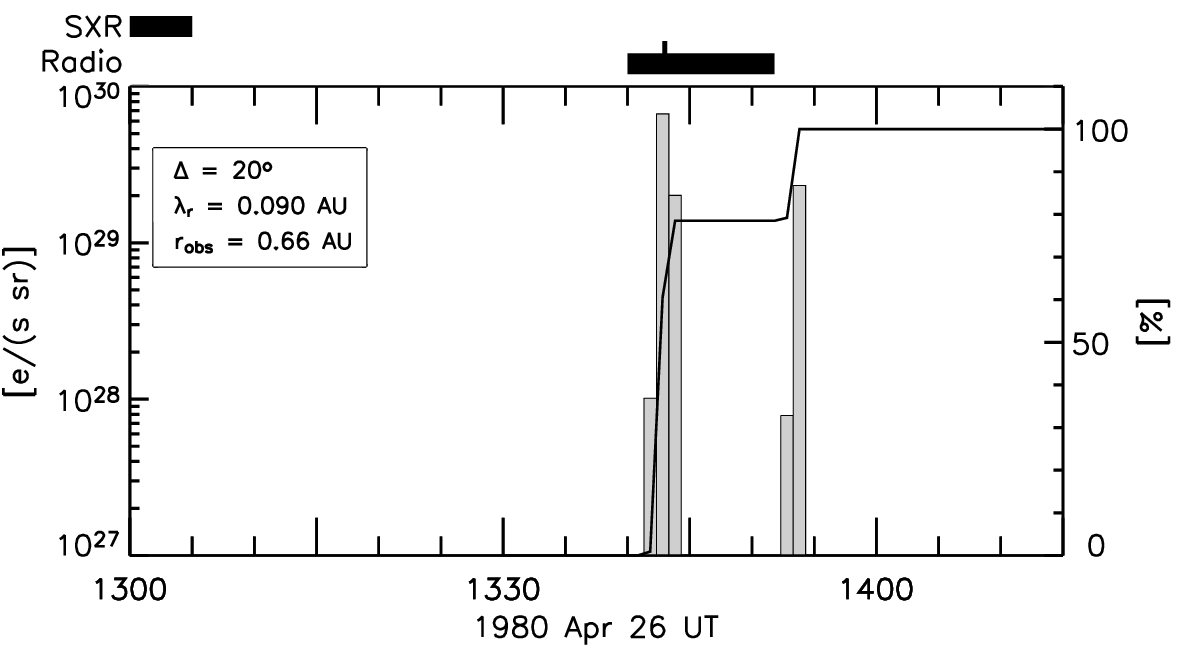}
\includegraphics[width=\columnwidth]{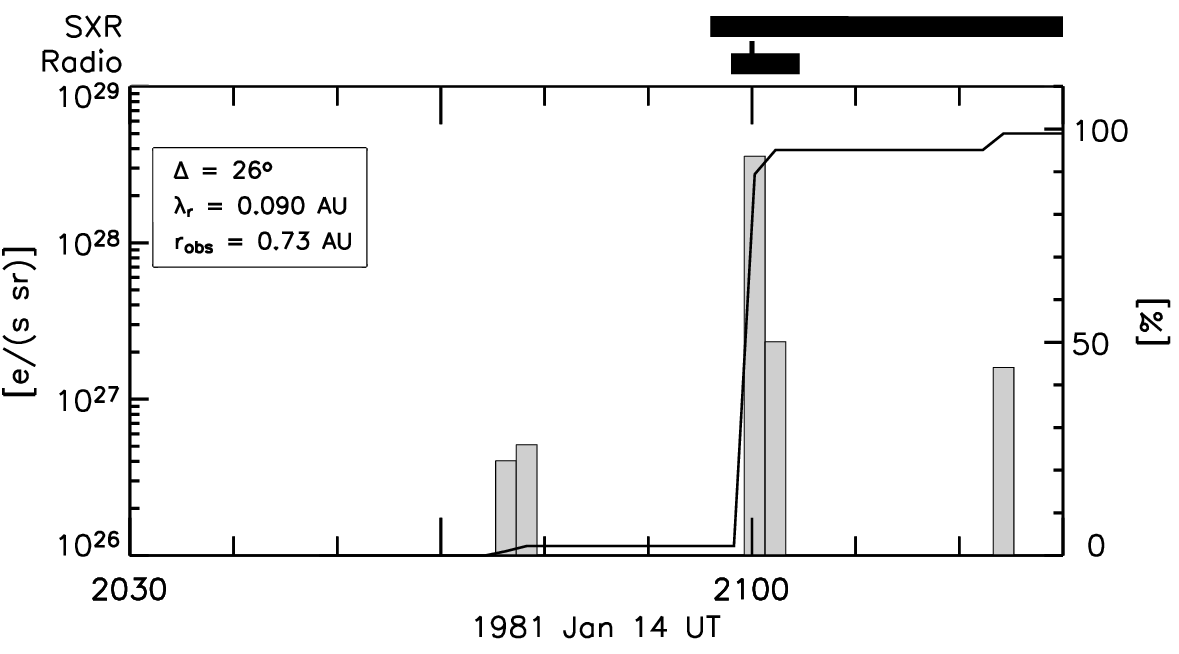}
\par\end{centering}
\caption{Same as in Fig.~\ref{5fig:Results_1inj} for those events with short ($<$30~min) injection profiles.}
\label{5fig:Results_allinj1}
\end{figure}

Figure~\ref{5fig:Results_allinj1} shows the injection profiles for those events with short ($<30$~min) durations. In these cases, the timing of the maximum release is consistent with the timing of the radio emission peak. Also, the 1976 March 21 event shows more than one peak in radio emission which is consistent with several injection episodes. The correspondence between the injection episodes and the SXR emission varies from event to event. For the event on 1980 April 26, as explained in Sect.~\ref{5sec:eventselection}, no SXR emission was reported in association with the onset time of the event, but we show instead the closest SXR emission reported, which clearly appears much earlier than the inferred electron release. Nevertheless, a short radio emission was found matching the injection start time and showing a peak that coincides accurately with the maximum injection. 

The duration of the electron release for these events seems consistent with flare emission. The associated flares have connection angles $\le 26^\circ$, which indicates that the open magnetic flux tubes through which electrons escape cover several tens of degrees in longitude on the source surface. \cite{Klein08} found that open field lines may connect the parent active region to the footpoint of the nominal Parker spiral, even when the parent active region is as far as $50^\circ$ away.

\begin{figure*}
\begin{centering}
\includegraphics[width=0.47\textwidth]{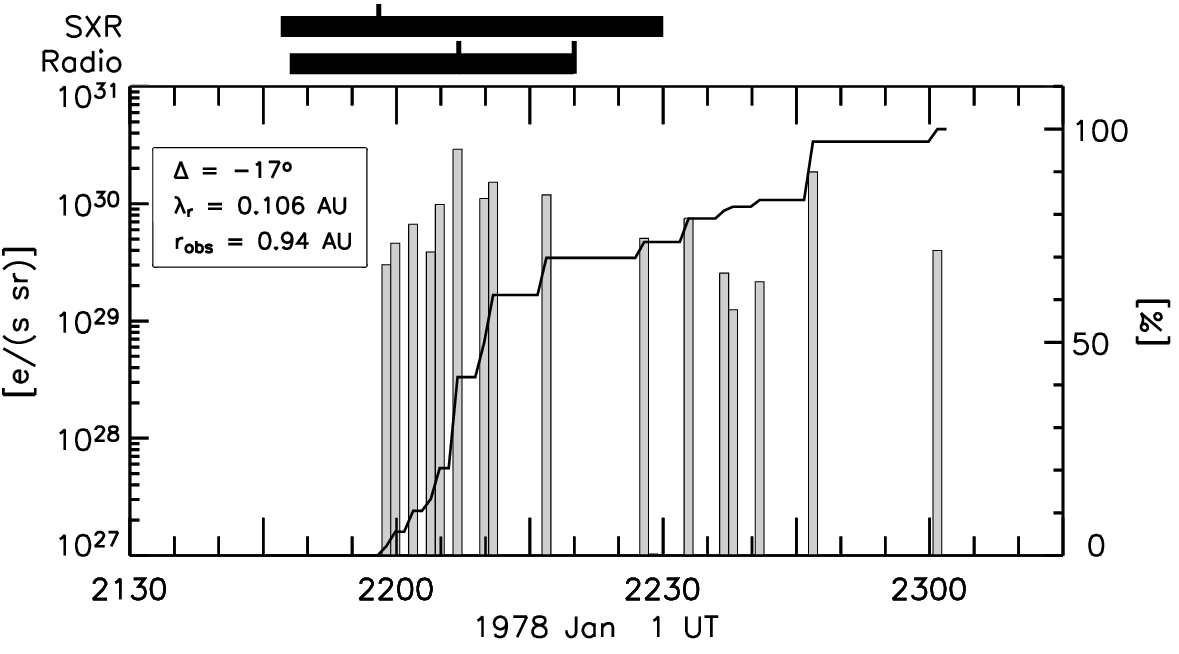}\hfill
\includegraphics[width=0.47\textwidth]{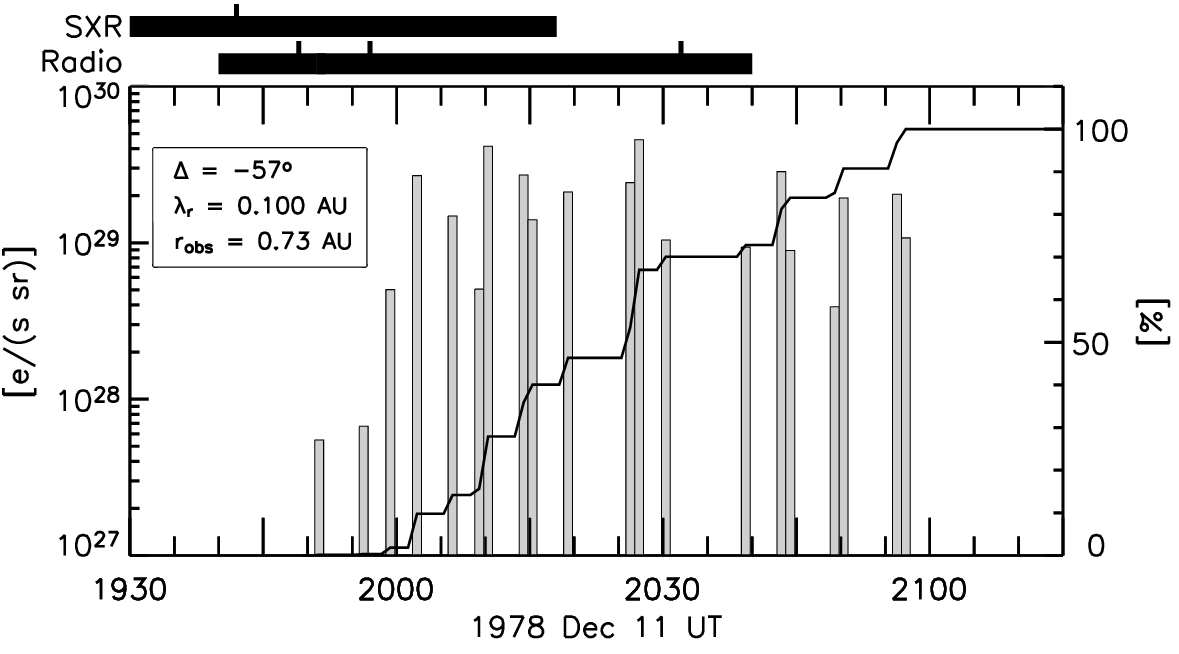}
\includegraphics[width=0.47\textwidth]{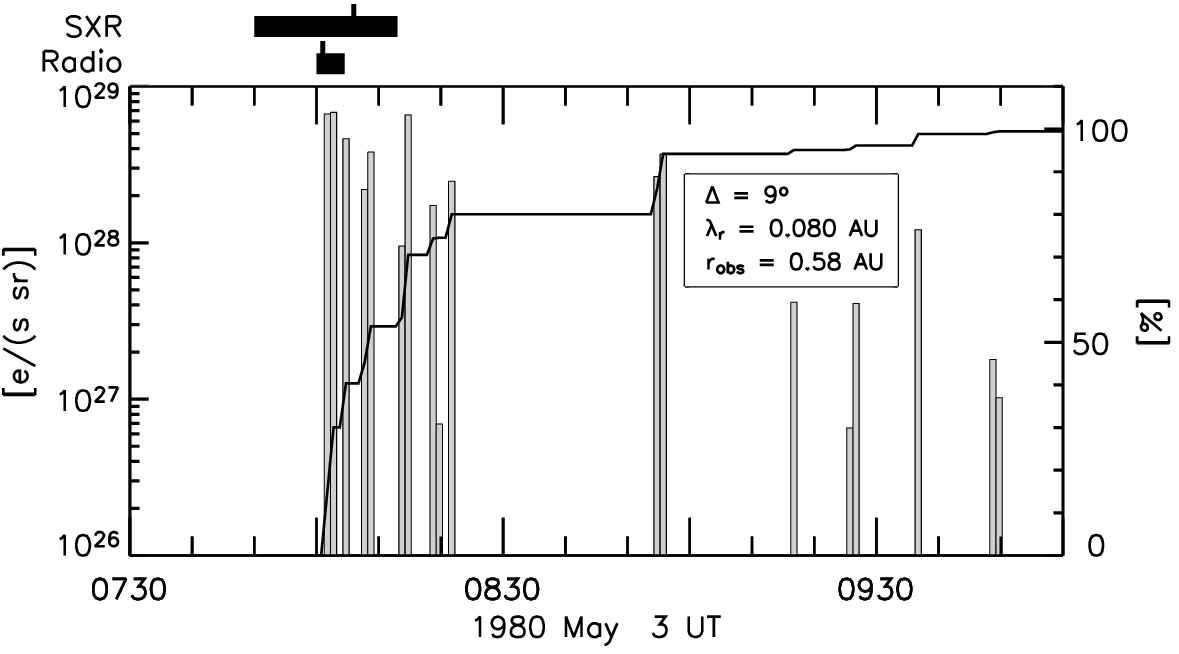}\hfill
\includegraphics[width=0.47\textwidth]{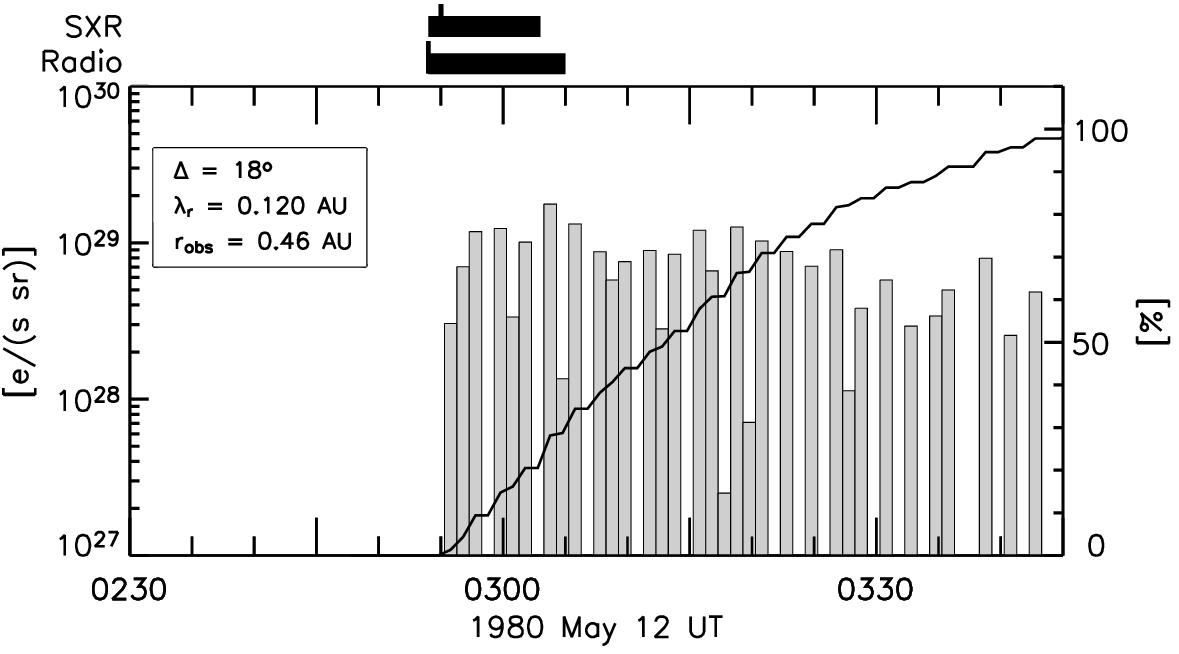}
\includegraphics[width=0.47\textwidth]{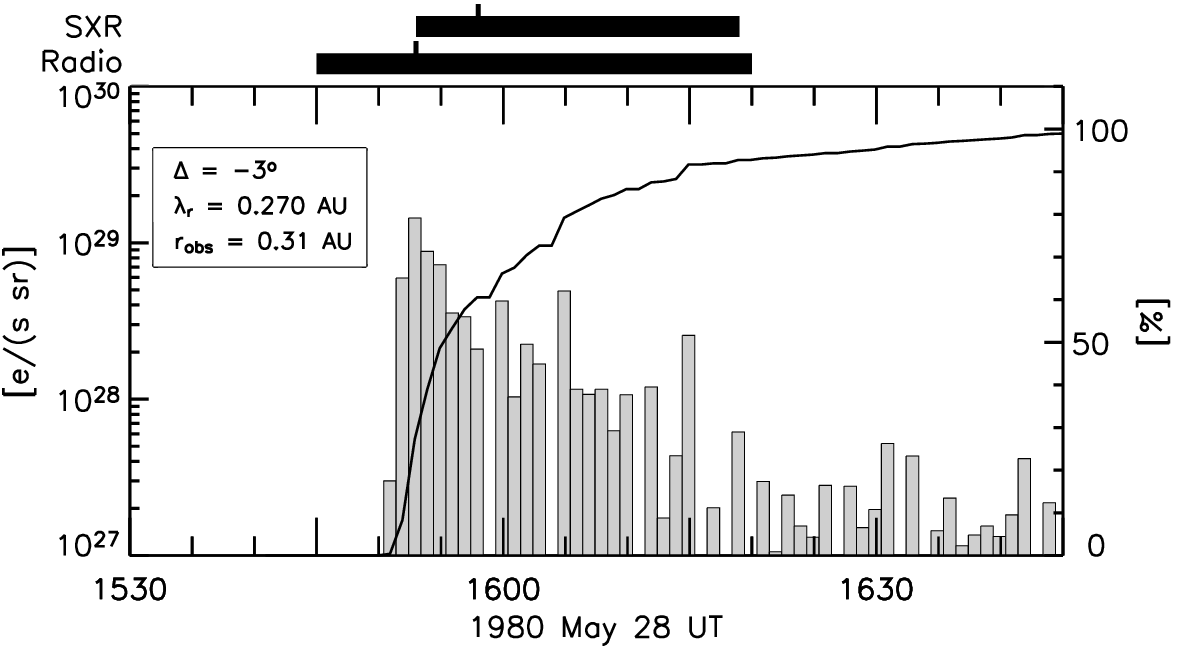}\hfill
\includegraphics[width=0.47\textwidth]{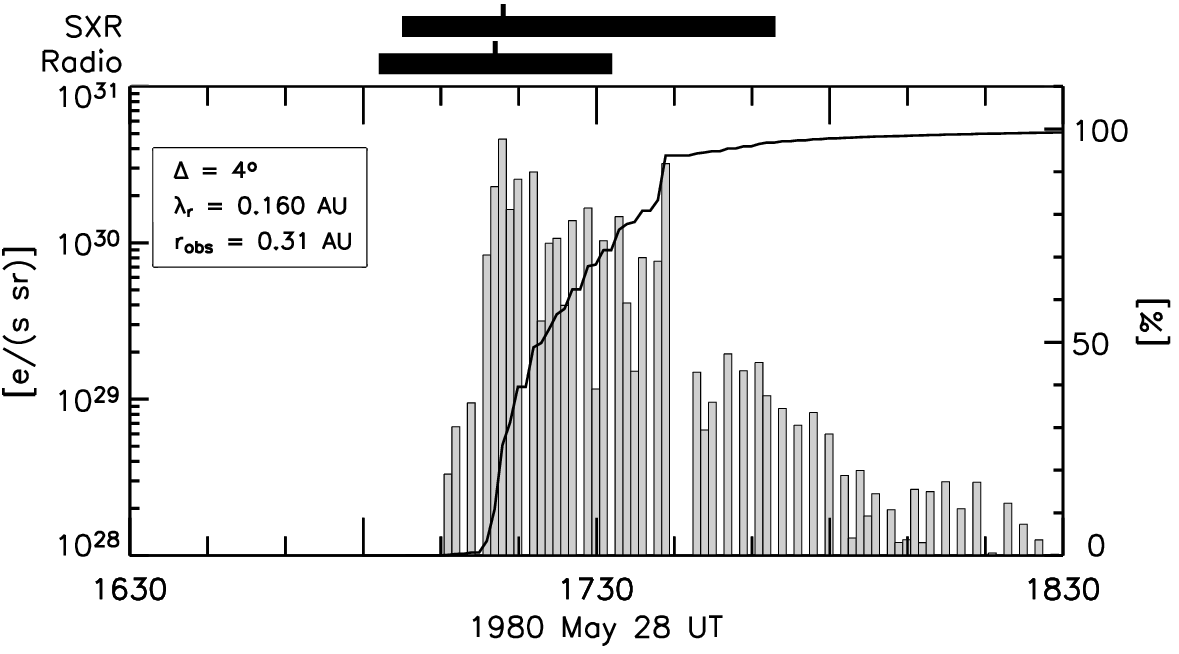}
\includegraphics[width=0.47\textwidth]{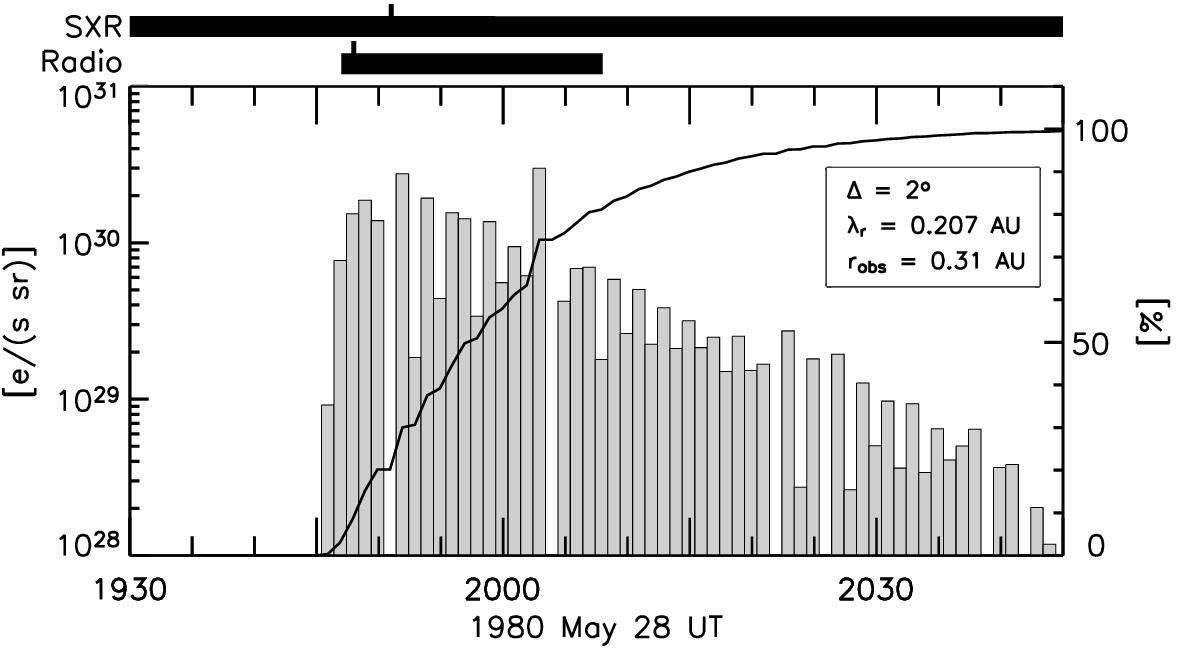}\hfill
\includegraphics[width=0.47\textwidth]{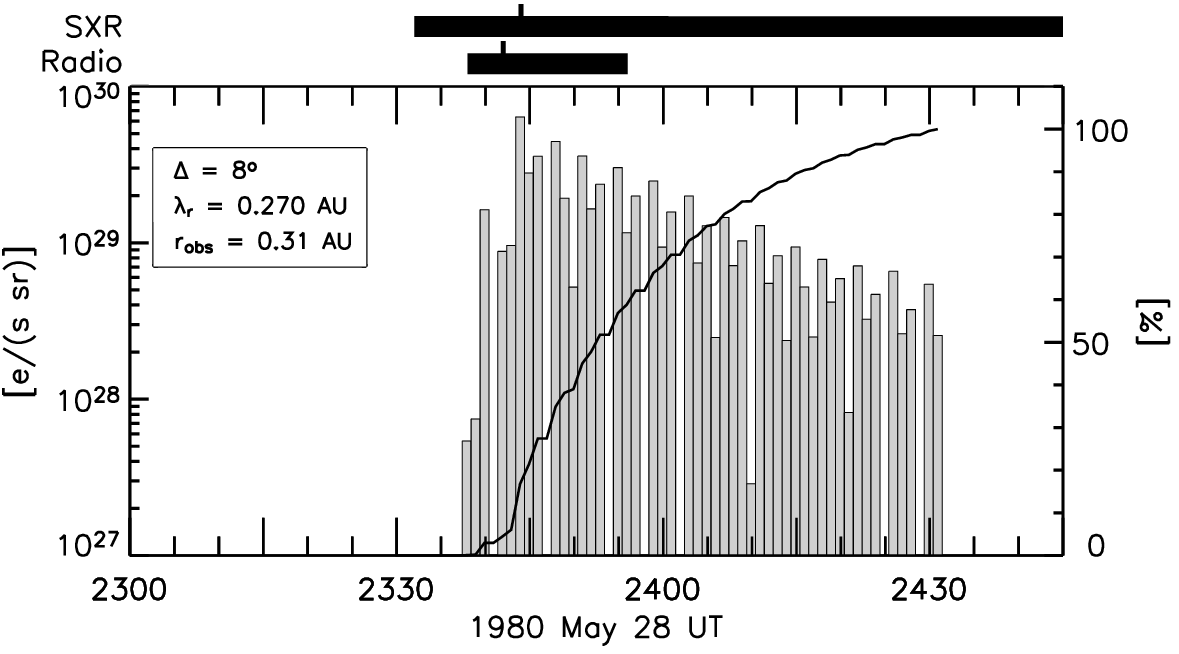}
\includegraphics[width=0.47\textwidth]{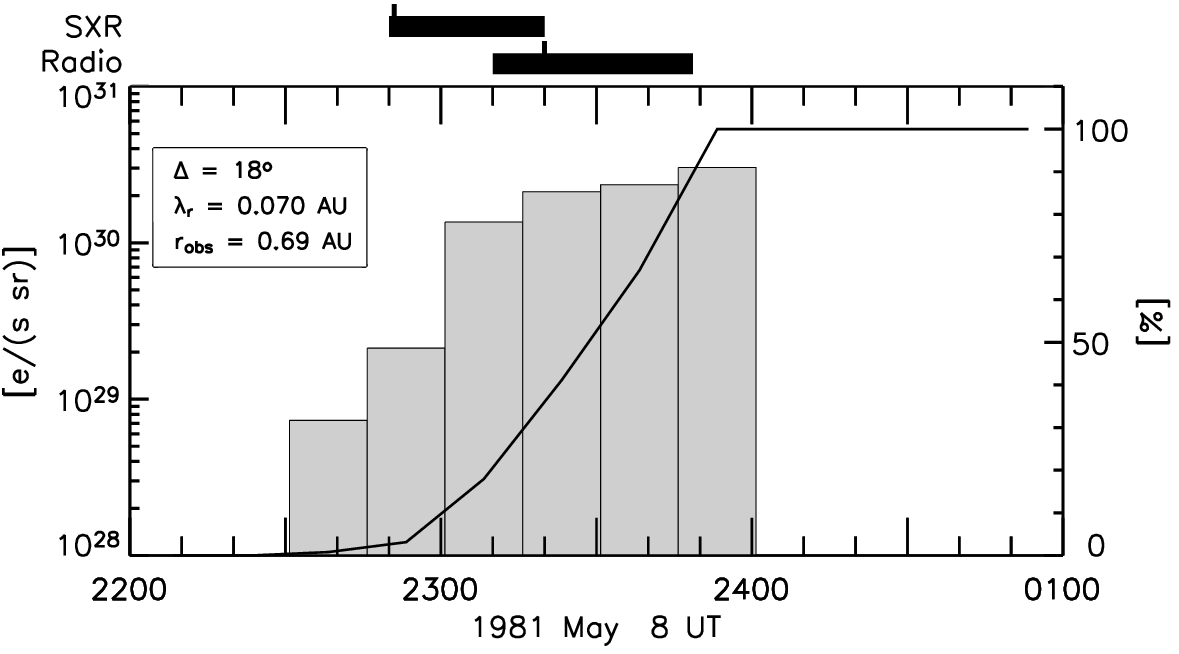}\hfill
\includegraphics[width=0.47\textwidth]{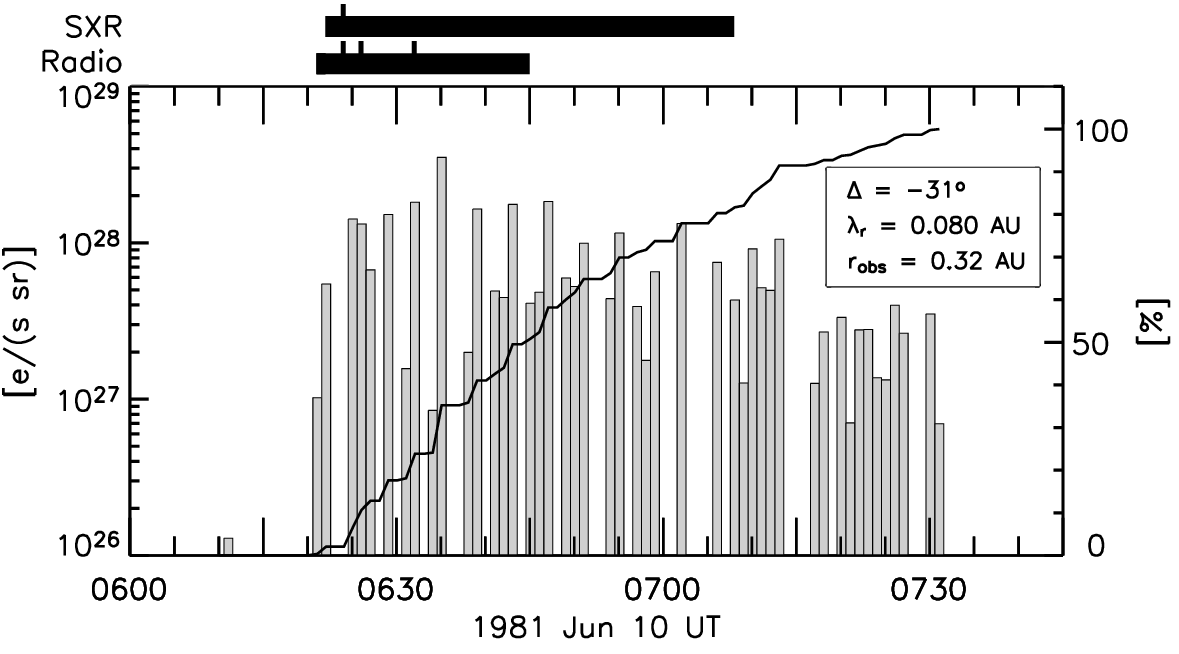}
\par\end{centering}
\caption{Same as in Fig.~\ref{5fig:Results_1inj} for those events with extended ($>$30~min) injection profiles.}
\label{5fig:Results_allinj2}
\end{figure*}

Figure~\ref{5fig:Results_allinj2} shows the injection profiles for those events with long ($>30$~min) durations. The event on 1980 May 3 shows a $\sim$20-min data gap just after the peak of the event (Fig.~\ref{5fig:Results_allfits1}). This gap results on an equivalent gap in the injection profile between 08:22 and 08:54~UT and a higher intensity of the injection right before and after the gap. We performed an analysis filling the data gap using simple linear interpolation and found no difference for the inferred value of the best mean free path. However, the inferred injection is continuous with lower intensities during the gap. 

Most (10) of the events in our sample show extended release episodes lasting at least an hour. In these cases, the beginning of the release appears before or at the peak in SXRs emission. The duration of the SXR emission does not seem related to the injection duration as three of the events (1980 May 3, 1980 May 12 and 1981 May 8) have the shortest SXR emission in the sample. Furthermore, the injection extends past the duration of the radio emission. Although the events observed on 1978 January 1 and 1978 December 11 show more than one radio peak consistent with several injection episodes.

We found no correlation between the source flare class and the duration of the inferred injection profile, being intense flares related to short injection profiles (such as the M9.9 flare on 1982 June 2, with $\sim$8~min injection duration) as well as weaker flares being connected to extended injection profiles (e.g. the C4.3 flare associated with the event on 1980 May 12, for which we inferred an injection profile lasting more than 45~min). Nevertheless, for the two events associated with the two strongest flares we obtained extended injection profiles lasting at least an hour.

\section{Discussion}
\subsection{Transport conditions and particle injection}
The sample of 15 events selected for this study suggests different electron transport conditions in the interplanetary medium with best-fit values of the radial mean free path between 0.02~AU and 0.27~AU. These values are in general small compared with the distance of the spacecraft to the Sun along the Archimedean spiral, which implies that the propagation is not scatter-free, and the $\lambda_{\parallel}/L$ ratios found indicate that the propagation occur in the focused and weak-focused regimes; therefore the focused-diffusion transport equation that we used is the appropriate framework to model these electron events. We noted that ten of the events in our sample show weak focusing conditions at the observer's position, four a transport regime clearly dominated by focusing and one event (1982 June 2) evolves under diffusive conditions. The main transport effect neglected by our model in the description of these SEP events is the diffusion of particles perpendicular to the mean IMF \citep[e.g.][]{StraussEtF15,Droge10}. \cite{StraussEtal17} and \cite{Droge16} studied the influence of the perpendicular processes on the longitudinal spreading of $\sim$100~keV electrons. These authors found that perpendicular diffusion is particularly important to explain the onset times and the anisotropies found in SEP events measured by spacecraft with distant magnetic connection to the solar source. Also, \cite{StraussEtal17} found that for heliocentric radial distances $\lesssim$0.7~AU (see their Fig.~3), the transport of SEPs is dominated by focusing ($\lambda_{\parallel} > L$), and hence SEPs propagate ballistically to the spacecraft, especially in cases where there is a good magnetic connection. Most of events in our sample are well-connected to the solar source, and, only in one case we found that the transport is diffusive (but it is well connected $\Delta = 4^{\circ}$). Hence, we expect no significant changes on the results in our sample of events in the case that perpendicular diffusion were considered in the modelling.

\begin{table*}[t]
\centering
\caption{Mean free path and injection type inferred by our study, \cite{Kallenrode92b}, \cite{Kallenrode93b} and \cite{Agueda16}. Red values indicate discrepancies with our result for the radial mean free path.}
\label{5tab:discussion1}
\resizebox{\textwidth}{!}{%
\begin{tabular}{ccccccccccccc}
\hline\hline
 &  &  & \multicolumn{2}{c}{\textbf{This study}} &  & \multicolumn{2}{c}{\textbf{Kallenrode et al. 1992}} &  & \textbf{Kallenrode 1993} &  & \multicolumn{2}{c}{\textbf{Agueda \& Lario 2016}} \\ \midrule
\textbf{Year} & \textbf{Date} & \textbf{Onset} & \textbf{$\mathbf{\lambda_{r}}$} & \textbf{Injection} & \textbf{} & \textbf{$\mathbf{\lambda_{r}}$} & \textbf{Injection} & \textbf{} & \textbf{$\mathbf{\lambda_{r}}$} & \textbf{} & \textbf{$\mathbf{\lambda_{r}}$} & \textbf{Injection} \\
\textbf{} & \textbf{} & \textbf{[UT]} & \textbf{[AU]} & \textbf{Type} & \textbf{} & \textbf{[AU]} & \textbf{Type} & \textbf{} & \textbf{[AU]} & \textbf{} & \textbf{[AU]} & \textbf{Type} \\ \hline
1980 & Apr 26 & 13:40 & 0.080 & Short &  & 0.15 & $\delta$ &  & 0.15 &  & - & - \\
1980 & May 3 & 08:00 & 0.080 & Extended &  & 0.05 & $\delta$ &  & 0.06 &  & - & - \\
1980 & May 12 & 02:51 & 0.120 & Extended &  & 0.10 & Reid-Axford &  & 0.15 &  & - & - \\
1981 & Jan 14 & 21:01 & 0.090 & Short &  & - & - &  & 0.10 &  & - & - \\
1981 & May 8 & 22:50 & 0.070 & Extended &  & - & - &  & \textcolor{red}{0.20} &  & - & - \\
1981 & Jun 10 & 06:16 & 0.080 & Extended &  & 0.05 & Reid-Axford &  & - &  & - & - \\
1982 & Jun 2 & 15:44 & 0.020 & Short &  & 0.02 & $\delta$ &  & - &  & - & - \\
1980 & May 28 a & 15:44 & 0.270 & Extended &  & - & - &  & - &  & 0.26 & Short \\
1980 & May 28 b & 17:04 & 0.160 & Extended &  & - & - &  & - &  & 0.14 & Extended \\
1980 & May 28 b & 19:38 & 0.207 & Extended &  & - & - &  & - &  & 0.18 & Extended \\
1980 & May 28 d & 23:34 & 0.270 & Extended &  & - & - &  & - &  & \textcolor{red}{0.20} & Extended \\ \hline
\end{tabular}%
}
\end{table*}

\cite{Kallenrode92b} studied a sample of 6 events (1978 April 11, 1980 April 26, 1980 May 3, 1980 May 12, 1981 June 10 and 1982 June 2), five of them contained in the sample of the present study. These include all but the 1978 April 11 event, that we discarded because of a gap in the magnetic field data during the rising phase of the event. \cite{Kallenrode92b} determined the electron transport conditions by fitting the averaged intensity and the anisotropy time profiles with the results of a focused transport model assuming an instantaneous $\delta$-injection or a Reid-Axford injection at the Sun and the nominal energy range of E03 (0.3\,--\,0.8~MeV). They estimated the uncertainty in the local values of $\lambda_{r}$ achieved by their method to be of the order of 50\%. They found small values of the radial mean free path, between 0.02~AU (1982 June 2) and 0.15~AU (1980 April 26) where our results show a range of 0.020\,--\,0.12~AU (for 1982 June 2 and 1980 May 12, respectively). The values of the radial mean free path obtained by \cite{Kallenrode92b} are summarised in Table~\ref{5tab:discussion1} and compared with the values inferred in this study. It can be seen that the values are consistent within the errors reported by \cite{Kallenrode92b}. Regarding the properties of the injection profile at the Sun, we found that for two of the events (1980 April 26 and 1982 June 2), \cite{Kallenrode92b} could fit the observations assuming a $\delta$-injection and the results of the inversion suggest short episodes as well. On the other hand, two of the events (1980 May 12, 1981 June 10) could not be fitted by a $\delta$-injection by \cite{Kallenrode92b}. In addition, their fit for the 1980 May 3 event failed reproducing the slower anisotropy decay suggesting also a longer injection duration. Consistently, for these three events we inferred extended injection profiles.

\cite{Kallenrode93b} also studied a sample of 27 proton and electron events observed by the Helios spacecraft. Five of them are part of our sample (1980 April 26, 1980 May 3 and 1980 May 12, already studied by \cite{Kallenrode92b}, and 1981 January 14 and 1981 May 8). \cite{Kallenrode93b} made use of a combination of the first two electron channels of E6 (0.3\,--\,2~MeV) and tried to fit the averaged intensity and the anisotropy time profiles with the results of an interplanetary transport model as done by \cite{Kallenrode92b}. They mention that the electron channels in the event on 1980 May 3 exhibit proton contamination in this range of energies, which is mainly due to the contribution of the second channel of E6, E08, with a higher response to protons than that of E03 (see Fig.~3 at \citealp{Bialk91}); hence, we can neglect proton cross-contamination in the E03 channel. Furthermore, as there is no energy dependence of the radial mean free path over this range of energies and we also considered in the energy response a similar range (from 0.25~MeV to 3.5~MeV) a direct comparison of the mean free path values obtained by \cite{Kallenrode93b} and those obtained in this study is possible. \cite{Kallenrode93b} found, in general, small values of the radial mean free path, between $<$0.02~AU and 0.35~AU, except for the event on 1978 April 28, for which they found $\lambda_r \geq$\,0.5~AU. For 15 of the events of their sample they found $\lambda_r < 0.2$~AU. The values of the radial mean free path obtained by \cite{Kallenrode93b} for the events present in our sample are compatible with the values we inferred, except for the 1981 May 8 event, for which we found a mean free path a factor $\sim$3 smaller (0.07~AU). The values are summarised in Table~\ref{5tab:discussion1}. 

\cite{WibberenzEtC06} studied a sample of impulsive electron events observed by Helios associated with short flares and type III bursts. They took into account the energy response \citep{Bialk91} and assumed diffusive transport conditions. These authors determined the radial mean free path by evaluating the time in the profile from the particle onset to the maximum intensity and compared it with the electron flight time between the Sun and the spacecraft. They analysed the event on 1976 March 21, which is included in our sample, and found $\lambda_r = 0.046$~AU, which agrees with our inferred value ($\lambda_r = 0.040$~AU).

Recent studies have applied inversion techniques to events detected by the Helios mission. \cite{Agueda16} presented a study of the four events observed on 1980 May 28, were they fitted the observed PADs with an exponential function in order to infer the electron transport conditions and the injection profile at the Sun by using the transport model by \cite{Agueda08} and assuming the nominal energy range of E03. The main differences between \cite{Agueda16} and the present study are the use of the energetic response from \cite{Bialk91} and the fact that we fitted the most direct form of directional data, that is the sectored intensities. For the four events on 1980 May 28, the values of $\lambda_{r}$ inferred in this study and by \cite{Agueda16} (see column 4 and 9 in Table~\ref{5tab:discussion1}, respectively) are very similar except for the 4th event, for which we derived a slightly larger value. The tiny differences found are explained by the different grid of $\lambda_{r}$-values tested and due to the difference in the assumed energy spectra.

The injection profiles inferred by \cite{Agueda16} are very similar to the ones inferred in the present study (see their Fig.~8). Since they use a secondary product (i.e. PADs obtained by fitting an exponential function to the sectored intensities) less affected by noise and where data gaps had been interpolated, their injections show smoother profiles. For the first three events (a, b and c in their Fig.~8), with a very similar value of the radial mean free path, we found smaller values of the maximum of the injection per energy unit, where \cite{Agueda16} found values of $5 \times 10^{29}$~e/(s~sr~MeV), $1 \times 10^{31}$~e/(s~sr~MeV), $6 \times 10^{30}$~e/(s~sr~MeV), respectively. For event d, they found $\lambda_{r} = 0.20$~AU and a value of the maximum injection of $1 \times 10^{30}$ e/(s sr MeV). Assuming $\gamma=2.0$, the same spectral index as \cite{Agueda16}, we obtained $2 \times 10^{28}$~e/(s~sr~MeV), $8 \times 10^{29}$~e/(s~sr~MeV), $6 \times 10^{29}$~e/(s~sr~MeV), $1 \times 10^{29}$~e/(s~sr~MeV), respectively for the first to the fourth event. The difference between the peak injection values found by \cite{Agueda16} and this study is mainly due to the fact that we took into account the energetic response of the E03 channel and due to the different approach on fitting the intensities. \cite{Agueda16} assumed electrons in the energy range of 0.3\,--\,0.8~MeV, while we assumed 0.25\,--\,3.5~MeV. This latter energy range makes the injection values in units of e/(s~sr~MeV) to become smaller, since the considered energy range is a factor 6.5 larger than the nominal energy range. Further, \cite{Agueda16} fitted an exponential function to the sectored intensities, which might yield to an overestimation of the intensities for the electrons propagating along the IMF with pitch angle 0$^{\circ}$; thus, implying higher inferred injection values.

A synthetic scenario mimicking the four SEP events in 1980 May was simulated by \cite{StraussEtal17}. These authors describe the transport of $\sim$100~keV electrons including the modelling of the diffusion of particles perpendicular to the IMF by assuming a field-line random walk process, formerly described by \cite{Jokipii66}. The resulting intensity-time profiles and first order anisotropies overall recover the results obtained by \cite{Agueda16} for the events observed both by Helios 1 at 0.3~AU and by IMP-8 near Earth. Therefore, our modelling results are in agreement of the results by \cite{StraussEtal17} and, as such, suggesting that the journey of particles from the Sun towards Helios 1 was mainly controlled by the parallel transport effects during these events. 

\begin{figure}
\begin{centering}
\vspace{-0.cm}
\includegraphics[width=\columnwidth]{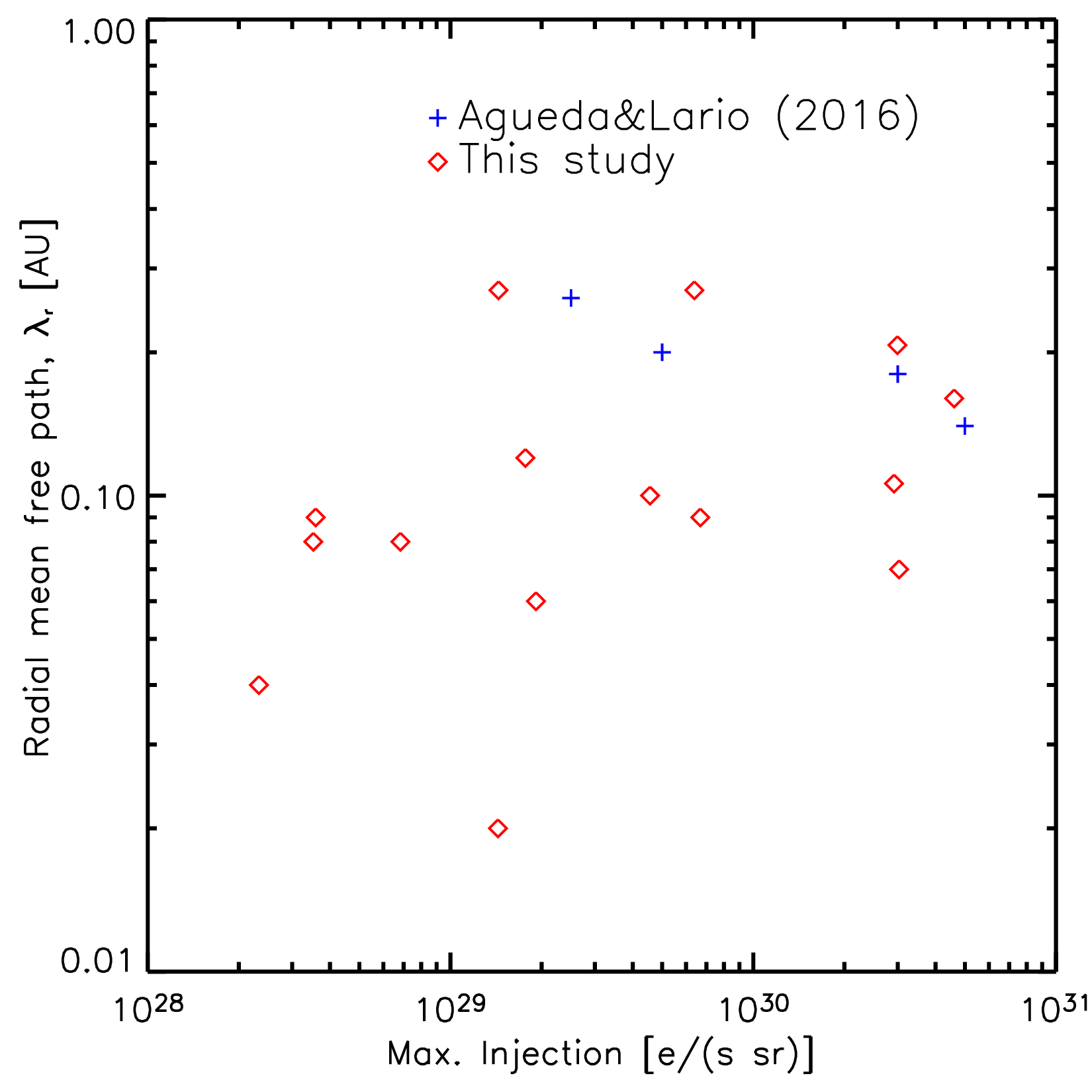}
\par\end{centering}
\caption{Radial mean free path vs. maximum electron injection at the Sun, i.e. the highest injection inferred for each event as shown in Figs.~\ref{5fig:Results_1inj}, \ref{5fig:Results_allinj1} and \ref{5fig:Results_allinj2}. Diamonds: values inferred in this study; crosses: values from \cite{Agueda16}.}
\label{5fig:disc2}
\end{figure}

Figure~\ref{5fig:disc2} shows the electron radial mean free path versus the maximum electron release at the Sun, where results are depicted from this study as well as from \cite{Agueda16} with just 4 events. These latter authors reported that the amount of interplanetary scattering undergone by the electrons seemed to be related to the highest electron injection at the Sun for the four events on 1980 May 28, in such a way that $\lambda_r$ decreases with increasing peak injection. However, with a larger number of events, we found no indication of a general relation between the maximum injection at the Sun and the value of the radial mean free path when considering the whole sample, in agreement with the results from \cite{Kallenrode92b}. 

\subsection{Particle release: duration and plausible processes}
Regarding the duration of the injection profile, we classified the events in our sample into short (when the release of particles lasts less than 30~min) or extended. Previous studies \citep{Agueda12b,Agueda13,Agueda14,Gomez-Herrero15,Pacheco17} found a similar dichotomy from the study of solar near-relativistic electron events observed by ACE, Wind, STEREO, and Ulysses, in different regions of the heliosphere. For example, \cite{Agueda14} studied the duration of the release processes of seven near-relativistic electron events observed at the near-Earth environment by the ACE and Wind spacecraft. They found that the electron release was produced either during short ($<$~30~min) or long ($>$~2~h) periods of time, agreeing with the results of our study. Also, \cite{Agueda12b} and \cite{Agueda13} studied four multi-spacecraft electron events observed by ACE and Ulysses, the latter being located at high latitudes in the heliosphere and at $\sim$2~AU. They found extended periods of particle release, lasting a few hours in 5 of the events, whereas the other 3 events presented long-duration intermittent sparse injection episodes (when the magnetic footpoint of the spacecraft laid at the opposite magnetic sector of the flare site, see \citealp{Agueda13} for further details). Finally, \cite{Gomez-Herrero15} studied the multi-spacecraft event on 2011 November 3. They modelled the $\sim$62\,--\,105~keV electron event observed by STEREO-A, ACE and STEREO-B (covering $\sim$300$^{\circ}$ in longitude) and derived an extended injection episode of several hours for the three spacecraft. The observation of a single CME and the observed anisotropies support the direct injection of particles at the three locations by an extended source, but a clear observational evidence of such a wide coronal and\slash or interplanetary shock was not found.

The fact that ten out of the 15 events in our list show extended injections points towards some mechanism allowing a continuous electron acceleration or a slow release of the electrons into interplanetary space. In a previous analysis of near-relativistic electron events observed by the ACE and Wind spacecraft, \cite{Agueda09a} and \cite{Agueda14} related short ($<$ 15~min) particle release episodes to flare processes, and they indicated as the most plausible scenario for extended injection episodes ($> 1$~hour) the injection of particles from coronal CME-driven shocks and/or reconnection processes behind the CMEs. In \cite{Agueda14}, they found that only for those events associated with type III radio bursts reaching the plasma line near the spacecraft a short flare-related injection episode was inferred, suggesting that magnetic connectivity plays an important role in space for short injection profiles. This was consistent with a scenario where electrons released during type III radio bursts not reaching the local plasma line never reach the observer due to the lack of magnetic connectivity. In addition, they concluded that the presence of type II radio bursts does not seem to be a discriminator between short and extended injections. These authors found that extended injections are related to different EM signatures of long particle acceleration in the corona (long decay SXR emission, type IV radio bursts, and time-extended microwave emission).

Other mechanisms to explain the observed extended particle injections were proposed by \cite{Klein10}. These authors studied in detail the EM emissions of a sample of 15 CME-less flares. These flares show bright SXR and microwaves bursts which indicate an efficient acceleration of the electrons in the flare. They found that no SEP event was connected to these flares as accelerated particles remained confined in the low corona, because magnetic field lines kept closed over the Sun's surface. The only CME-less flare associated with a weak SEP event near the Earth environment showed type II radio emission, suggesting that a coronal shock (not related to a CME) was the source of the accelerated electrons. In addition, \cite{Klein10} analysed 3 eruptive flares (i.e. associated with CMEs) occurring few hours after the CME-less flares. These eruptive flares showed SEP events. \cite{Klein10} pointed out that an easy conclusion that could be drawn from their analysis is that CMEs are needed to open magnetic field lines in order for the electrons to escape into interplanetary space. However, \cite{Klein10} found that this is not the scenario for the three eruptive events in their study. They stated that even if the CME had opened the coronal magnetic field to allow particles access to interplanetary space along open flux tubes along, the observed SEP events associated with type III bursts were detected too early, during the impulsive phase of the flares, and thus CMEs did not have still enough time to open the coronal magnetic field around the particle source. Therefore, to explain the origin of these SEPs, they suggest the scenario where both CMEs and type III bursts are triggered from the surroundings of active regions with open magnetic field lines which connect that site with the interplanetary medium. 

\cite{Dresing18} also found that a long-lasting electron injection profile was needed to explain the widespread event on 2013 December 26. They pointed out that the shock front propagating into the interplanetary medium could play a role explaining that extended injection. They suggested two possible scenarios, (i) the extended shock is accelerating the observed electrons when propagating into the interplanetary medium, or (ii) a leakage process coming from a magnetic trap where particles are gradually released giving the same result as a long injection. They pointed out that the shock will hardly explain the high-energy particles observed, so they propose possible scenarios where particles are early accelerated in the corona and, while some of the electrons are directly injected into open field lines, a fraction of them are trapped by a closed magnetic field region that only allows a slow electron leakage towards the flux tube connecting the trap with the observer. This trap could be suddenly opened by solar activity in the corona, releasing abruptly the trapped particles it contains.

\begin{figure}[t]
\begin{centering}
\vspace{-0.cm}
\includegraphics[width=\columnwidth]{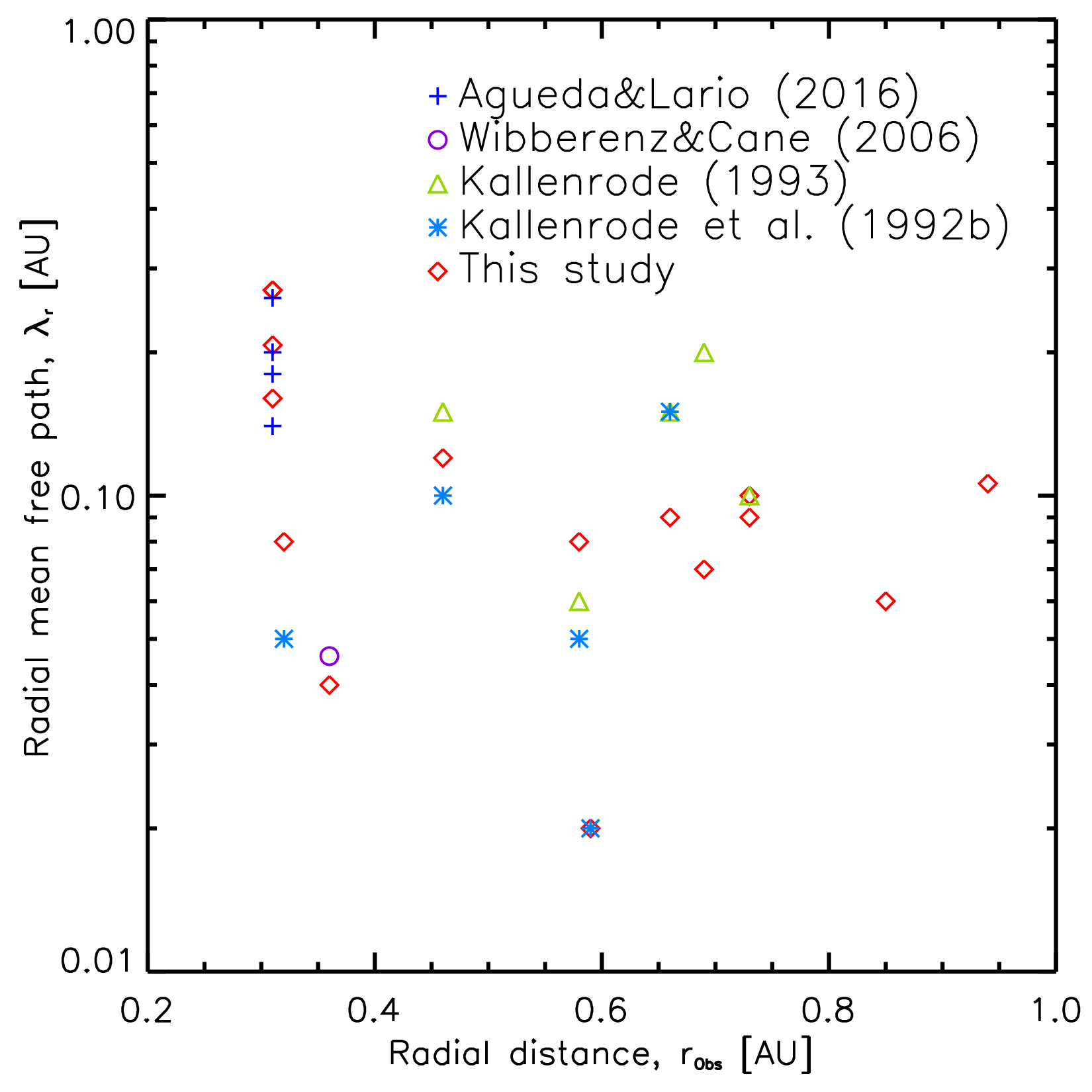}
\par\end{centering}
\caption{Radial mean free path vs. radial distance to the Sun. Red diamonds: values inferred in this study; blue crosses: values from \cite{Agueda16}; purple circle: \cite{WibberenzEtC06}; olive green triangles: values from \cite{Kallenrode93b}; light blue asterisks: values from \cite{Kallenrode92b}.}
\label{5fig:disc1}
\end{figure}

\begin{figure*}[t]
\begin{centering}
\vspace{-0.cm}
\includegraphics[width=0.45\textwidth]{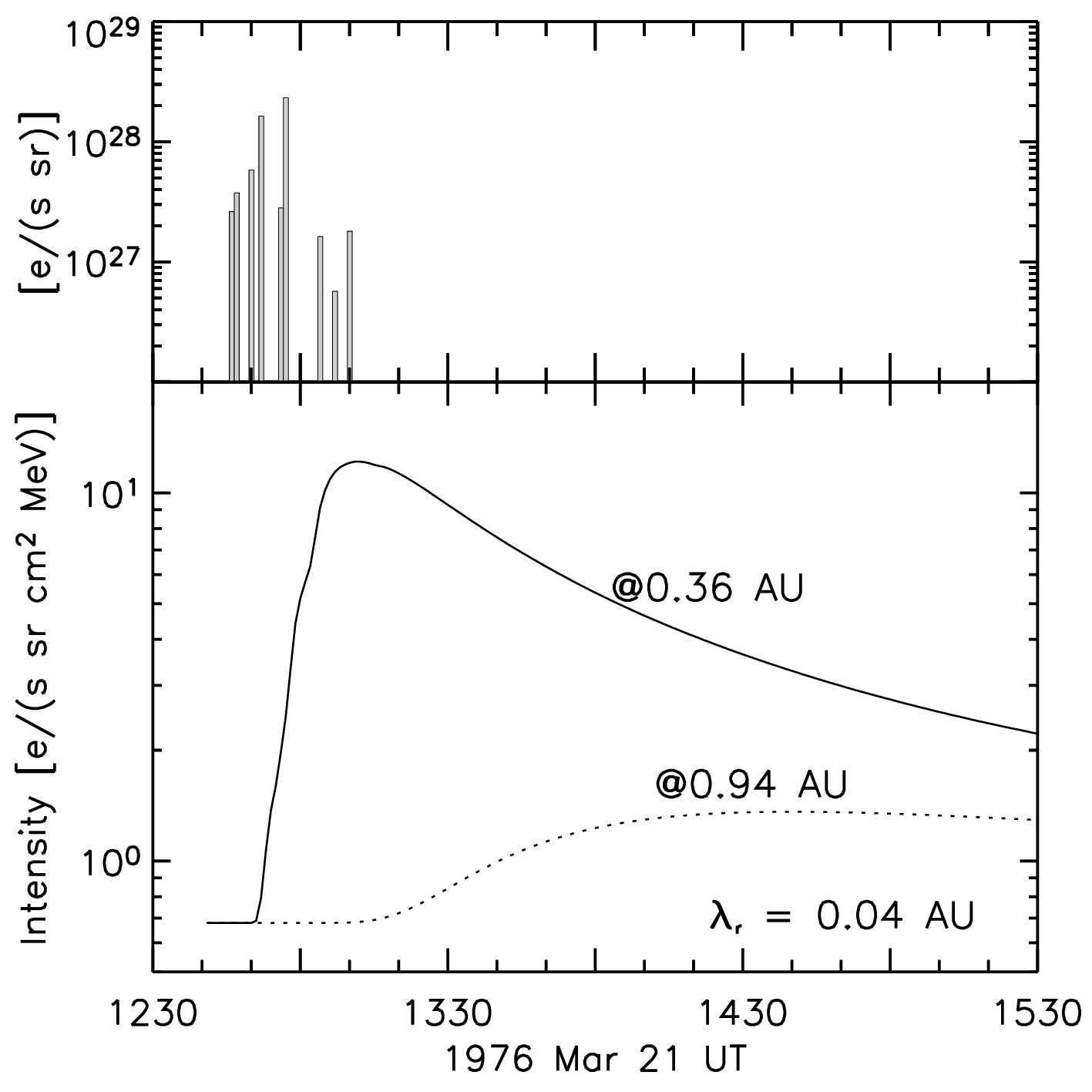}\hspace{2em}
\includegraphics[width=0.45\textwidth]{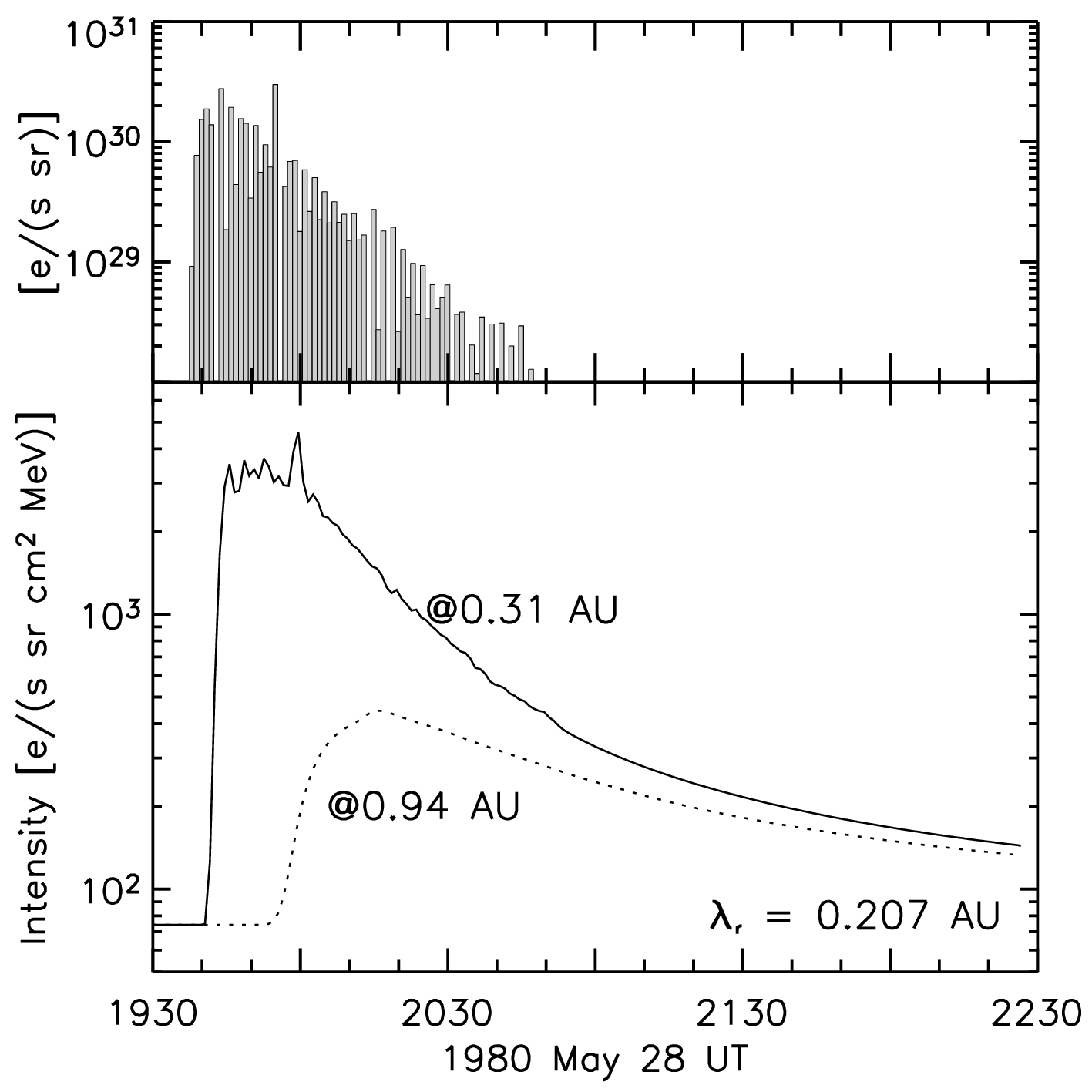}
\par\end{centering}
\caption{Events on 1976 March 21 (left) and 1980 May 28 (right) modelled at different radial distances. The upper panel shows the injection profile at the Sun. The lower panel shows the omni-directional modelled electron event at the Helios location (solid curve) and at 0.94~AU (dashed curve).}
\label{5fig:1au}
\end{figure*}

Therefore, the influence of magnetic structures and the importance of the flare sites being connected to open magnetic field lines has been proved to be a relevant factor for the electron release timing and duration \citep{Klein10}. Then, alternatively to coronal CME-driven shocks, it is reasonable to suggest that extended electron injections can be due to scenarios where the flare site is not well connected to open magnetic field structures in the corona. So even if there is a partial escape of the injected particles coinciding in time with the radio emissions, a bulk of electrons is kept magnetically trapped until they reach open field lines and are gradually leaked into the flux tube connecting with the observer.

\subsection{Intensity profiles and mean free path variation with the heliocentric radial distance}
Finally, we analysed the variation of $\lambda_{r}$ with the radial distance to the observer location. Figure~\ref{5fig:disc1} shows this variation for the events in this study, and those obtained by \cite{Kallenrode92b}, \cite{Kallenrode93b}, \cite{WibberenzEtC06} and \cite{Agueda16} for the events in the present list. We found no evidence of a radial dependence of the radial mean free path, which agrees with the conclusions of \cite{Kallenrode92b}. We expect that observers at small heliocentric distances to the Sun will observe events showing any value of the mean free path, at least between 0.03 and 0.27 AU (as seen in Fig.~\ref{5fig:disc1}), when analysing their transport conditions. For this reason, it will be important to have the full-angular-coverage data from missions like the Parker Solar Probe and Solar Orbiter available in order to be able to disentangle the PADs.

\cite{Kallenrode93b} found a weak dependence of $\lambda_{r}$ with the heliocentric radial distance below 0.5~AU for protons. We did not find such a trend in our sample of electron events (see Fig.~\ref{5fig:disc1}). On the other hand, \cite{Kallenrode93b} suggests that such radial dependence could stem from a bias in the selection of the events, that hinders from choosing events with a diffusive profile near 1~AU. We have further inspected this point. 

Figure~\ref{5fig:1au} shows, for the event on 1976 March 21 (left) and the third event on 1980 May 28 at $\sim$19:30 (right), the inferred electron release profiles at the Sun (top panels) and the electron profiles (lower panels) obtained by convolving the injection profiles with the Green's functions of particle transport for the best fit value of $\lambda_r$ at the Helios location (solid curves) and at a radial distance of 0.94~AU (dashed curves). The profiles show that these two events, clearly observable in the inner heliosphere, appear barely above the background when observed close to 1~AU, especially in the case of the event on 1976 March 21, due to the diffusive transport conditions in the interplanetary medium characterised by the same $\lambda_{r}$ for all helioradii. Therefore, the selection bias may be twofold: (i) some SEP events, especially those evolving under strong diffusive transport conditions, may not be observable at 1~AU because the intensities they show at these distances may remain below the background level and (ii) some SEP events may be instead observed at 1~AU but exhibiting, even under focused transport conditions, low intensity levels that render their modelling difficult given the low statistics of the measurements. Hence, small to middle size (in terms of their peak intensity) SEP events, like those in our sample that are detected at radial distances close to the Sun, might be undetected at larger distances.

\section{Summary}

The full inversion approach presented in this study represents a step forward with respect to previous analysis of Helios observations, as for the first time both energy and angular responses have been taken into account in order to develop a more accurate approach to fit the directional distributions of electrons observed in-situ. 

We scanned the full Helios mission looking for the best-observed electron events to model and we found a sample of 15 events fulfilling the selection criteria. Then, we modelled the angular and the energetic response of the E6 instrument on board Helios and computed a sample of Green's functions for the different transport scenarios, given by different values of the mean free path, taking into account several parameters such as the solar wind speed, the energy spectrum and the radial distance between the Sun and the spacecraft. For each event in our sample, we inferred the injection profile and the value of the radial mean free path that best fitted the observations.

The results suggest values of the radial mean free path between 0.02~AU and 0.27~AU. When we compared the computed parallel mean free paths to the focusing lengths they agree, in general, with weak-focused transport ($0.1 < \lambda_{\parallel}/L < 1$ for ten cases). Four events (the four consecutive events on 1980 May 28) show higher values of the radial mean free path (0.16\,--\,0.27~AU), and suggest focused transport ($\lambda_{\parallel}/L \geq 1$). Only one event (1982 June 2) presents diffusive transport results ($\lambda_{\parallel}/L \leq 0.1$) with $\lambda_{r} = 0.020$~AU.

We compared the obtained values of the radial mean free path and maximum injection with those reported by previous studies \citep{Kallenrode92b, Kallenrode93b, Agueda16} and found that our results were compatible with them for all cases except for the result on 1981 May 8 \citep{Kallenrode93b}, which differs a factor $\sim$3 with our inferred mean free path. We also compared the duration of the events studied by \cite{Kallenrode92b} finding that two of them could be fitted by using short episodes of $\delta$-injections and other three, for which their fit failed adjusting the slow anisotropy decay, we inferred extended injection profiles.

Regarding the injection profiles, we found two separated groups depending on the duration of the injection. We found five short injection profiles (lasting less than 30~min) and ten extended injection profiles (lasting more than 30~min). The value of the maximum injection takes values from $1.7 \times 10^{28}$~[e~(s~sr$)^{-1}$] to $3.1 \times 10^{30}$~[e~(s~sr$)^{-1}$]. The peak and duration of the inferred electron release histories match, in general, with radio and soft x-ray emissions extracted from the literature. We suggested that extended injection profiles can be explained by either coronal CME-driven shocks or complex magnetic structures trapping the electrons and allowing a slow release over a long time period, as discussed by \cite{Klein10} and \cite{Dresing18}.

We found no dependence between the radial mean free path and the radial distance between the Sun and the observer. We compared the modelled profiles at small radial distances with those modelled profiles close to 1~AU and concluded that diffusive events associated with relatively small injection profiles may not be observable at 1~AU. According to this and together with the fact that in our sample we find events observed close to the Sun with a rather wide range of values for $\lambda_r$, we might expect that SEP events to be observed by the Parker Solar Probe and especially by Solar Orbiter, that will travel to similar heliocentric radial distances to those of the Helios orbit, show a large variety of transport conditions. Hence sectored data as it will be provided by the EPD instrument of Solar Orbiter \citep{Rodriguez-Pacheco19} and EPI-Lo and EPI-Hi on board Parker Solar Probe \citep{McComas16} may be important to infer both the transport effects at play in SEP events and the particle release histories. Also, in order to improve our understanding of the electron interplanetary transport conditions (including the role of cross-field transport processes in SEP events in general, e.g. \citealp{StraussEtal17}) and the release processes at the Sun, it will be crucial to have multi-spacecraft observations from several radial distances. This will allow us to study differences in the transport conditions over the heliosphere and characterise the angular extent of their solar sources.

\begin{acknowledgements}
This work was developed under the MINECO predoctoral grants BES-2014-067894 and EEBB-I-17-12302, cofunded by the European Social Fund. The work at University of Barcelona was partly supported by the Spanish MINECO under the project AYA2016-77939-P with partial support by the European Regional Development Fund (ERDF/FEDER). Funding of this work was also partially provided by the Spanish MINECO under the project MDM-2014-0369 of ICCUB (Unidad de Excelencia "Mar\'ia de Maeztu"). DL was partially supported by NASA-LWS grant NNX15AD03G and NASA-HGI grant NNX16AF73G. The authors thank B. Sanahuja and R. Vainio for their valuable comments.
\end{acknowledgements}

\bibliographystyle{aa}
\bibliography{Bibliography_helios}

\begin{thebibliography}{43}
\expandafter\ifx\csname natexlab\endcsname\relax\def\natexlab#1{#1}\fi

\bibitem[{{Agueda}(2008)}]{Agueda08Tesis}
{Agueda}, N. 2008, PhD thesis, University of Barcelona,
  http://hdl.handle.net/10803/749

\bibitem[{{Agueda} {et~al.}(2014){Agueda}, {Klein}, {Vilmer},
  {Rodr{\'{\i}}guez-Gas{\'e}n}, {Malandraki}, {Papaioannou}, {Subir{\`a}},
  {Sanahuja}, {Valtonen}, {Dr{\"o}ge}, {Nindos}, {Heber}, {Braune}, {Usoskin},
  {Heynderickx}, {Talew}, \& {Vainio}}]{Agueda14}
{Agueda}, N., {Klein}, K.-L., {Vilmer}, N., {et~al.} 2014, \aap, 570, A5

\bibitem[{{Agueda} \& {Lario}(2016)}]{Agueda16}
{Agueda}, N. \& {Lario}, D. 2016, \apj, 829, 1

\bibitem[{{Agueda} {et~al.}(2012{\natexlab{a}}){Agueda}, {Lario}, {Ontiveros},
  {Kilpua}, {Sanahuja}, \& {Vainio}}]{Agueda12b}
{Agueda}, N., {Lario}, D., {Ontiveros}, V., {et~al.} 2012{\natexlab{a}},
  \solphys, 281, 319

\bibitem[{{Agueda} {et~al.}(2009){Agueda}, {Lario}, {Vainio}, {Sanahuja},
  {Kilpua}, \& {Pohjolainen}}]{Agueda09a}
{Agueda}, N., {Lario}, D., {Vainio}, R., {et~al.} 2009, \aap, 507, 981

\bibitem[{{Agueda} \& {Vainio}(2013)}]{AguedaEtV13}
{Agueda}, N. \& {Vainio}, R. 2013, J. Space Weather Space Clim., 3, A10

\bibitem[{Agueda {et~al.}(2013)Agueda, Vainio, Dalla, Lario, \&
  Sanahuja}]{Agueda13}
Agueda, N., Vainio, R., Dalla, S., Lario, D., \& Sanahuja, B. 2013, \apj, 765,
  83

\bibitem[{{Agueda} {et~al.}(2008){Agueda}, {Vainio}, {Lario}, \&
  {Sanahuja}}]{Agueda08}
{Agueda}, N., {Vainio}, R., {Lario}, D., \& {Sanahuja}, B. 2008, \apj, 675,
  1601

\bibitem[{{Agueda} {et~al.}(2012{\natexlab{b}}){Agueda}, {Vainio}, \&
  {Sanahuja}}]{Agueda12}
{Agueda}, N., {Vainio}, R., \& {Sanahuja}, B. 2012{\natexlab{b}}, \apjs, 202,
  18

\bibitem[{{Aran} {et~al.}(2018){Aran}, {Agueda}, {Afanasiev}, \&
  {Sanahuja}}]{Aran2018}
{Aran}, A., {Agueda}, N., {Afanasiev}, A., \& {Sanahuja}, B. 2018, Charged
  Particle Transport in the Interplanetary Medium, ed. O.~E. {Malandraki} \&
  N.~B. {Crosby} (Springer International Publishing), 63--78

\bibitem[{{Beeck} \& {Wibberenz}(1986)}]{Beeck86}
{Beeck}, J. \& {Wibberenz}, G. 1986, \apj, 311, 437

\bibitem[{{Bialk} {et~al.}(1991){Bialk}, {Dr{\"{o}}ge}, \& {Heber}}]{Bialk91}
{Bialk}, M., {Dr{\"{o}}ge}, W., \& {Heber}, B. 1991, International Cosmic Ray
  Conference, 3, 764

\bibitem[{{Dresing} {et~al.}(2018){Dresing}, {G{\'o}mez-Herrero}, {Heber},
  {Klassen}, {Temmer}, \& {Veronig}}]{Dresing18}
{Dresing}, N., {G{\'o}mez-Herrero}, R., {Heber}, B., {et~al.} 2018, \aap, 613,
  A21

\bibitem[{{Dr\"oge} {et~al.}(1993){Dr\"oge}, {Achatz}, {Wanner},
  {Schlickeiser}, \& {Wibberenz}}]{Droge93}
{Dr\"oge}, W., {Achatz}, U., {Wanner}, W., {Schlickeiser}, R., \& {Wibberenz},
  G. 1993, \apjl, 407, L95

\bibitem[{{Dr{\"{o}}ge} {et~al.}(2016){Dr{\"{o}}ge}, {Kartavykh}, {Dresing}, \&
  {Klassen}}]{Droge16}
{Dr{\"{o}}ge}, W., {Kartavykh}, Y.~Y., {Dresing}, N., \& {Klassen}, A. 2016,
  \apj, 134, 1

\bibitem[{{Dr{\"o}ge} {et~al.}(2010){Dr{\"o}ge}, {Kartavykh}, {Klecker}, \&
  {Kovaltsov}}]{Droge10}
{Dr{\"o}ge}, W., {Kartavykh}, Y.~Y., {Klecker}, B., \& {Kovaltsov}, G.~A. 2010,
  \apj, 709, 912

\bibitem[{{Fr{\"a}nz} \& {Harper}(2002)}]{Franz02}
{Fr{\"a}nz}, M. \& {Harper}, D. 2002, \planss, 50, 217

\bibitem[{Gardini {et~al.}(2011)Gardini, Laurenza, \& Storini}]{Gardini11}
Gardini, A., Laurenza, M., \& Storini, M. 2011, Adv. Space Res., 47, 2127

\bibitem[{{G{\'o}mez-Herrero} {et~al.}(2015){G{\'o}mez-Herrero}, {Dresing},
  {Klassen}, {Heber}, {Lario}, {Agueda}, {Malandraki}, {Blanco},
  {Rodr{\'i}guez-Pacheco}, \& {Banjac}}]{Gomez-Herrero15}
{G{\'o}mez-Herrero}, R., {Dresing}, N., {Klassen}, A., {et~al.} 2015, \apj,
  799, 55

\bibitem[{{Heber} {et~al.}(2018){Heber}, {Agueda}, {B{\"u}tikofer}, {Galsdorf},
  {Herbst}, {K{\"u}hl}, {Labrenz}, \& {Vainio}}]{Heber18}
{Heber}, B., {Agueda}, N., {B{\"u}tikofer}, R., {et~al.} 2018, in Astrophysics
  and Space Science Library, Vol. 444, Solar Particle Radiation Storms
  Forecasting and Analysis, ed. O.~E. {Malandraki} \& N.~B. {Crosby}, 179--199

\bibitem[{Jokipii(1966)}]{Jokipii66}
Jokipii, J.~R. 1966, \apj, 146, 480

\bibitem[{{Kallenrode}(1993)}]{Kallenrode93b}
{Kallenrode}, M.-B. 1993, \jgr, 98, 19

\bibitem[{{Kallenrode} {et~al.}(1992{\natexlab{a}}){Kallenrode}, {Cliver}, \&
  {Wibberenz}}]{Kallenrode92a}
{Kallenrode}, M.-B., {Cliver}, E.~W., \& {Wibberenz}, G. 1992{\natexlab{a}},
  \apj, 391, 370

\bibitem[{{Kallenrode} {et~al.}(1992{\natexlab{b}}){Kallenrode}, {Wibberenz},
  \& {Hucke}}]{Kallenrode92b}
{Kallenrode}, M.-B., {Wibberenz}, G., \& {Hucke}, S. 1992{\natexlab{b}}, \apj,
  394, 351

\bibitem[{{Klein} {et~al.}(2008){Klein}, {Krucker}, {Lointier}, \&
  {Kerdraon}}]{Klein08}
{Klein}, K.-L., {Krucker}, S., {Lointier}, G., \& {Kerdraon}, A. 2008, \aap,
  486, 589

\bibitem[{{Klein} {et~al.}(2010){Klein}, {Trottet}, \& {Klassen}}]{Klein10}
{Klein}, K.-L., {Trottet}, G., \& {Klassen}, A. 2010, \solphys, 263, 185

\bibitem[{{Kunow} {et~al.}(1975){Kunow}, {Wibberenz}, {Green},
  {M{\"u}ller-Mellin}, {Witte}, \& {Hempe}}]{Kunow75}
{Kunow}, H., {Wibberenz}, G., {Green}, G., {et~al.} 1975, Raumfahrtforschung,
  19, 253

\bibitem[{{Kunow} {et~al.}(1977){Kunow}, {Witte}, {Wibberenz}, {Hempe},
  {Mueller-Mellin}, {Green}, {Iwers}, \& {Fuckner}}]{Kunow77}
{Kunow}, H., {Witte}, M., {Wibberenz}, G., {et~al.} 1977, Journal of Geophysics
  Zeitschrift Geophysik, 42, 615

\bibitem[{{Lario} {et~al.}(2006){Lario}, {Kallenrode}, {Decker}, {Roelof},
  {Krimigis}, {Aran}, \& {Sanahuja}}]{Lario06}
{Lario}, D., {Kallenrode}, M.-B., {Decker}, R.~B., {et~al.} 2006, \apj, 653,
  1531

\bibitem[{{McComas} {et~al.}(2016){McComas}, {Alexander}, {Angold}, {Bale},
  {Beebe}, {Birdwell}, {Boyle}, {Burgum}, {Burnham}, {Christian}, {Cook},
  {Cooper}, {Cummings}, {Davis}, {Desai}, {Dickinson}, {Dirks}, {Do}, {Fox},
  {Giacalone}, {Gold}, {Gurnee}, {Hayes}, {Hill}, {Kasper}, {Kecman}, {Klemic},
  {Krimigis}, {Labrador}, {Layman}, {Leske}, {Livi}, {Matthaeus}, {McNutt},
  {Mewaldt}, {Mitchell}, {Nelson}, {Parker}, {Rankin}, {Roelof}, {Schwadron},
  {Seifert}, {Shuman}, {Stokes}, {Stone}, {Vandegriff}, {Velli}, {von
  Rosenvinge}, {Weidner}, {Wiedenbeck}, \& {Wilson}}]{McComas16}
{McComas}, D.~J., {Alexander}, N., {Angold}, N., {et~al.} 2016, \ssr, 204, 187

\bibitem[{{Musmann} {et~al.}(1975){Musmann}, {Neubauer}, {Maier}, \&
  E.}]{Musmann75}
{Musmann}, G., {Neubauer}, F.~M., {Maier}, A., \& E., L. 1975,
  Raumfahrtforschung, 19, 232

\bibitem[{{Pacheco} {et~al.}(2017){Pacheco}, {Agueda}, {G{\'o}mez-Herrero}, \&
  {Aran}}]{Pacheco17}
{Pacheco}, D., {Agueda}, N., {G{\'o}mez-Herrero}, R., \& {Aran}, A. 2017, J.
  Space Weather Space Clim., 7, A30

\bibitem[{{Porsche}(1975)}]{Porsche75}
{Porsche}, H. 1975, Raumfahrtforschung, 19, 223

\bibitem[{{Reames} {et~al.}(1996){Reames}, {Barbier}, \& {Ng}}]{Re96}
{Reames}, D.~V., {Barbier}, L.~M., \& {Ng}, C.~K. 1996, \apj, 466, 473

\bibitem[{{Rodr{\'i}guez-Pacheco} {et~al.}(2019){Rodr{\'i}guez-Pacheco},
  {Wimmer-Schweingruber}, {Mason}, {Ho}, {Sánchez-Prieto}, {Prieto},
  {Mart\'{i}n}, {Seifert}, {Andrews}, {Kulkarni}, L., \&
  {Boden}}]{Rodriguez-Pacheco19}
{Rodr{\'i}guez-Pacheco}, J., {Wimmer-Schweingruber}, R.~F., {Mason}, G.~M.,
  {et~al.} 2019, Submitted to \aap

\bibitem[{{Roelof}(1969)}]{Roelof69}
{Roelof}, E.~C. 1969, in Lectures in High-Energy Astrophysics, ed.
  {{{\"O}gelman}, H. and {Wayland}, J.~R.}, 111

\bibitem[{{Ruffolo} {et~al.}(1998){Ruffolo}, {Khumlumlert}, \&
  {Youngdee}}]{Ruffolo98}
{Ruffolo}, D., {Khumlumlert}, T., \& {Youngdee}, W. 1998, \jgr (Space Physics),
  1981 [\eprint{9806097v1}]

\bibitem[{{Schwenn} \& {Marsch}(1990)}]{Heliosbook90}
{Schwenn}, R. \& {Marsch}, E. 1990, {Physics of the Inner Heliosphere I.
  Large-Scale Phenomena.} (Springer-Verlag, Berlin, Heidelberg), 103

\bibitem[{{Schwenn} \& {Marsch}(1991)}]{Heliosbook91}
{Schwenn}, R. \& {Marsch}, E. 1991, {Physics of the Inner Heliosphere II.
  Particles, Waves and Turbulence.} (Springer, Berlin, Heidelberg), 152

\bibitem[{{Schwenn} {et~al.}(1975){Schwenn}, {Rosenbauer}, \&
  {Miggenrieder}}]{Schwenn75}
{Schwenn}, R., {Rosenbauer}, H., \& {Miggenrieder}, H. 1975,
  Raumfahrtforschung, 19, 226

\bibitem[{{Strauss} \& {Fichtner}(2015)}]{StraussEtF15}
{Strauss}, R.~D. \& {Fichtner}, H. 2015, \apj, 801, 29

\bibitem[{{Strauss} {et~al.}(2017){Strauss}, {Dresing}, \&
  {Engelbrecht}}]{StraussEtal17}
{Strauss}, R.~D.~T., {Dresing}, N., \& {Engelbrecht}, N.~E. 2017, \apj, 837, 43

\bibitem[{{Wibberenz} \& {Cane}(2006)}]{WibberenzEtC06}
{Wibberenz}, G. \& {Cane}, H.~V. 2006, \apj, 650, 1199

\end{thebibliography}
\clearpage
\begin{appendix}

\section{Fits of the events}

\begin{figure*}[b!]
\begin{centering}
\vspace{-0.cm}
\includegraphics[height=0.188\textheight]{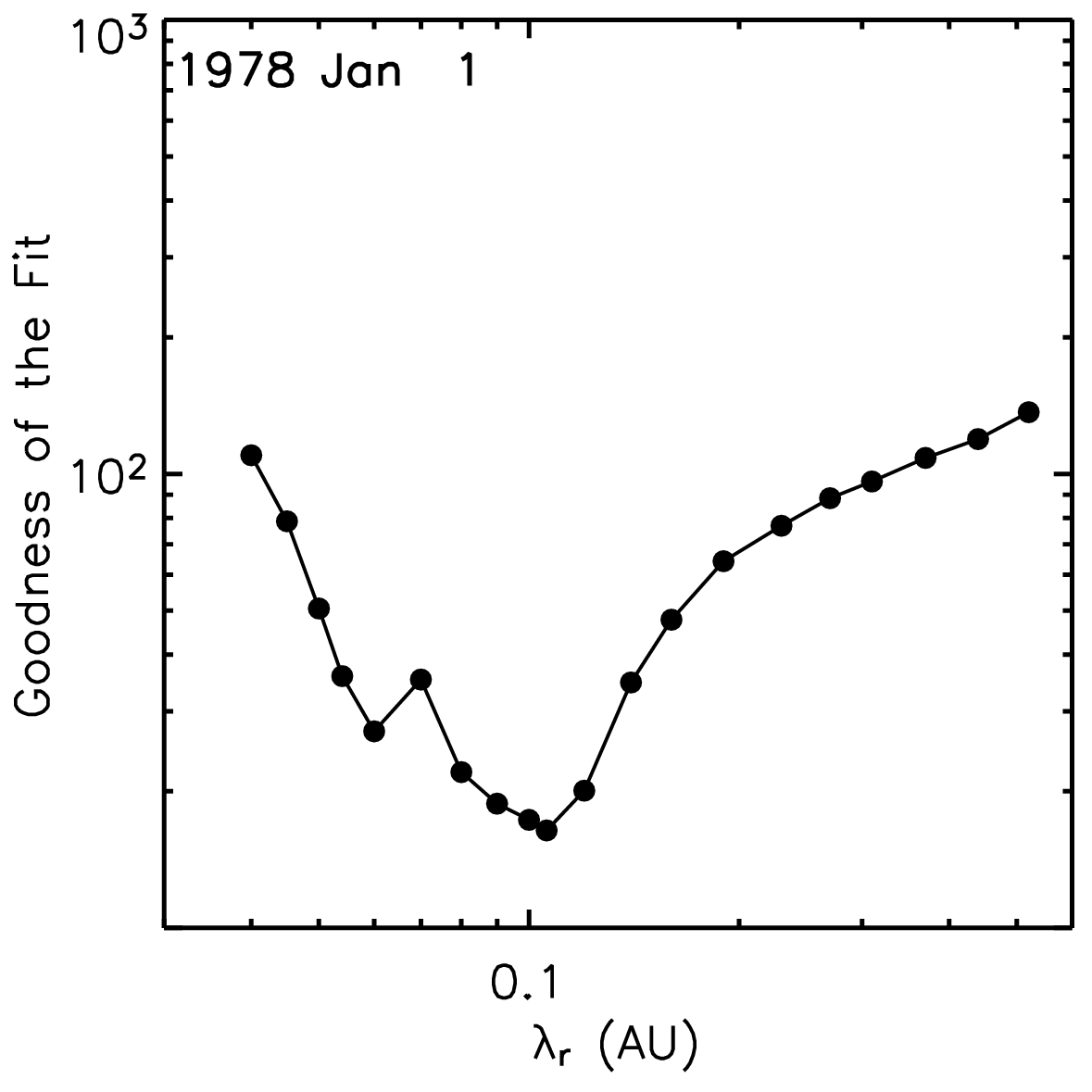}
\includegraphics[height=0.188\textheight]{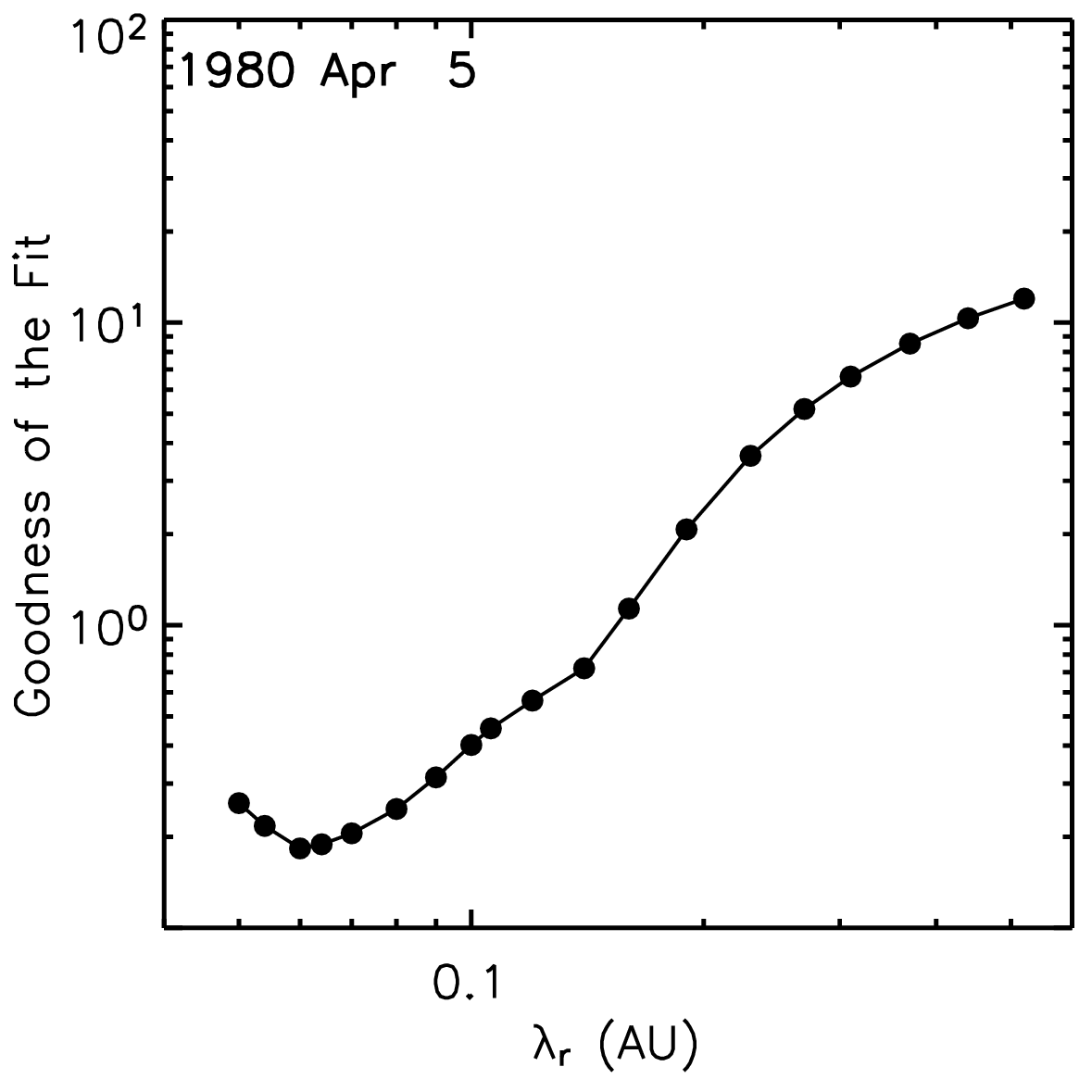}
\includegraphics[height=0.188\textheight]{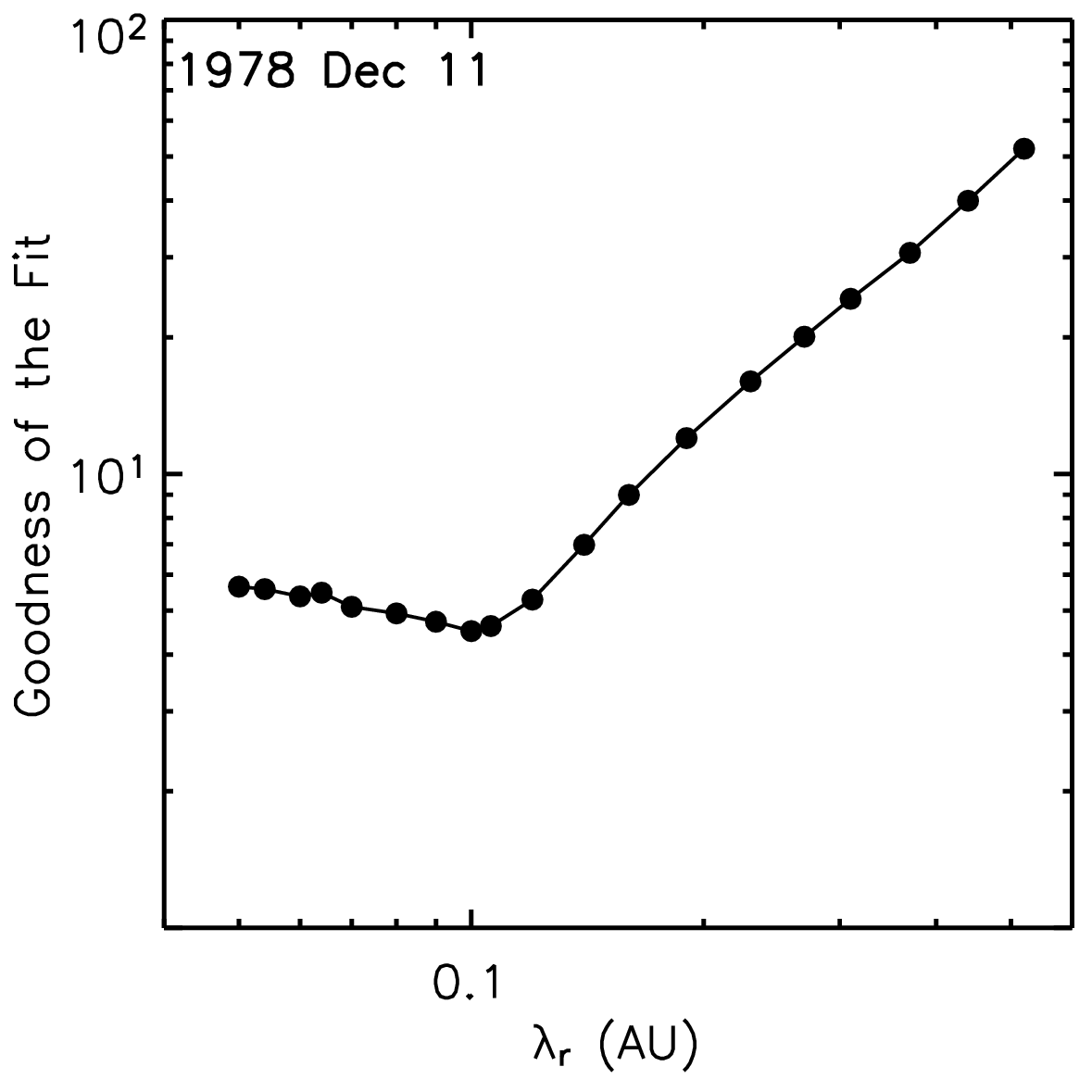}
\includegraphics[height=0.188\textheight]{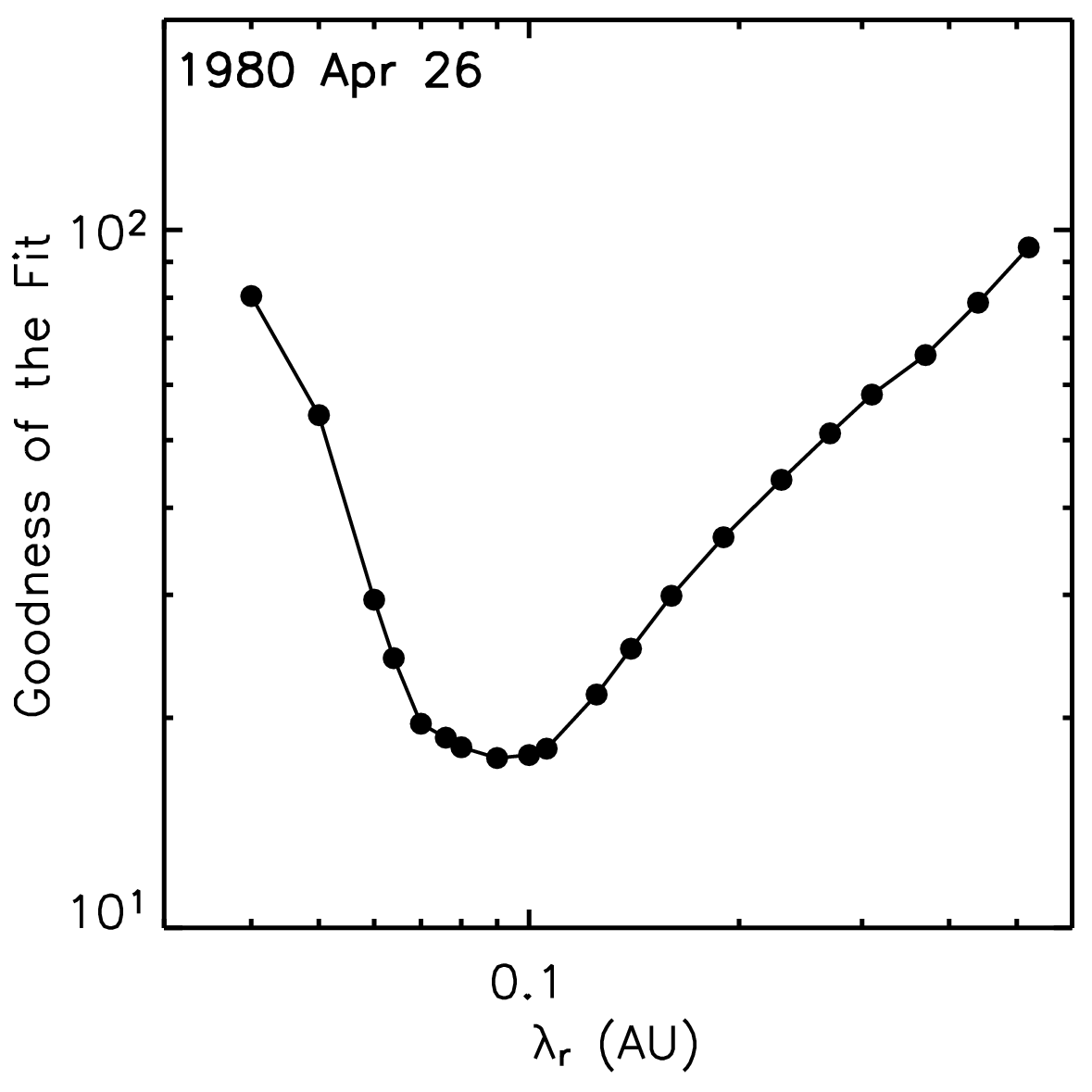}
\includegraphics[height=0.188\textheight]{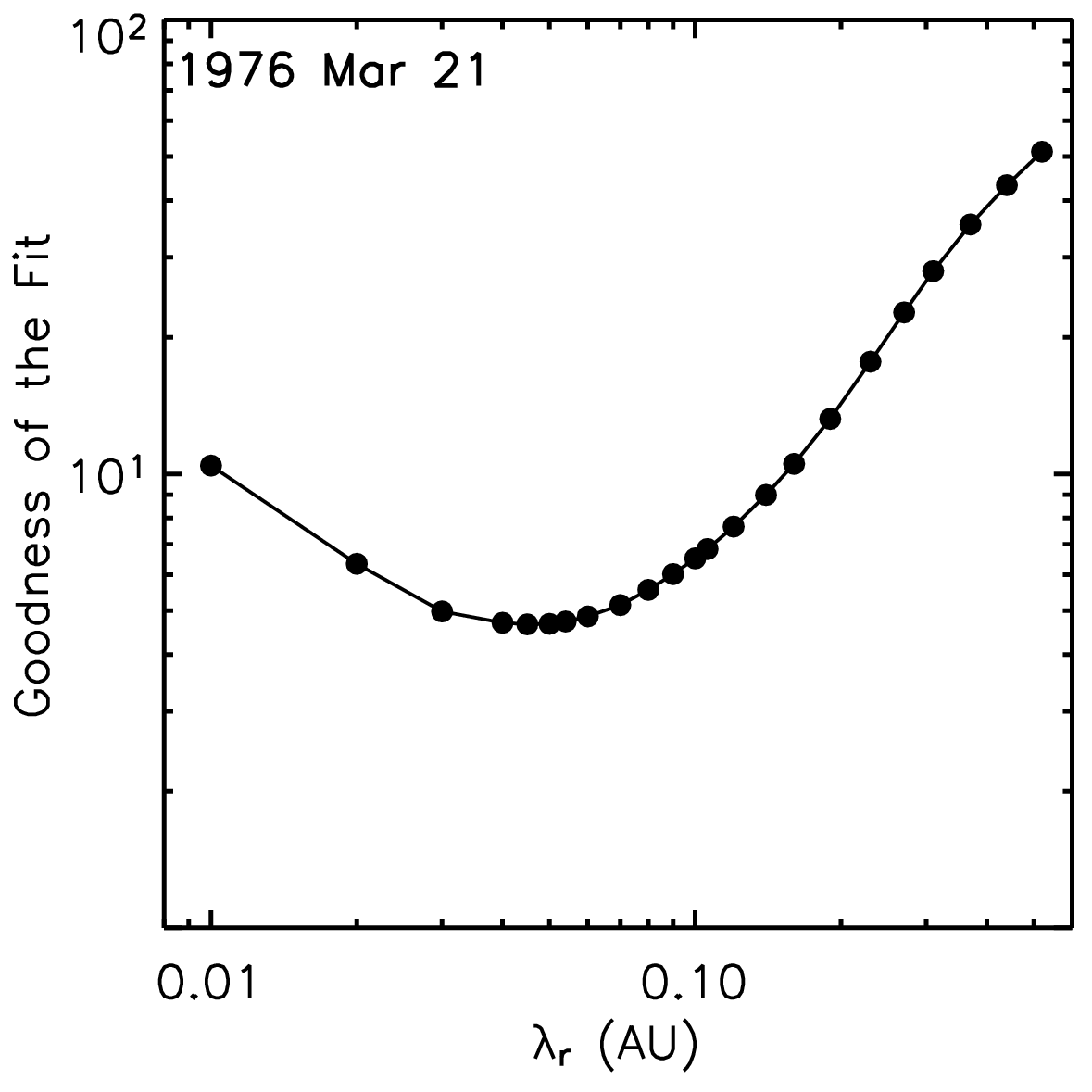}
\includegraphics[height=0.188\textheight]{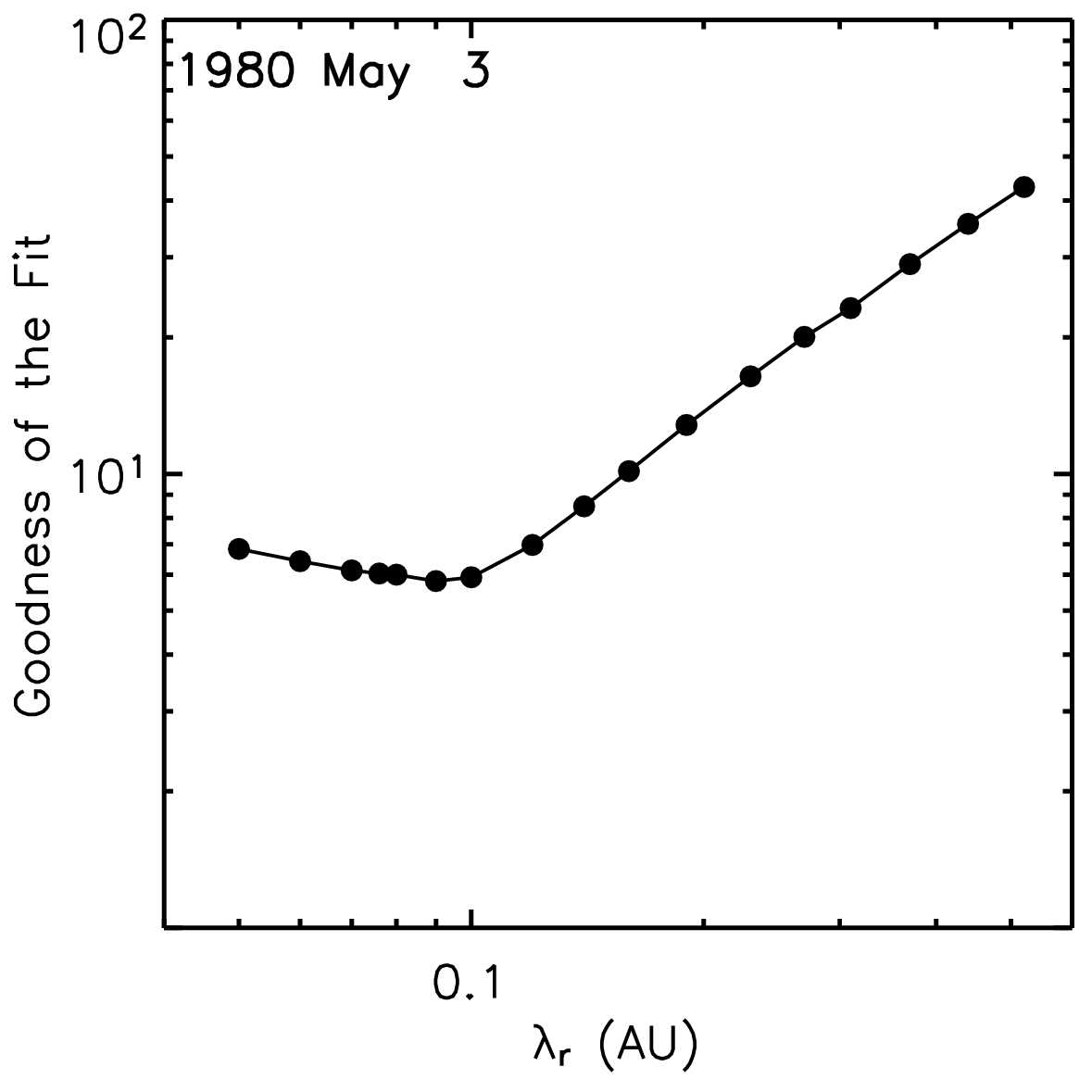}
\includegraphics[height=0.188\textheight]{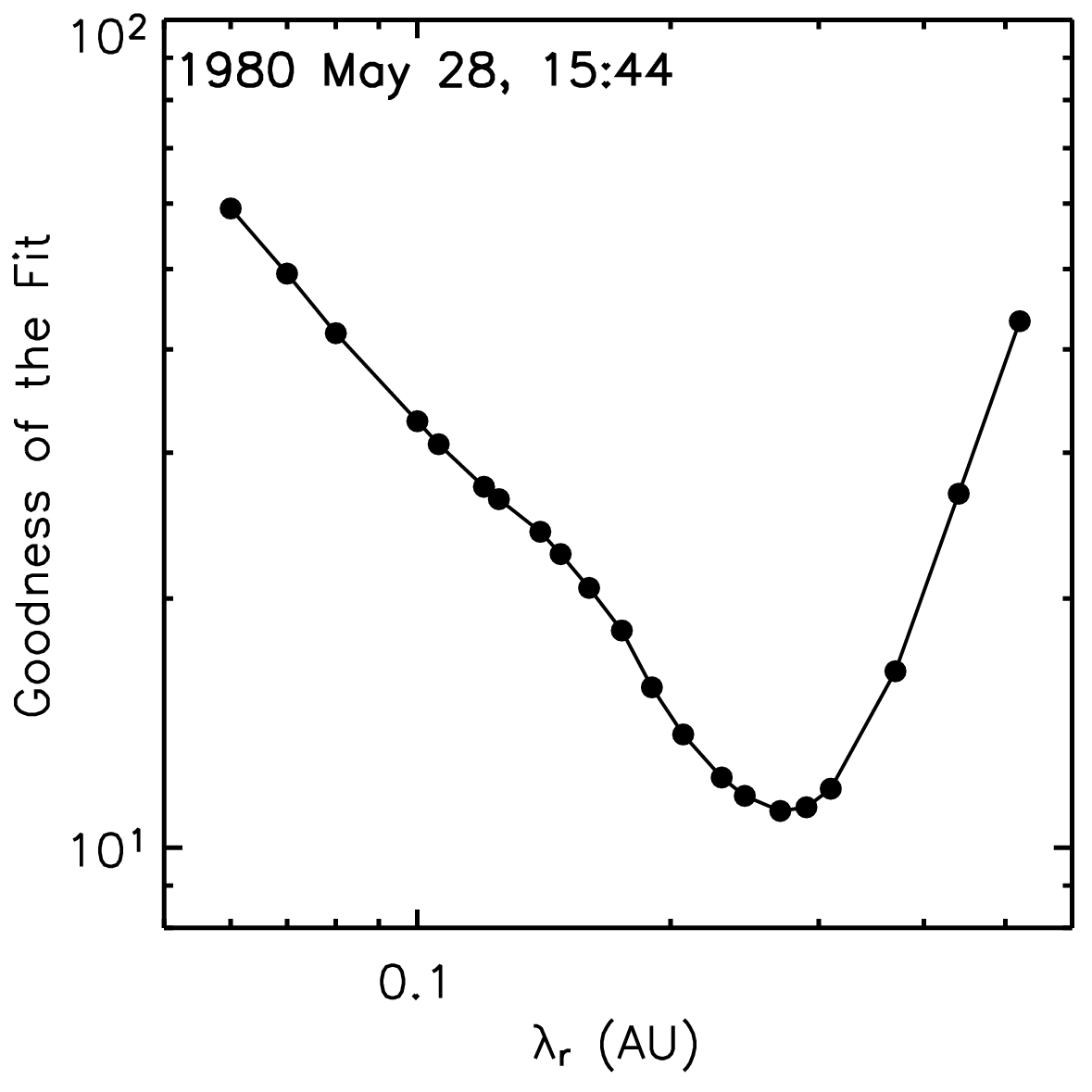}
\includegraphics[height=0.188\textheight]{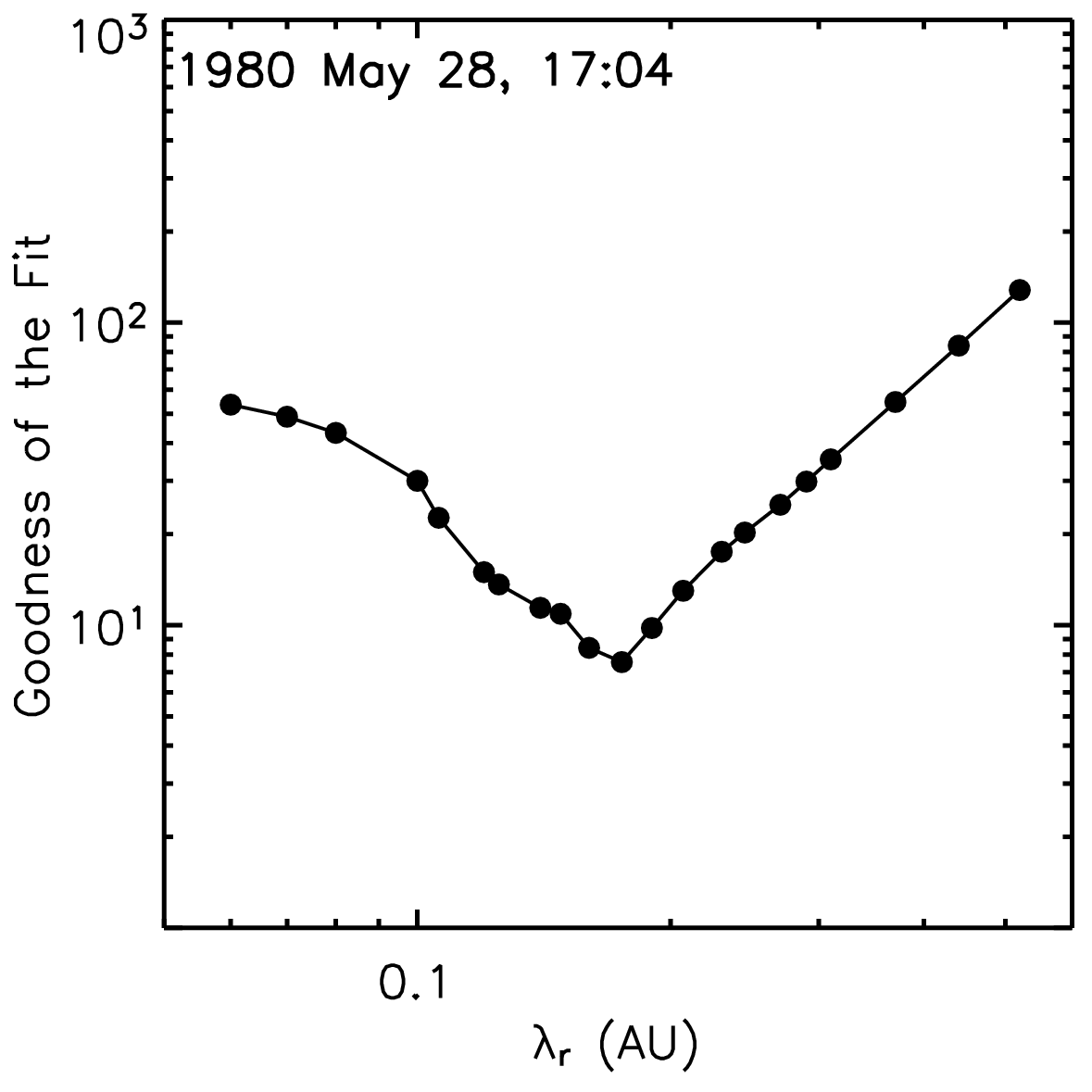}
\includegraphics[height=0.188\textheight]{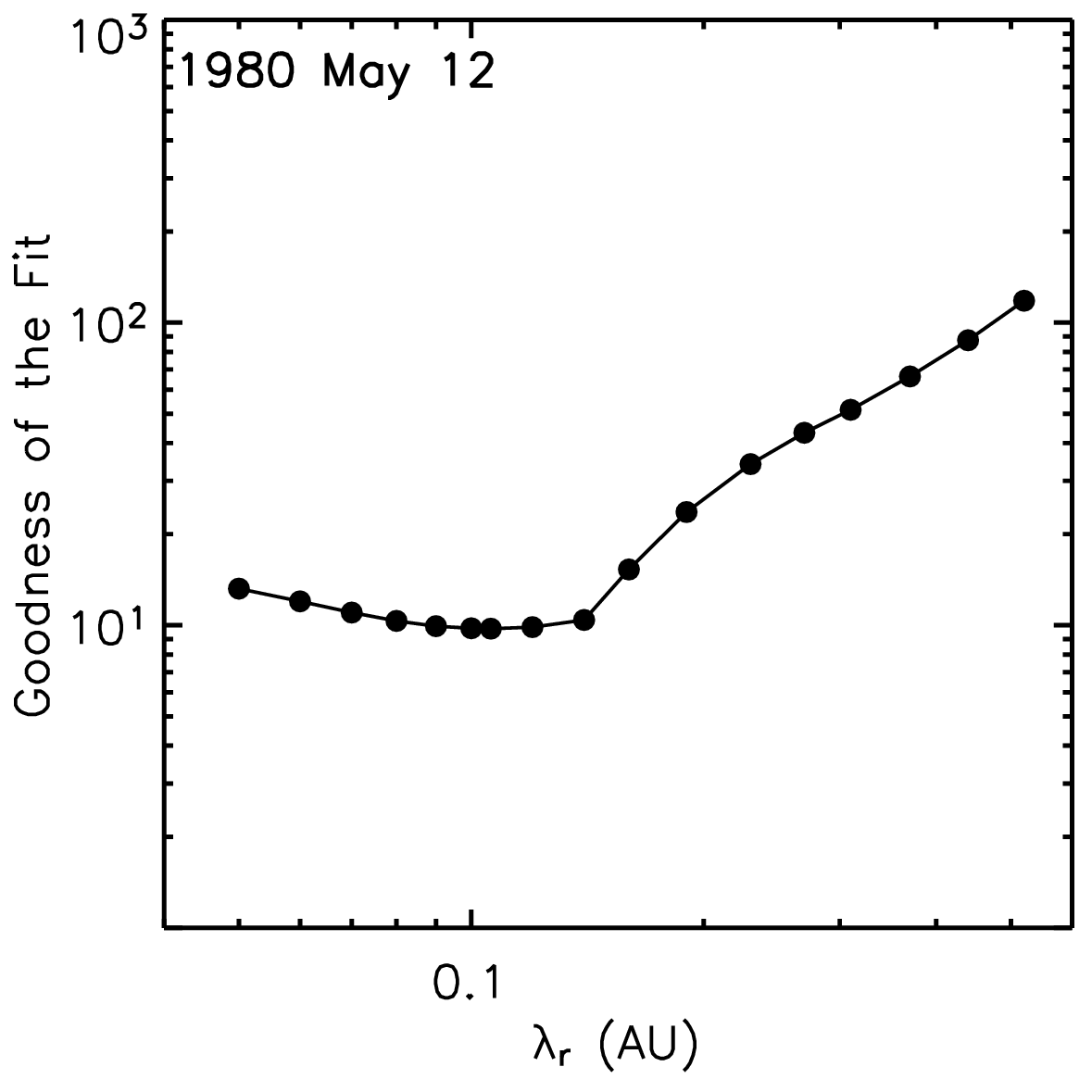}
\includegraphics[height=0.188\textheight]{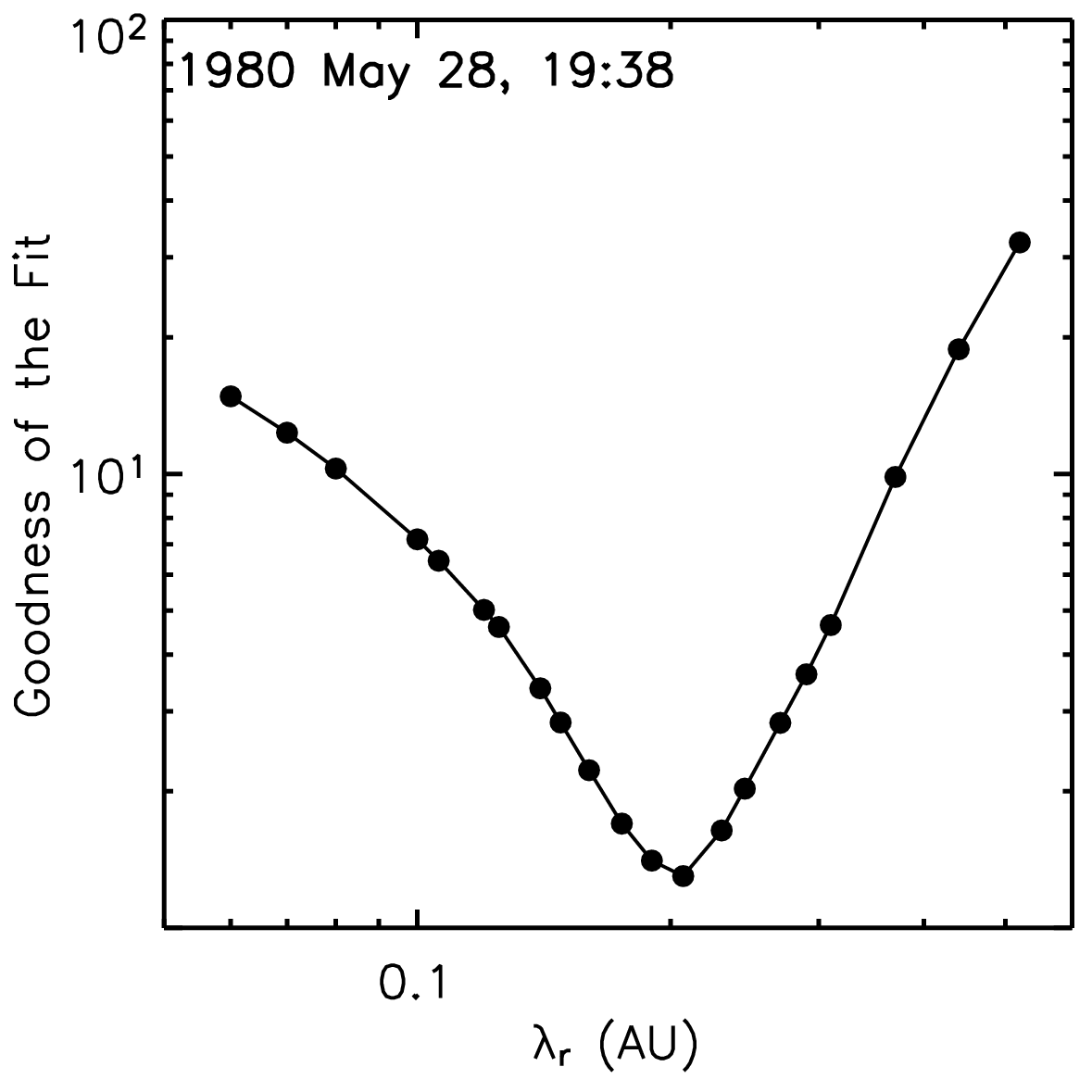}
\includegraphics[height=0.188\textheight]{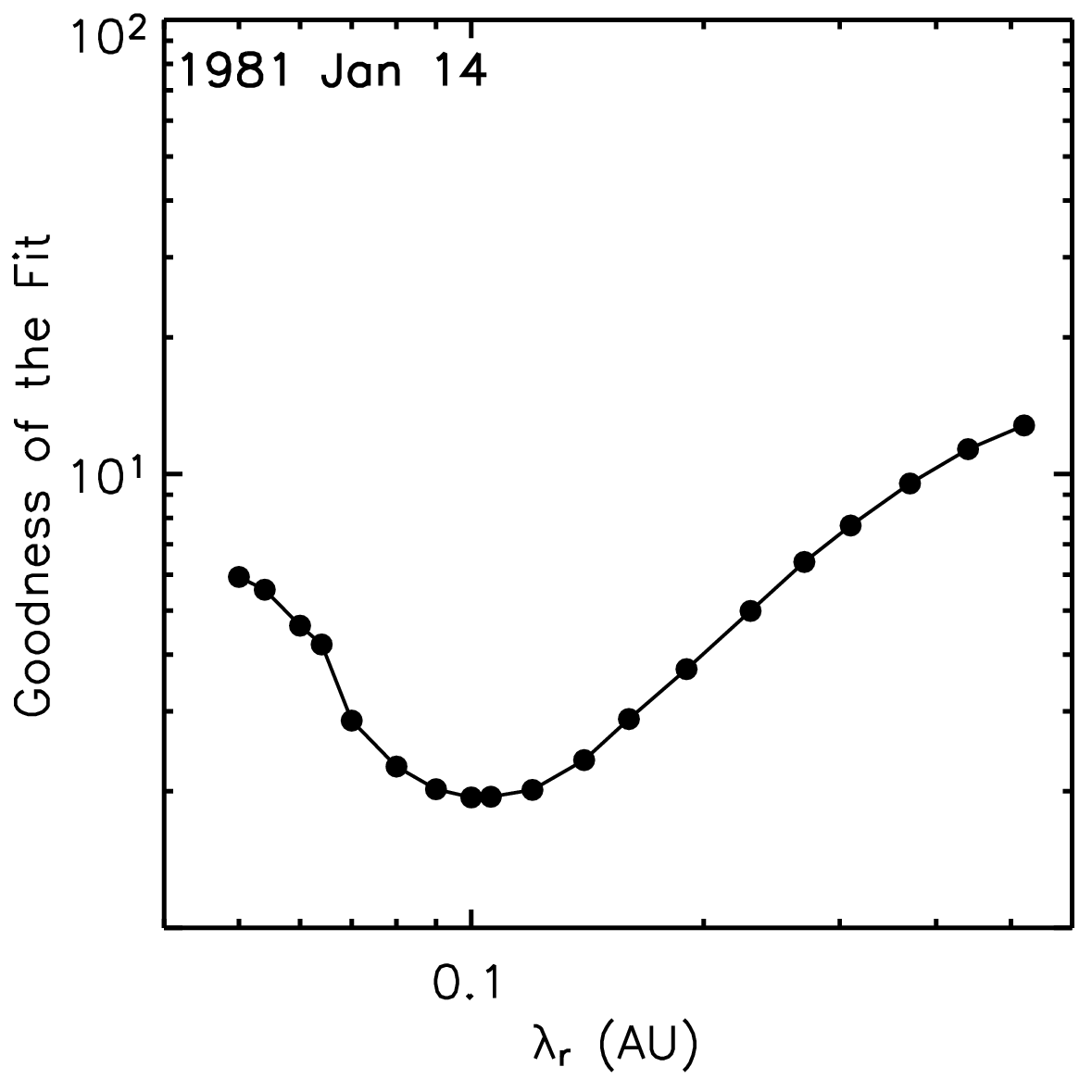}
\includegraphics[height=0.188\textheight]{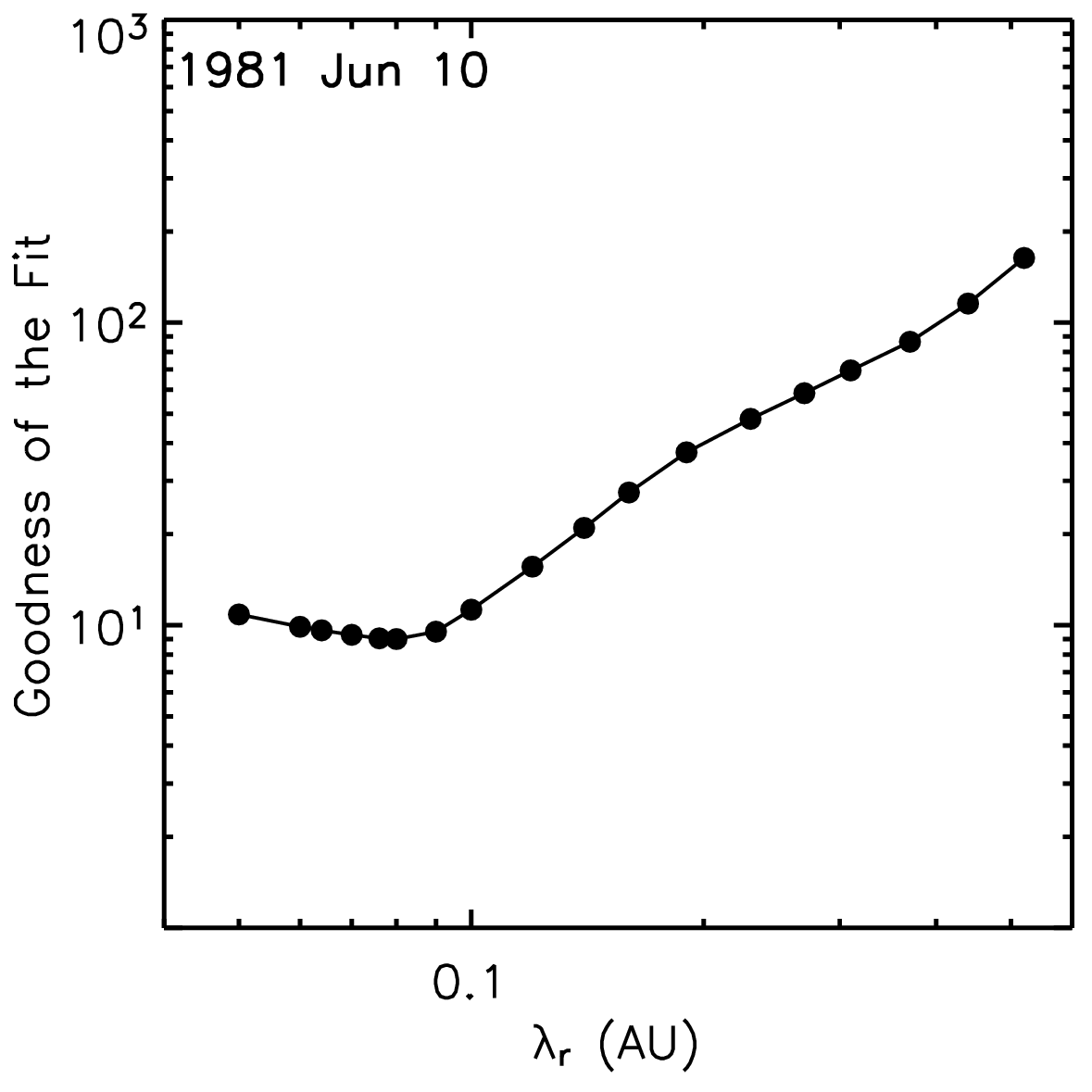}
\includegraphics[height=0.188\textheight]{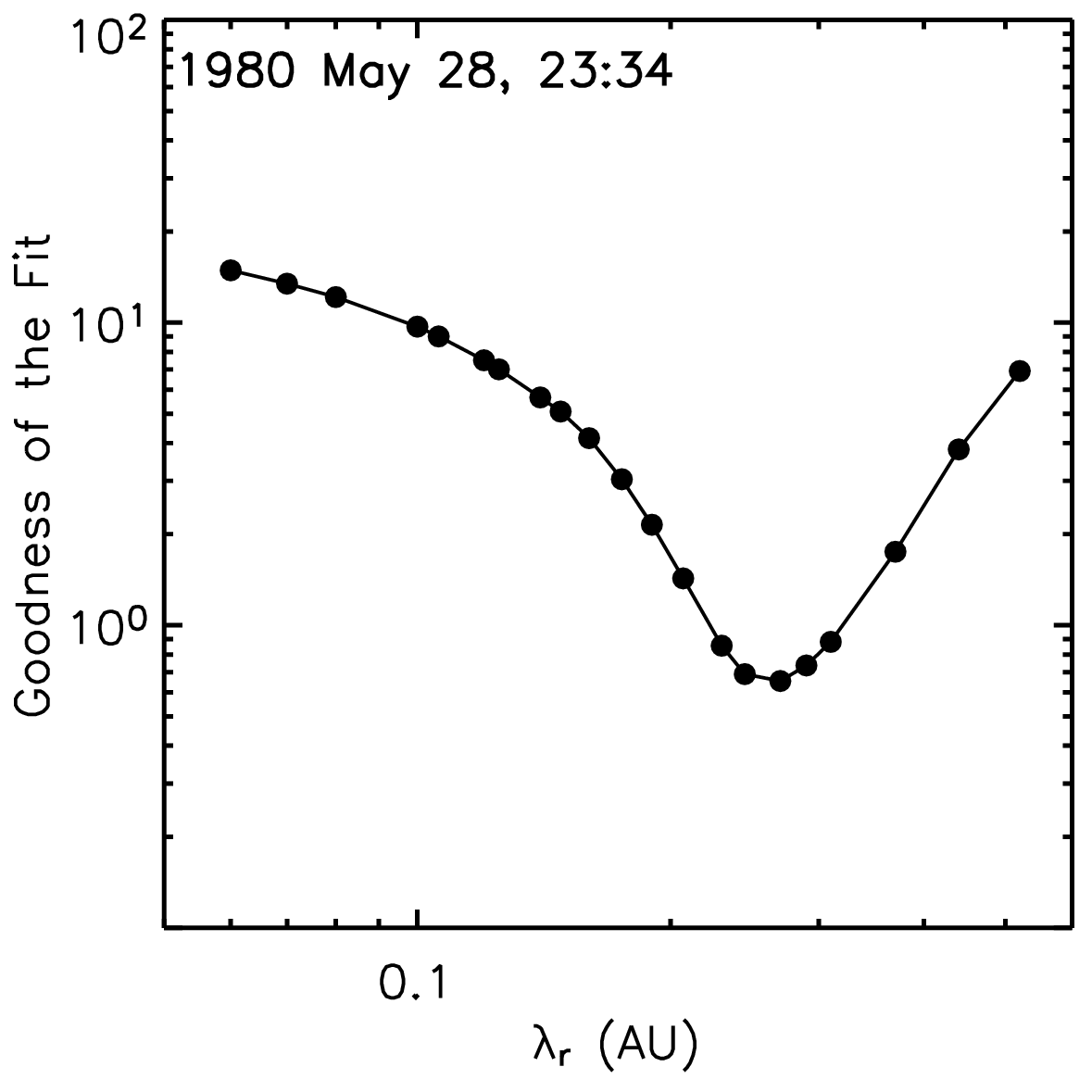}
\includegraphics[height=0.188\textheight]{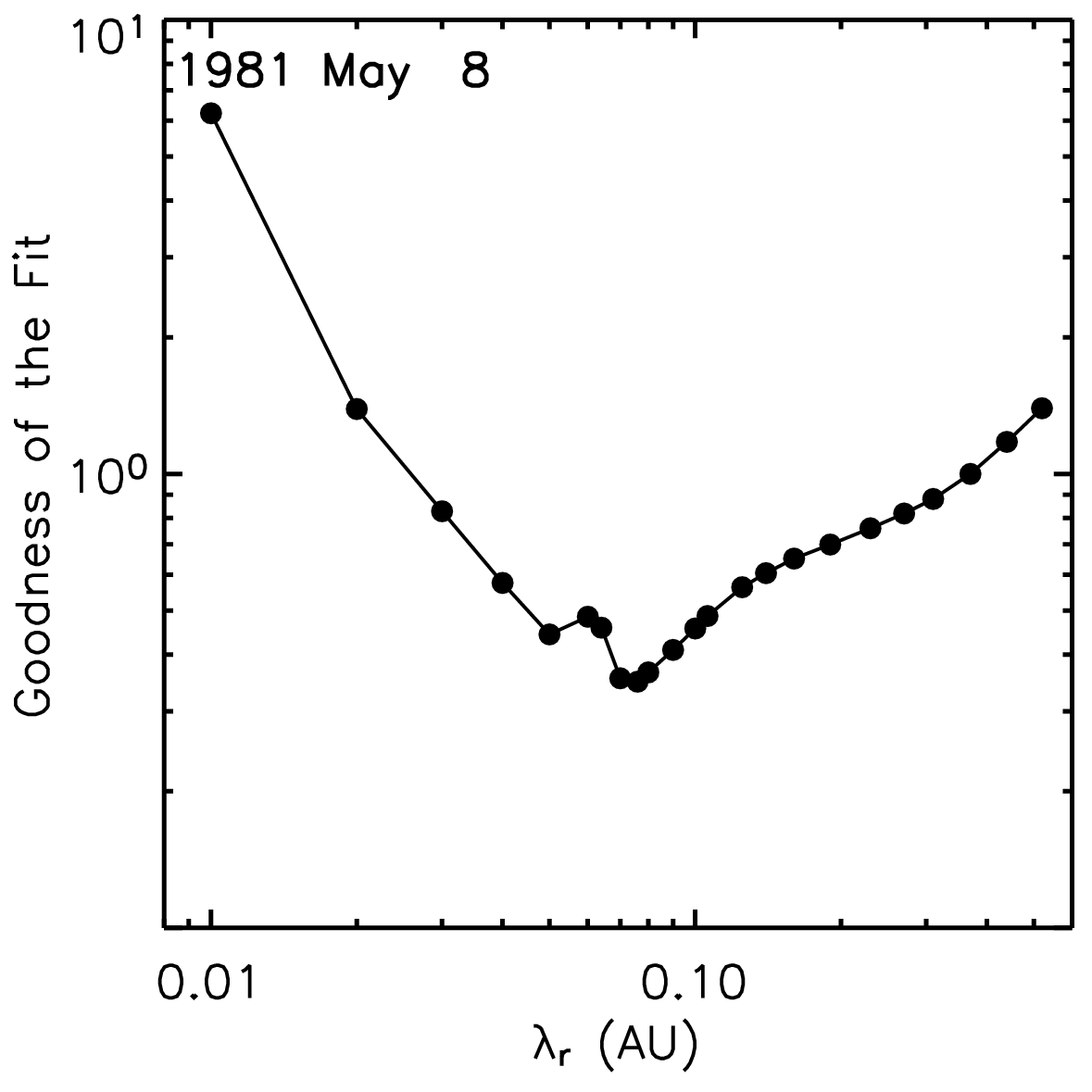}
\includegraphics[height=0.188\textheight]{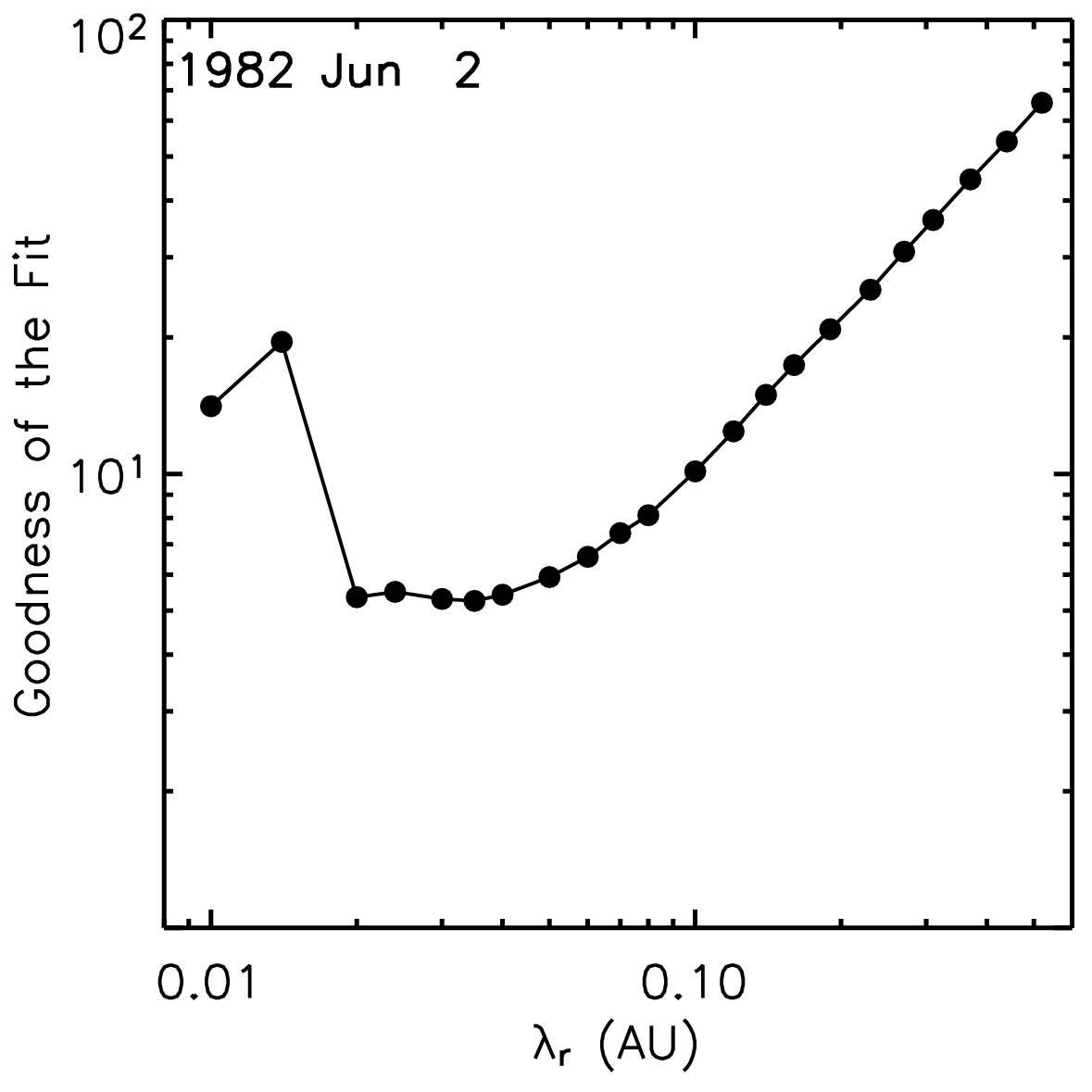}
\par\end{centering}
\caption{Goodness of the fit for every value of $\lambda_{r}$ tested. Most of the events show a clear minimum around the best fit value.}
\label{5fig:GOF}
\end{figure*}

\begin{figure*}
\begin{centering}
\includegraphics[height=0.325\textheight]{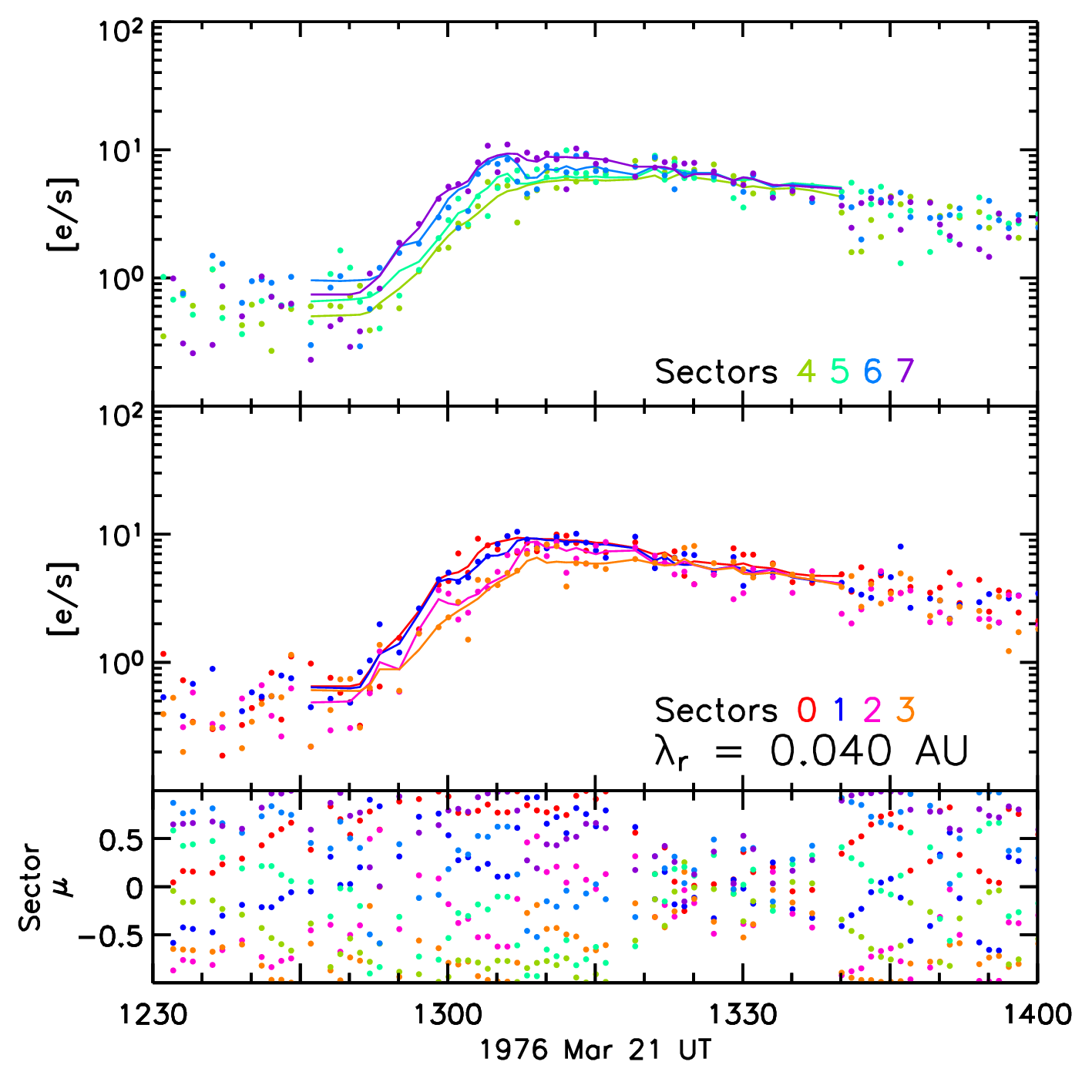}
\includegraphics[height=0.325\textheight]{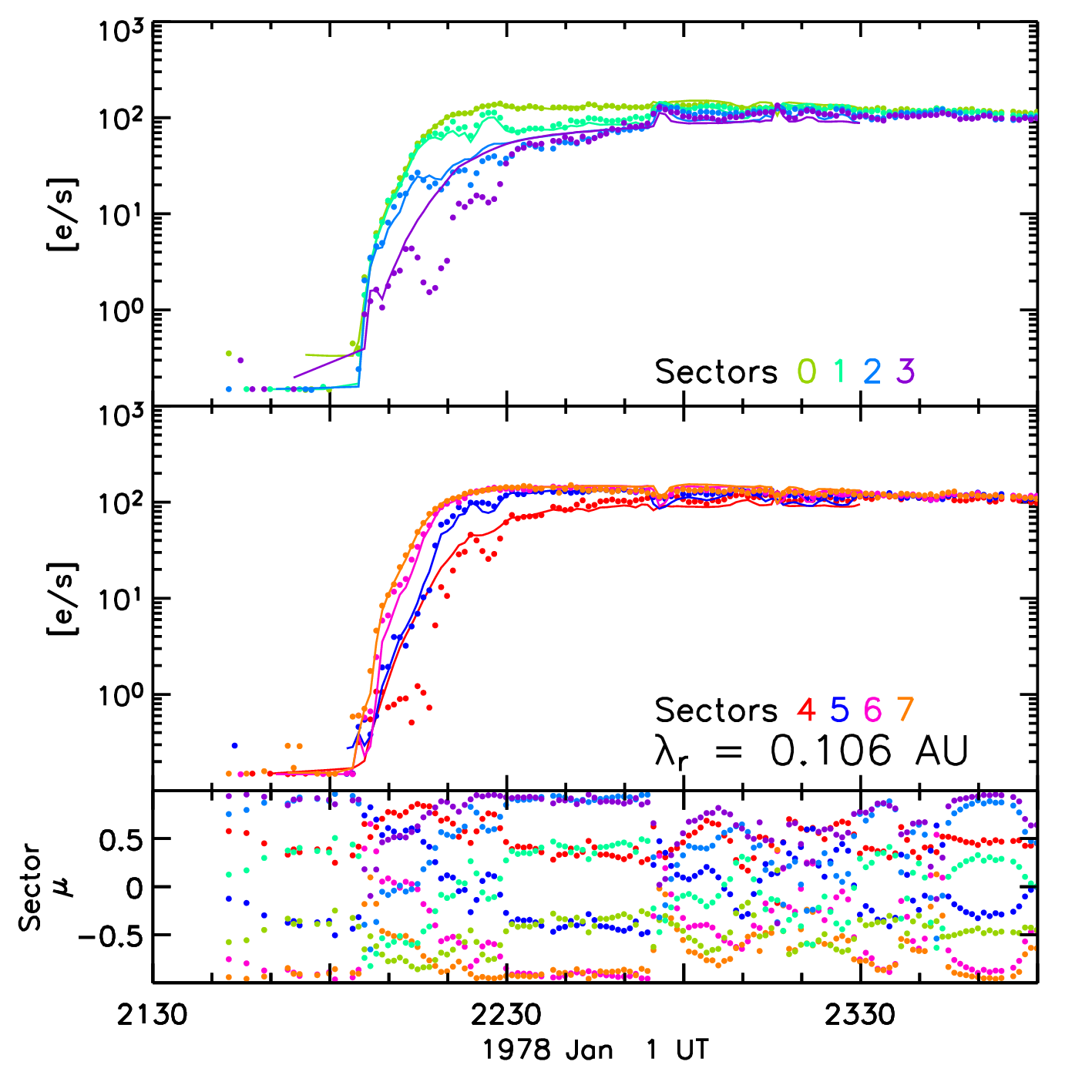}
\includegraphics[height=0.325\textheight]{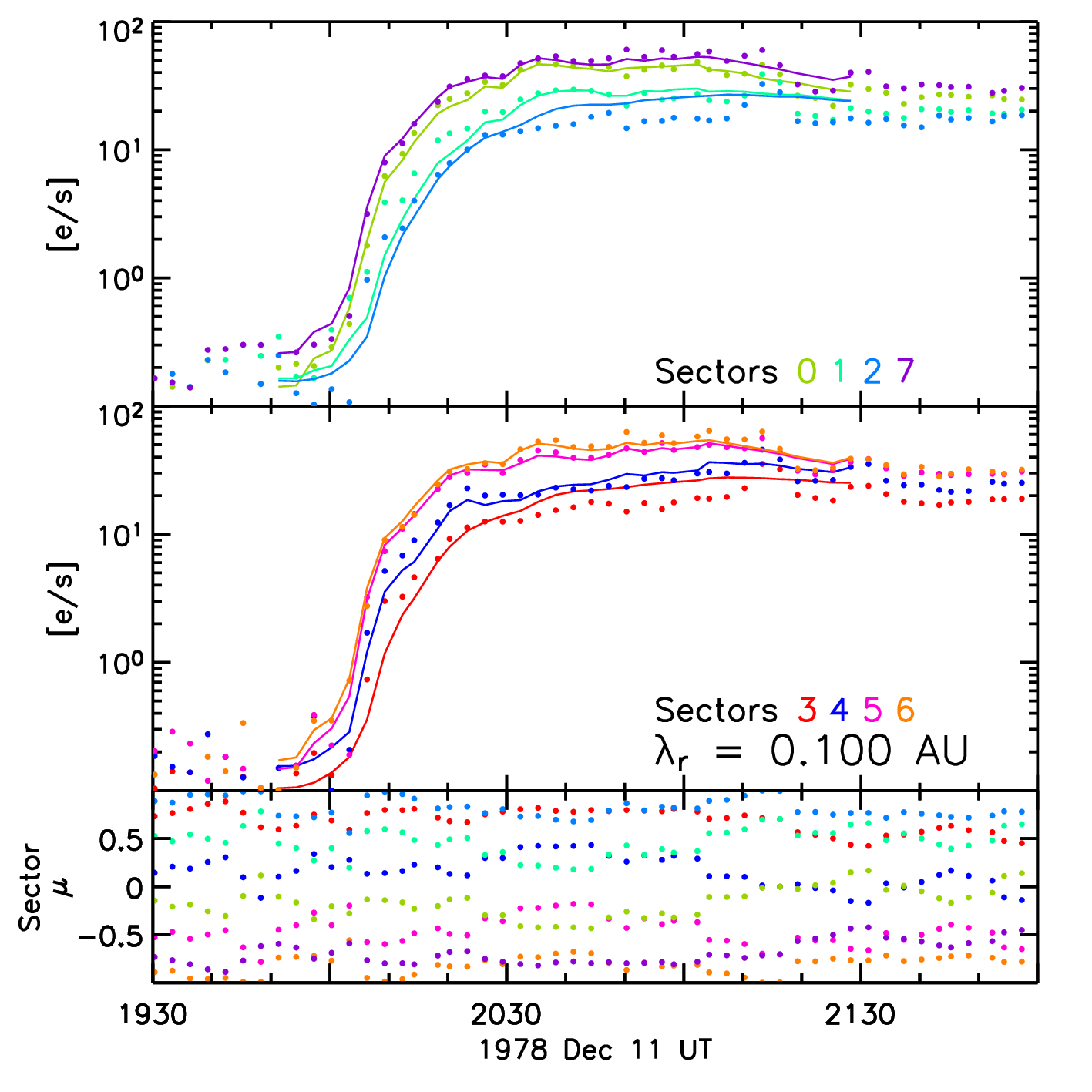}
\includegraphics[height=0.325\textheight]{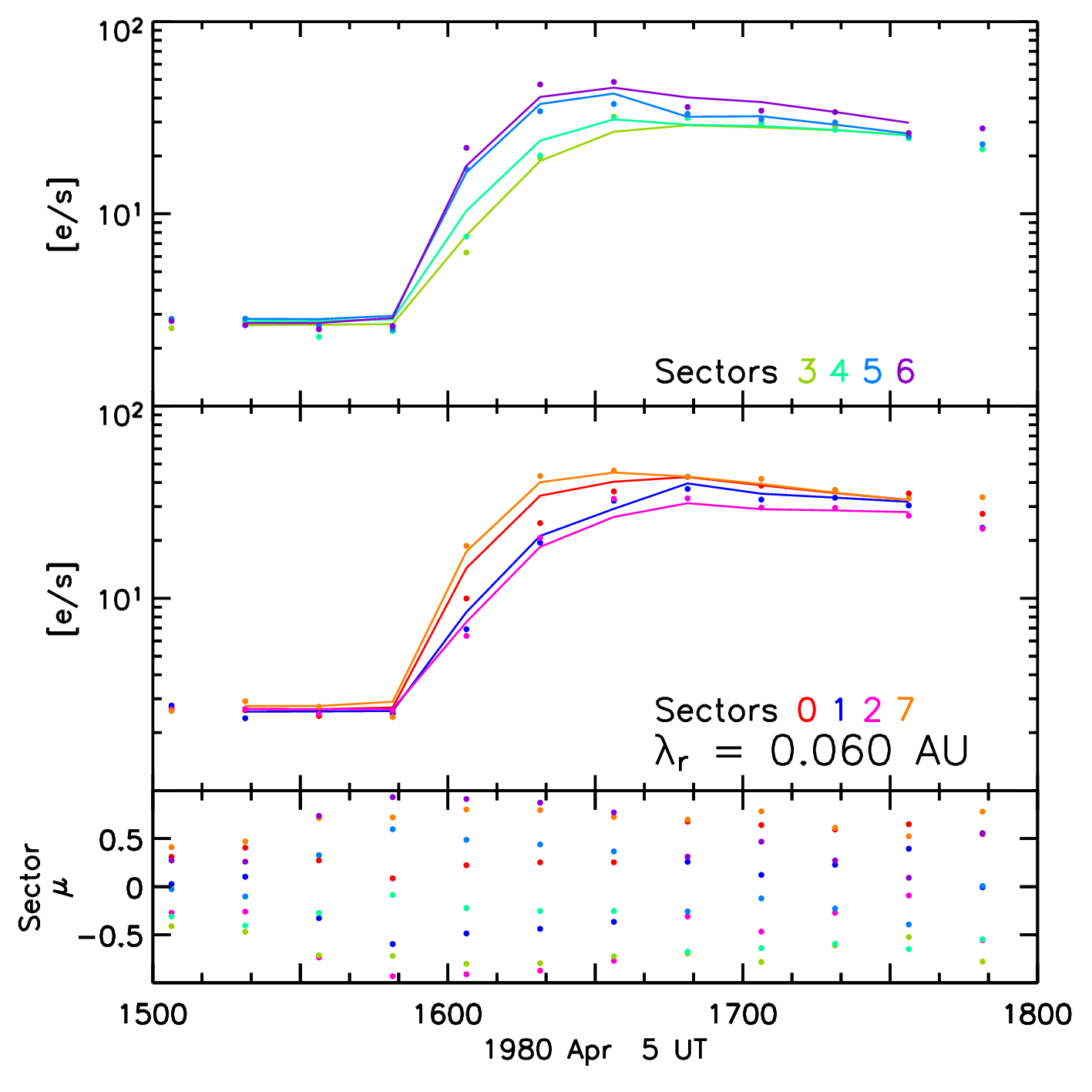}
\includegraphics[height=0.325\textheight]{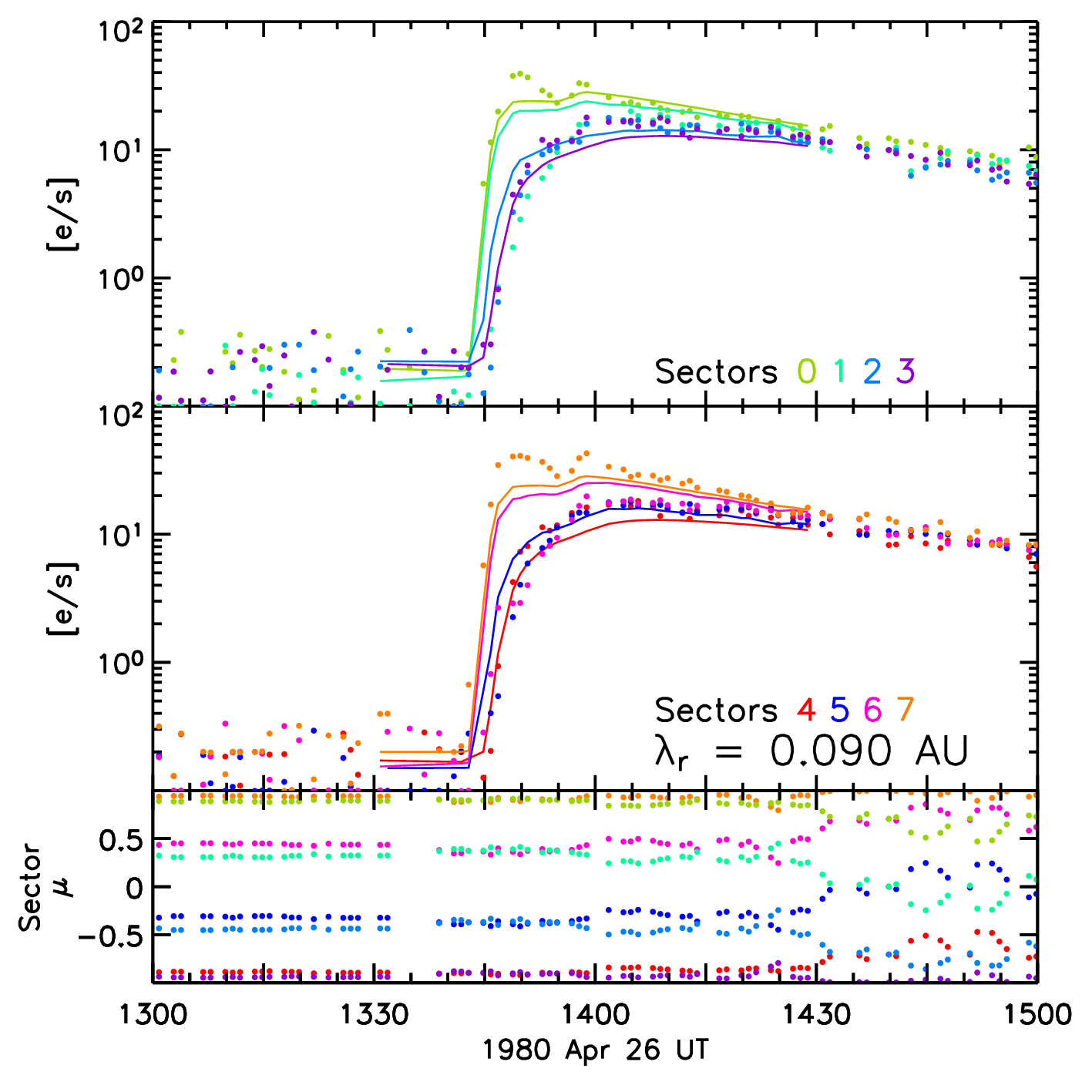}
\includegraphics[height=0.325\textheight]{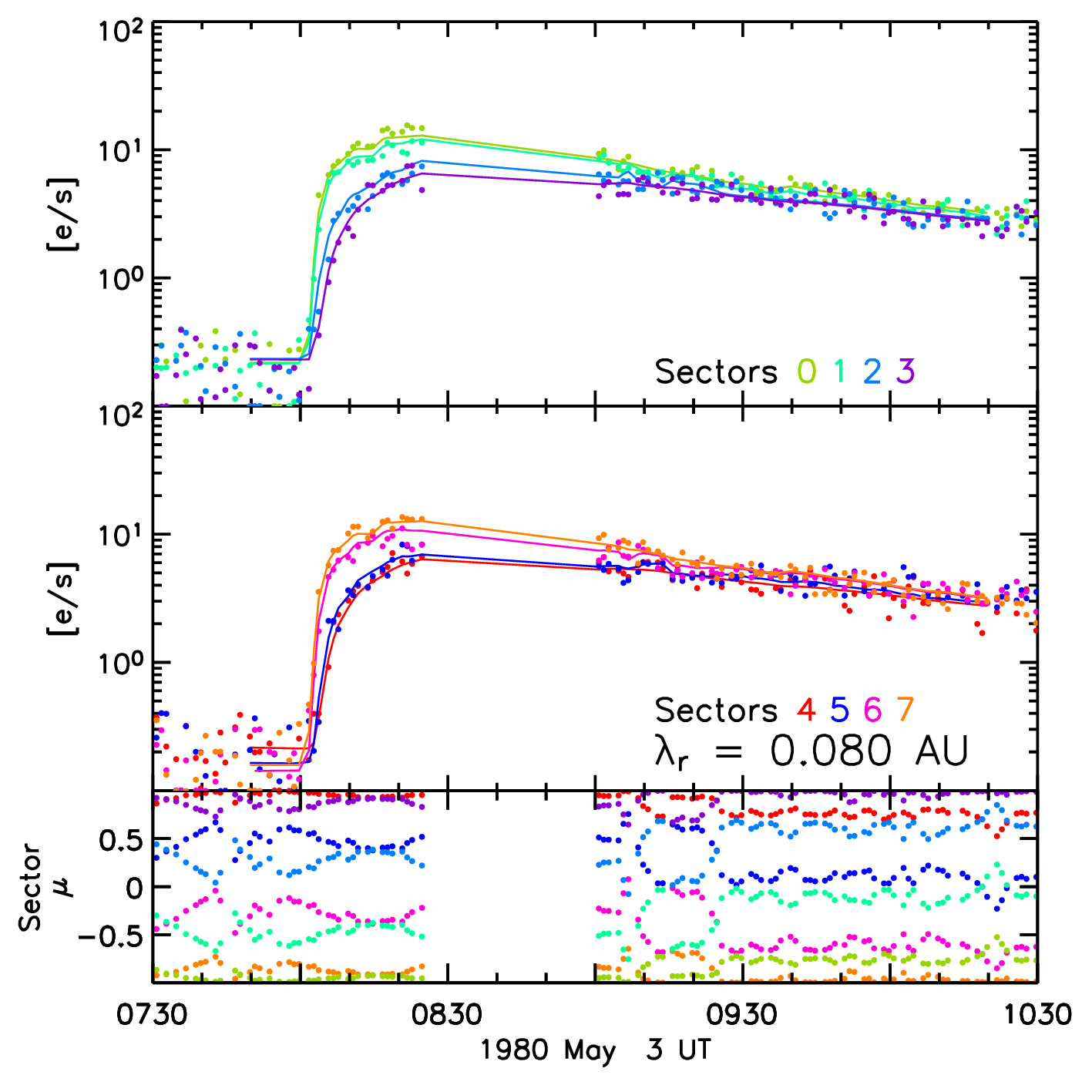}
\par\end{centering}
\caption{Solar relativistic electron events observed by Helios. Top two panels: observational sectored data (dots) and model predictions (coloured curves). Bottom panel: electron pitch-angle cosine observed by the midpoint clock-angle of each sector with the same colour code.}
\label{5fig:Results_allfits1}
\end{figure*}

\begin{figure*}
\begin{centering}
\includegraphics[height=0.325\textheight]{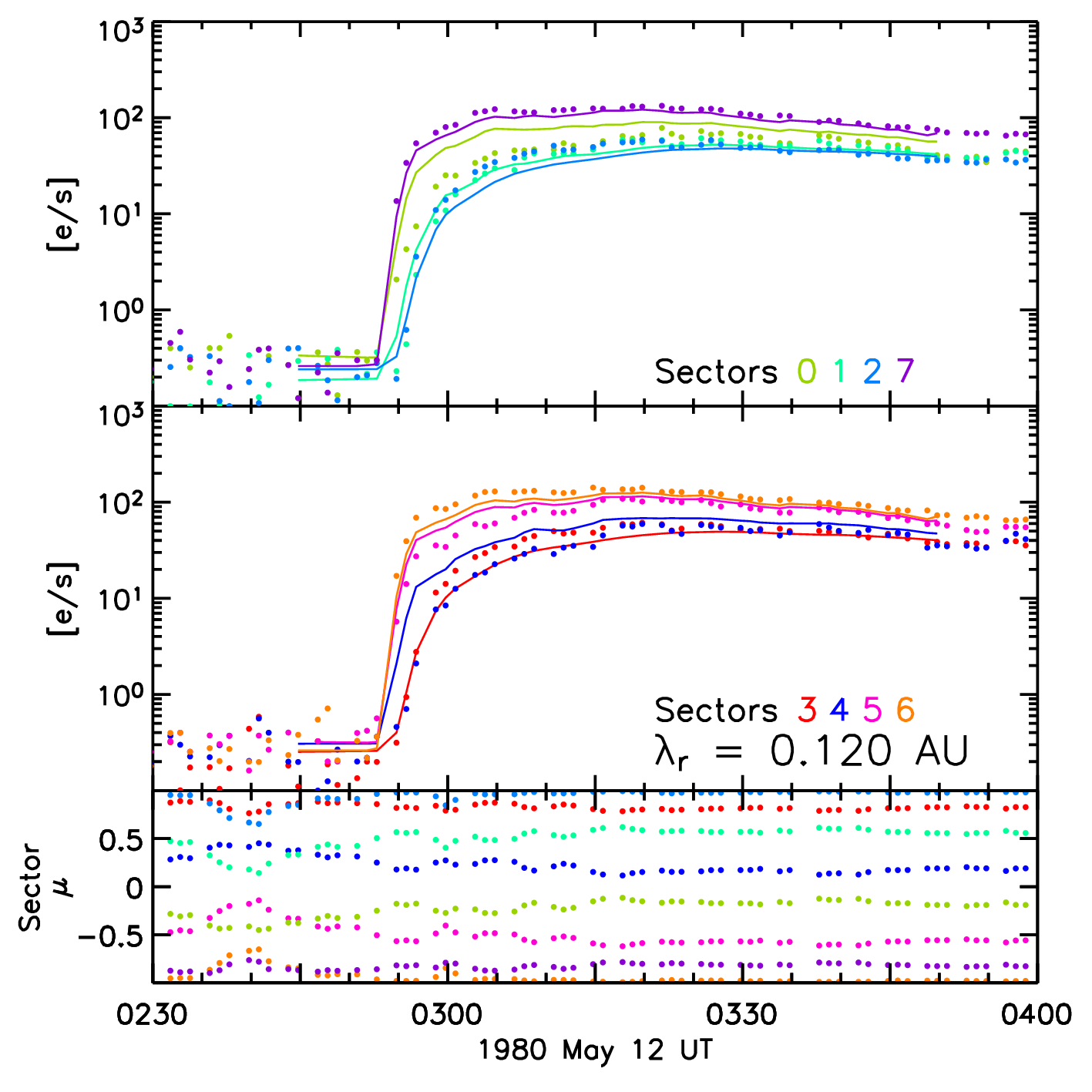}
\includegraphics[height=0.325\textheight]{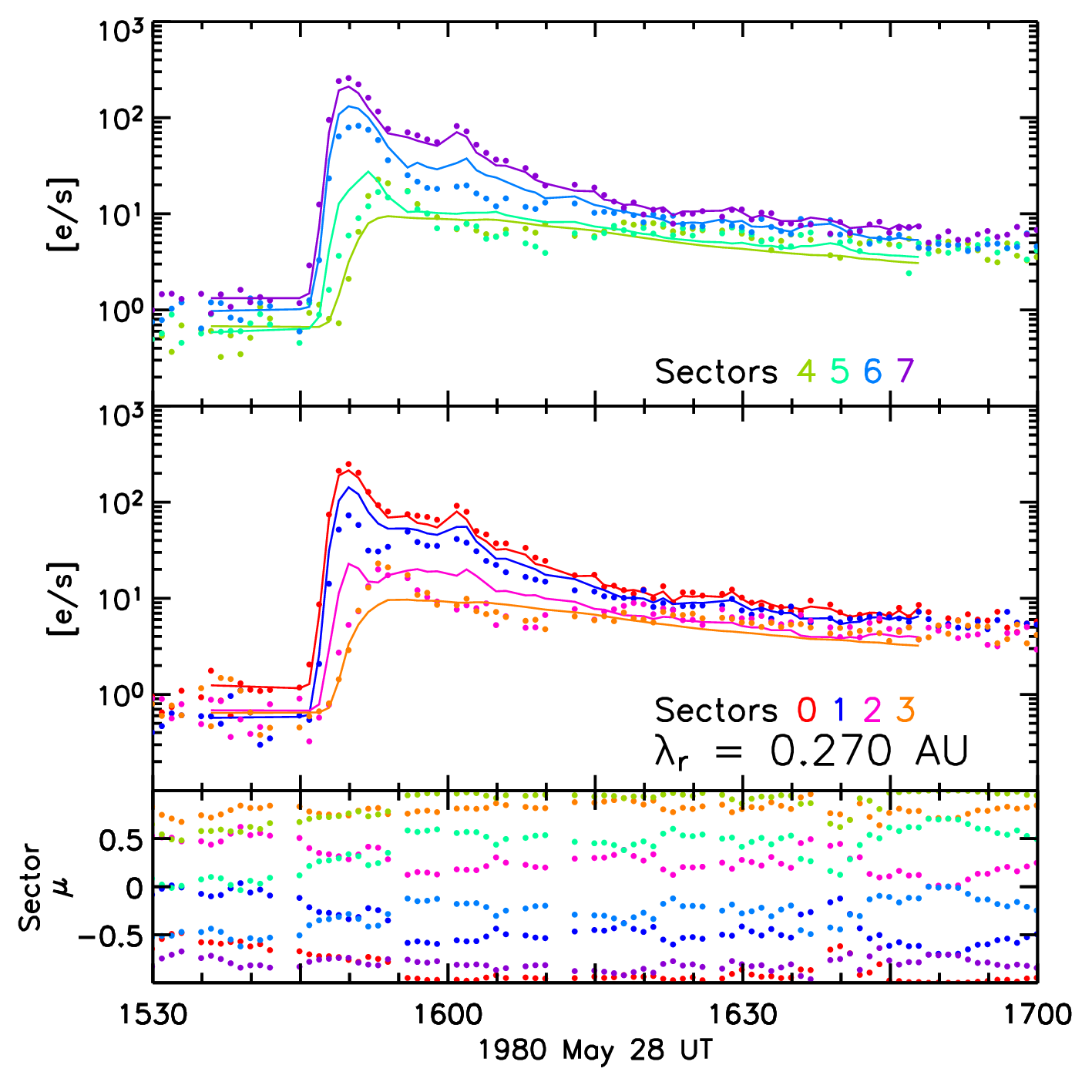}
\includegraphics[height=0.325\textheight]{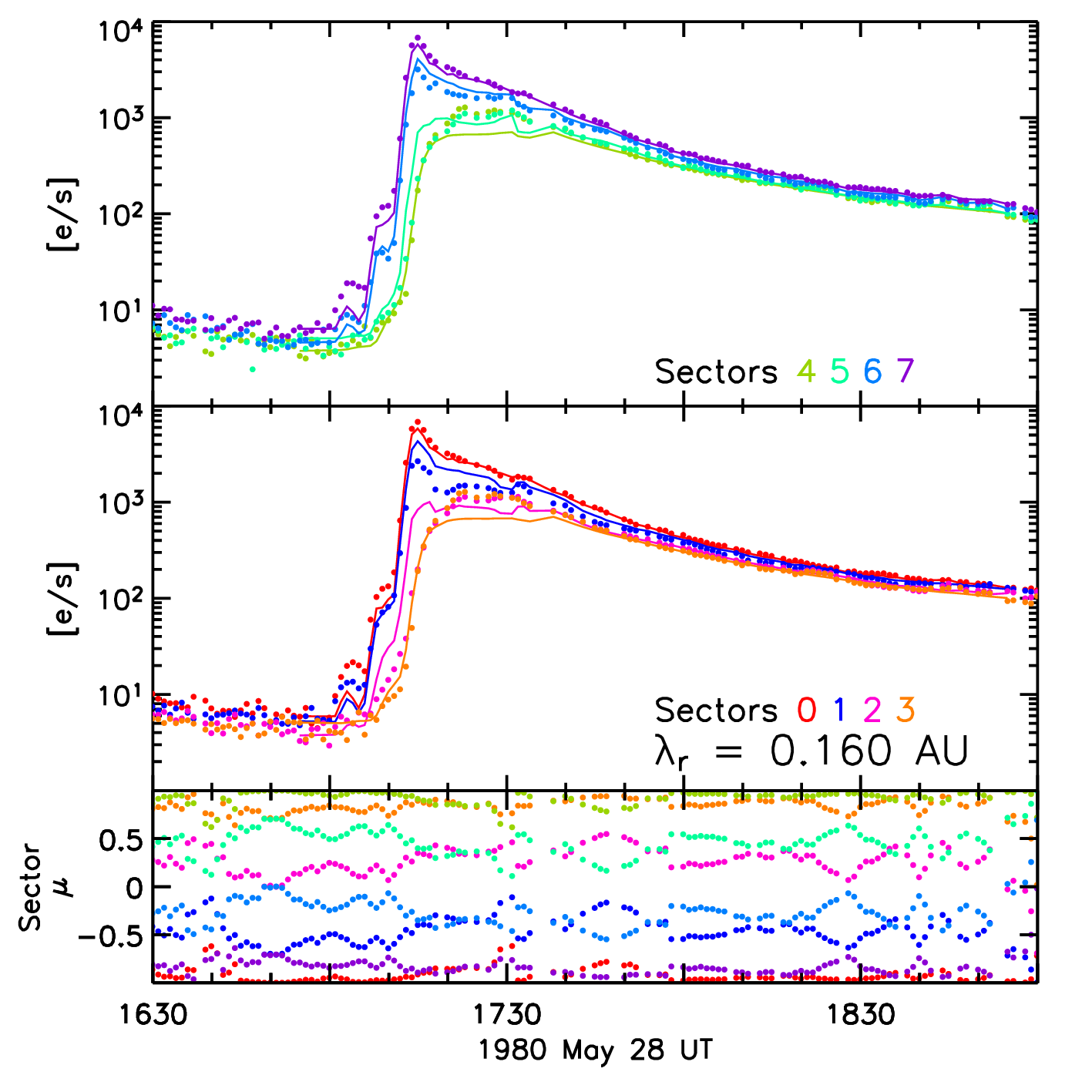}
\includegraphics[height=0.325\textheight]{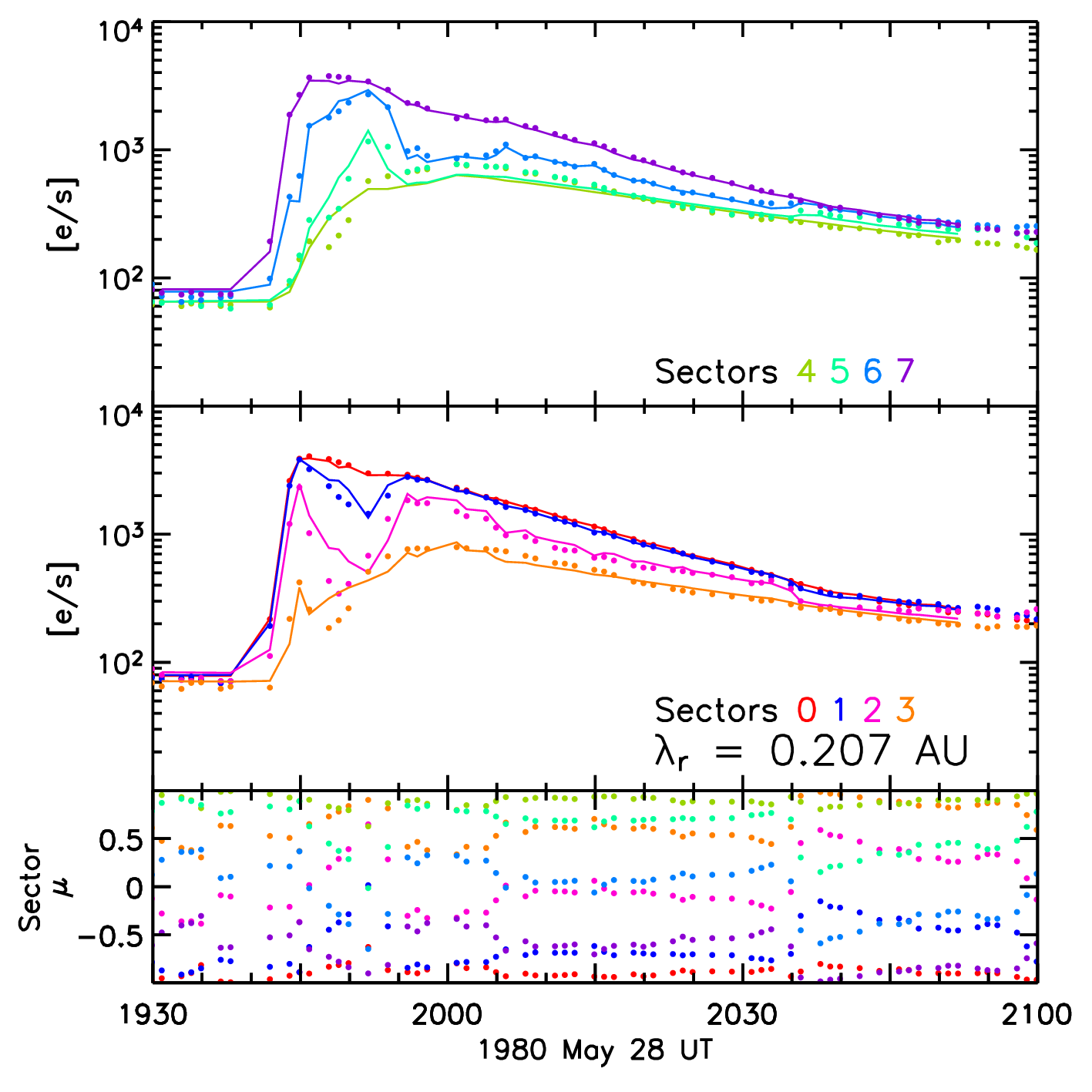}
\includegraphics[height=0.325\textheight]{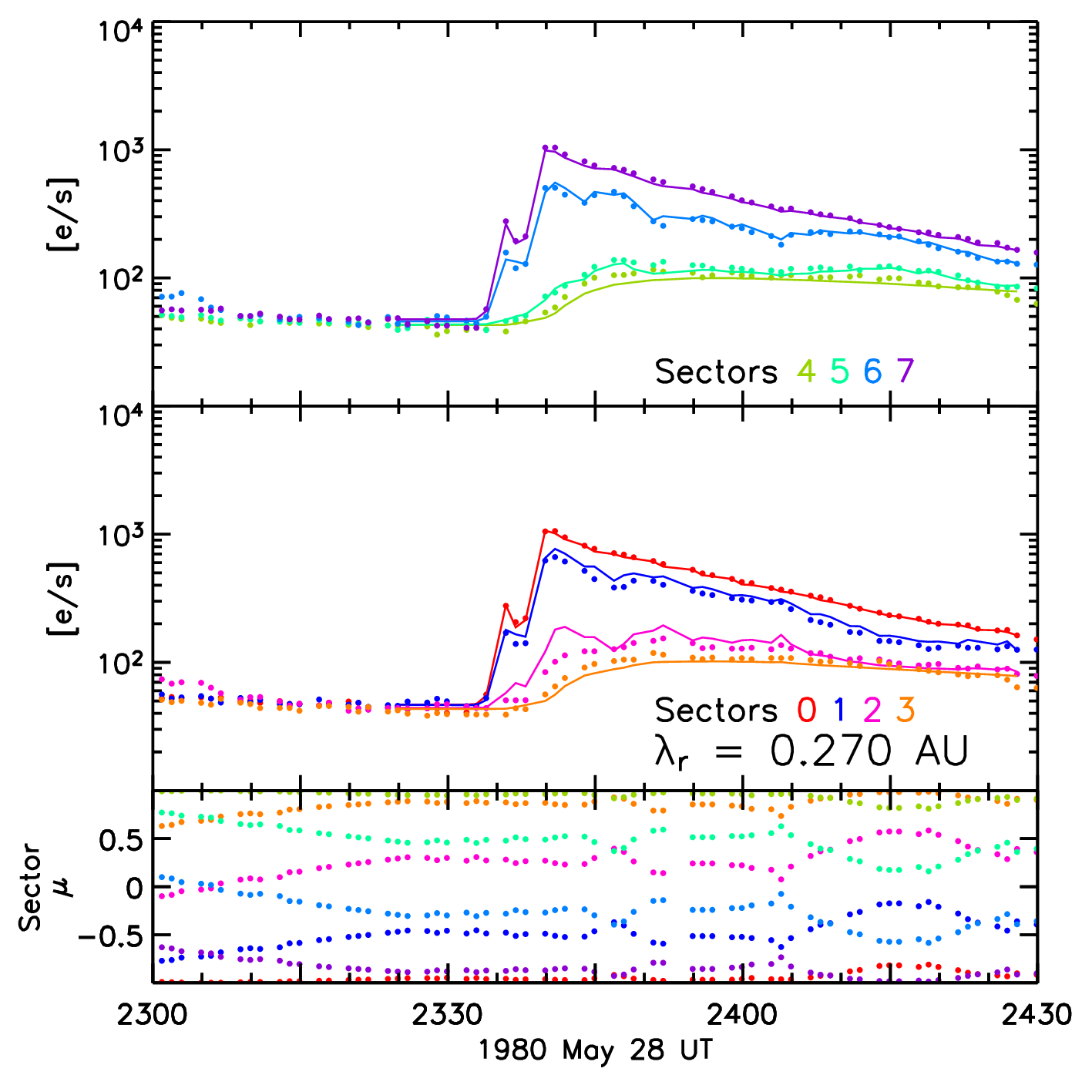}
\includegraphics[height=0.325\textheight]{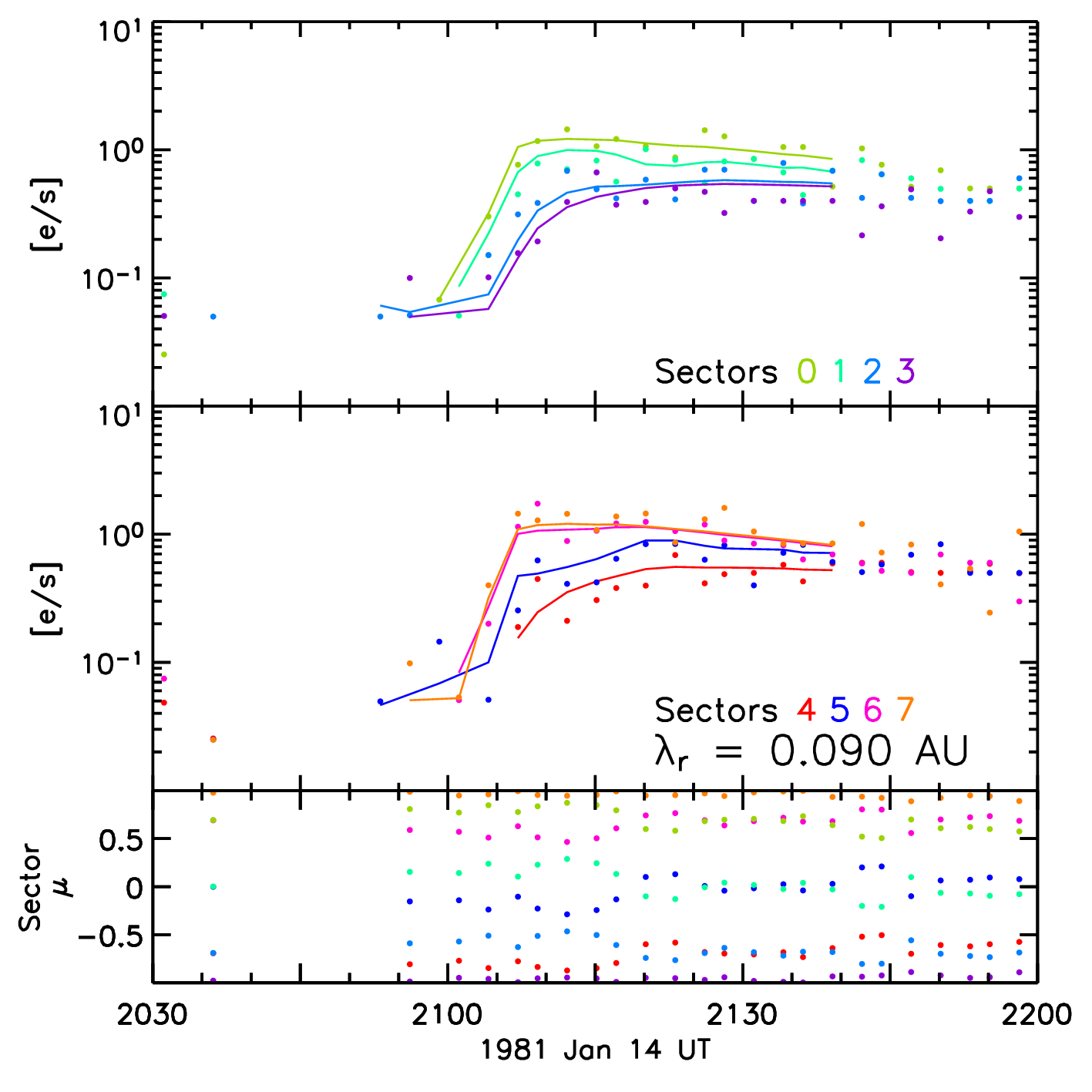}
\par\end{centering}
\caption{See caption in Figure~\ref{5fig:Results_allfits1} for details.}
\label{5fig:Results_allfits2}
\end{figure*}

\begin{figure*}
\begin{centering}
\includegraphics[height=0.325\textheight]{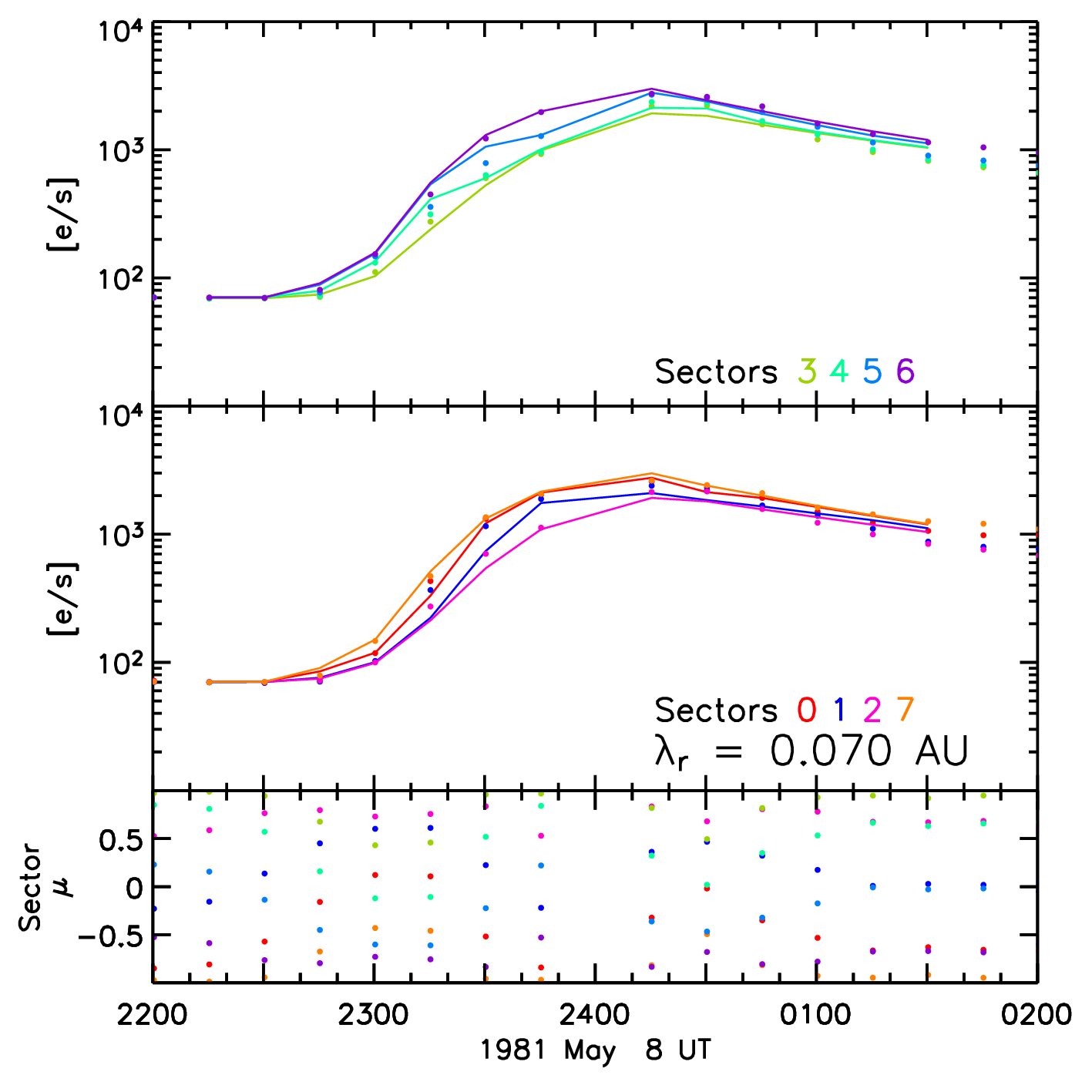}
\includegraphics[height=0.325\textheight]{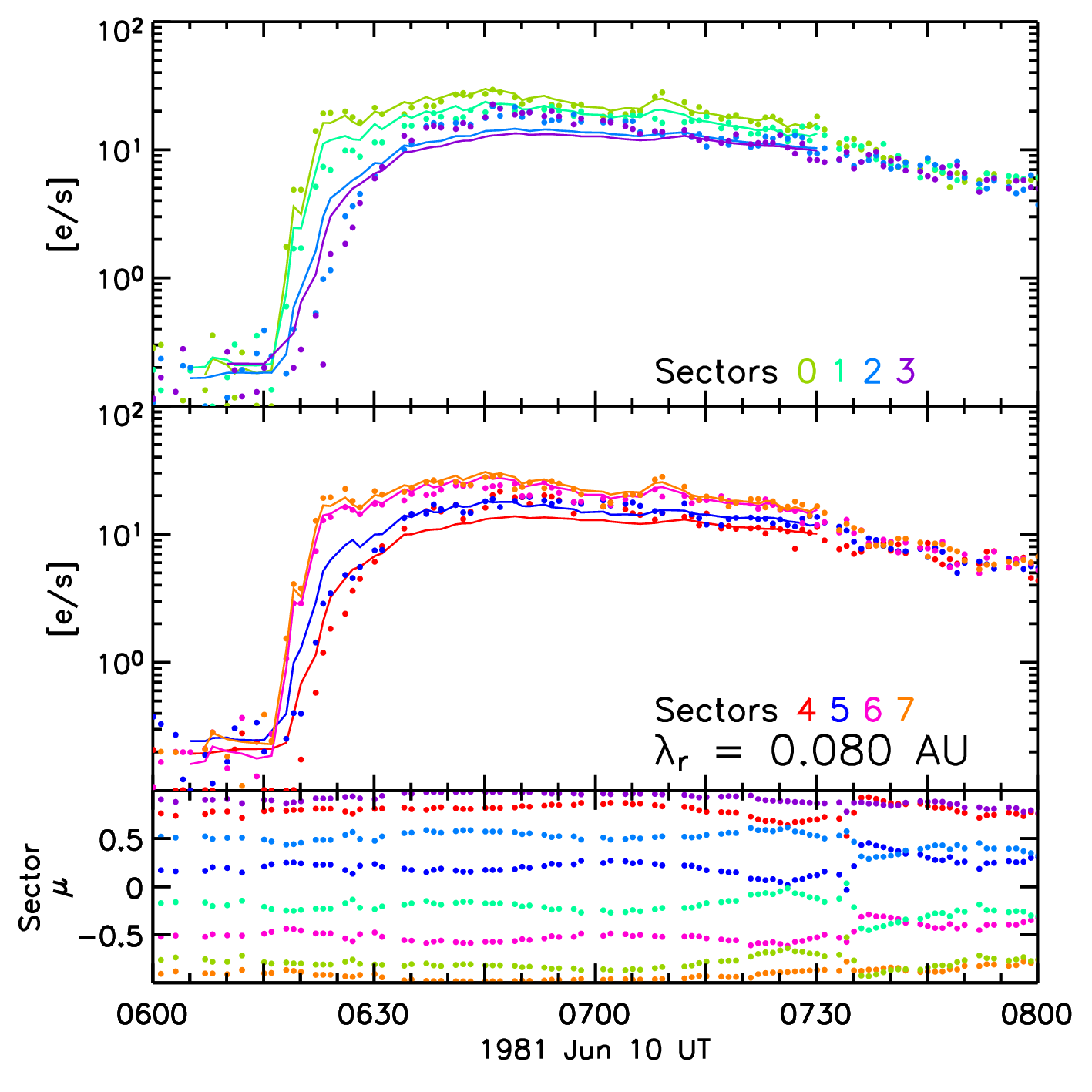}
\par\end{centering}
\caption{See caption in Figure~\ref{5fig:Results_allfits1} for details.}
\label{5fig:Results_allfits3}
\end{figure*}

\end{appendix}

\end{document}